\documentclass[11pt,a4paper]{article}

\usepackage[a4paper,textwidth=6.55in,top=0.95in,bottom=1in]{geometry}
\usepackage[T1]{fontenc}
\usepackage{lmodern}
\usepackage{etoolbox}
\usepackage{amsmath,amssymb,mathtools,amsthm}
\usepackage{mathrsfs}

\usepackage{array}
\usepackage{graphicx}
\usepackage{float}
\usepackage{needspace}
\usepackage{tikz}
\usetikzlibrary{arrows.meta,positioning,calc,fit}
\usepackage{xcolor}
\usepackage{microtype}
\usepackage{cite}
\usepackage[hidelinks]{hyperref}

\allowdisplaybreaks[2]
\numberwithin{equation}{section}

\newcommand{\Tr}{\operatorname{Tr}}
\newcommand{\rank}{\operatorname{rank}}
\newcommand{\ket}[1]{|#1\rangle}
\newcommand{\bra}[1]{\langle#1|}

\newcommand{\papertitle}{Microscopic Organization of BPS Black-Hole States\texorpdfstring{\\[0.1em]}{ }in the D1--D5--P CFT}
\title{\papertitle}
\hypersetup{
  pdftitle={\papertitle},
  pdfauthor={Ji-Seong Chae}
}

\begin{document}
\begin{center}
{\LARGE \bfseries \papertitle\par}
\vspace{1.2cm}
{\large\bfseries Ji-Seong Chae\footnote{jiseongchae17@gmail.com}\par}
\vspace{0.2cm}
{\em Department of Physics, Hanyang University, Seoul 04763, Korea\par}
\end{center}

\begin{abstract}
We study BPS states in the D1--D5 CFT through their fortuity, cycle
structure, and probe response.  Explicit fortuitous states with both long-
and short-string geometries show that the cycle partition does not determine
fortuity state by state.  At large $N$, fixed-charge counting nevertheless
selects a long string.  In the open uncondensed region, an entropy-carrying
exact non-graviton subspace concentrates on one cycle of winding $N-O(1)$.
Assuming subexponential monotone dimension, the same holds for the harmonic
fortuitous complement.  Beyond the critical line, the free ensemble develops
a short-string condensate.  To test how this cycle structure appears in
observables, we study finite-winding probes.  The long-cycle scalar correlator approaches the extremal-BTZ response at fixed time, while finite winding is resolved at \(t=O(\sqrt N)\), the exact circle recurs at \(t=O(N)\), and states with the same cycle geometry remain distinguishable by their spectral weights.
\end{abstract}

\clearpage
\tableofcontents
\clearpage

\section{Introduction}
\label{sec:introduction}

The D1--D5--P system provides a microscopic account of black-hole entropy.
A bound state of branes reproduces the
Bekenstein--Hawking entropy of a five-dimensional rotating black hole
\cite{StromingerVafa,BMPV}.  A long effective string carrying fractionated
momentum accounts for that entropy and for near-extremal radiation
\cite{MaldacenaSusskind,MaldacenaStromingerGreybody,HassanWadiaEffectiveString,MartinecLongStrings},
and the near-horizon decoupling limit turns the counting problem into a
question about a two-dimensional conformal field theory
\cite{Maldacena:1997re,DavidMandalWadiaReview}.  Logarithmic corrections to
the same entropy are reproduced by a one-loop computation in the
near-horizon geometry \cite{SenLogBMPV}.

That a long string suffices to carry the entropy is not the same statement
as that the entropy-carrying states are long strings.  At fixed total
charges, the symmetric-orbifold Hilbert space decomposes into sectors
labelled by cycle partitions.  The leading Cardy estimate assigns the same
entropy to these partitions when momentum and spin are shared in proportion
to winding; it does not establish equality of their exact multiplicities.
Whatever selects a long string must therefore distinguish the partitions
beyond this leading estimate.  We determine that distinction, its change
at short-string condensation, and the part that survives the marginal
deformation.

The orbifold point is a distinguished locus of the D1--D5 moduli space
\cite{Seiberg:1999xz,Larsen:1999uk}, and its tensionless-string
interpretation has been developed in detail
\cite{Eberhardt:2018ouy,Eberhardt:2019ywk,EberhardtPartitionFunctions};
symmetric-product universality and its limits organize what such a free
description can and cannot reproduce
\cite{Benjamin:2015vkc,BelinBintanjaCastroKnop}.  Moving away from that
locus lifts states that are BPS only at the orbifold point, and conformal perturbation theory
has been developed to the point where the lifting of large classes of states
can be computed
\cite{Keller:2019suk,Benjamin:2021zkn,Guo:2019pzk,GuoMathurLevelOne,GuoMathurHigherLevels,Guo:2022ifr,Hughes:2023apl,Hughes:2023fot},
including systematic treatments of the deformed symmetric orbifold and its
anomalous dimensions
\cite{Fiset:2022erp,GaberdielGopakumarNairz,Frolov:2023pjw,Gaberdiel:2024nge,Gaberdiel:2025smz,Gaberdiel:2026jor}.

On the gravity side, the same charges support a large family of smooth
horizonless solutions.  Multiply wound strings and their Lunin--Mathur
geometries
\cite{Lunin:2001fv,Lunin:2001jy,Lunin:2002iz,Kanitscheider:2007wq,BalasubramanianKrausShigemori},
superstrata and their generalizations
\cite{Bena:2015bea,Bena:2016ypk,Bena:2017xbt,Shigemori:2020yuo,Bena:2022rna,BenaHeidmannMontenWarner},
and the supergravity counting of such configurations
\cite{Rychkov:2005ji,Krishnan:2015vha,Shigemori:2019orj,MayersonShigemori}
provide concrete reference families for microscopic comparisons.  Their
holographic interpretation uses orbifold correlators and BPS lifting
methods
\cite{LuninMathurOrbifold,LuninMathurThreePoint,Avery,Burrington:2015mfa,GavaNarain}
and precision holography for the corresponding states
\cite{Baggio:2012rr,Kanitscheider:2006zf,Giusto:2015dfa,Giusto:2019qig,Rawash:2021pik,Giusto:2018ovt,Ceplak:2021wzCz}.
The elementary process that changes a cycle partition is the twist
deformation, whose action has been computed explicitly
\cite{CarsonHamptonMathurTurtonTwist,CarsonHamptonMathurTurton,Carson:2015ohj,Carson:2016uwf,CarsonMathurTurtonBogoliubov},
while its continuum interpretation raises the separate question of
unitary implementation of a bosonic Bogoliubov transformation
\cite{ShaleBoson}.

For BPS classes that survive at a given rank, the fortuity program asks a
different question.  Can they continue to higher ranks?  A class is monotone
if it continues to all higher ranks and fortuitous if it does not.  This
distinction grew out of studies of $\mathcal N=4$ super Yang--Mills
\cite{Chang:2013fba,ChangLinWords,ChangLinFortuity,Choi:2022caq,ChoiQuantumMicrostates,FollowingBlackHoleStates}
and has since been extended to several other settings
\cite{Chang:2024lxt,Belin:2025hsg,BehanPipoloABJM,deMelloKoch:2025ngs,deMelloKoch:2025cec,Chen:2025sum,deMelloKochRajaRudra}.
For the D1--D5 system the first-order supercharge cohomology and the
higher-rank continuation criterion have been constructed explicitly
\cite{ChangLinZhang,ChangZhang,GiustoInglisRusso}, and the supergravity
sector inside that cohomology has been identified and enlarged
\cite{HughesShigemori,ToweringGravitons}.  In parallel, the statistics of
the surviving spectrum has been probed through chaos and free-probability
diagnostics
\cite{Chen:2024oqv,Zhang:2026jnf,BelinBPSChiral,BelinFuLaRocca}.
Fortuity concerns continuation in the rank; the cycle partition is a
different microscopic datum, and the two need not determine one another.
Complementary collective constructions implement finite-$N$ relations in
invariant variables through null-state reductions and operator-algebra
representations
\cite{CollectiveFiniteNReduction,BilocalFiniteNHilbert,BilocalFiniteNAlgebra}.

Our individual-state tests show that cycle geometry does not determine
fortuity.  The continuation tests for the stated charge sectors produce
equal-charge pairs with the same partition and opposite continuation
properties, as well as fortuitous families built entirely from short
cycles.  The ensemble nevertheless selects a long string.  In the open
uncondensed region, the single-cycle contribution at the fixed-charge
saddle has a winding-dependent prefactor proportional to $w^{-3}$, rather
than the $1/w$ weight of a uniformly random permutation
\cite{FordCycleType}.  The resulting distribution lies in the convergent
regime of weighted partitions \cite{StuflerConvergent,StuflerUnlabelled}.
The total winding outside the longest cycle remains bounded in probability
as $N$ grows.  This is a statement about normalized multiplicities, not a
constraint on every state.

The cycle distribution changes when the aligned singly wound Ramond ground
states condense.  Their occupation is of order $\sqrt N$ on the critical
line and of order $N$ in the condensed interior.  The latter reproduces
the enigmatic phase of the free orbifold studied alongside competing
supergravity configurations in Ref.~\cite{BenaMoulting}.  The excited long
string carries the leading entropy and has charges on the critical line;
the condensate carries the excess winding and spin.  The entropy and its
first derivatives agree across the transition, but its second derivative
jumps.  The modified elliptic genus and its refinements retain protected
information despite the fermion zero modes of $T^4$
\cite{MaldacenaMooreStrominger,MaldacenaStromingerExclusion,DMVV,deBoer:1998us,Benjamin:2016pil,Hughes:2025oxu,Hughes:2026qqn}.
The modified-index seed cancels the double pole associated with the two
aligned bosonic occupations, while the protected asymptotic retains the
BMPV exponent.  The standard theta-function transformation
\cite{NISTTheta} supplies the modular input to this comparison.

In the uncondensed region, the protected lower bound and the free count
have the same exponential and polynomial scale.  They imply long-string
concentration for an exact BPS subspace transported regularly back to the
orbifold point, but do not fix its full distribution over cycle
partitions.  The exact non-graviton subspace retains both the black-hole
entropy scale and this concentration after generalized-gravity directions
are excluded.  Under a subexponential bound on the monotone dimension, the
harmonic fortuitous complement carries the same entropy and concentration.
The opposite conclusion for the entire monotone sector requires the
stronger assumption on its representatives stated below.  Thus the
explicit tower result, the dimension assumption, and the additional bound
on cycle support have distinct roles.

The selected cycle lengths determine the finite circles on which
interactions and probes propagate.  The horizon-mirage
problem asks how a finite-volume, horizonless microscopic state can
reproduce a BTZ-like correlator over an intermediate time range, and when
finite-circle recurrences or state-dependent spectral weights reveal the
discrete dynamics behind the apparent horizon \cite{BelinBintanjaCastroKnop}. The interpretation is also informed by
work on emergent horizons and emergent time
\cite{LeutheusserLiuEmergent,LeutheusserLiuCausal,GesteauLiuStringy}, by
low-temperature modes that become soft in near-extremal and near-BPS
AdS$_3$ backgrounds \cite{BacCastroJainNearBPS}, and by the separation
between finite-circle recurrences and much longer many-body time scales
\cite{BarbonRabinovici}.  We do not assume a long cycle for the probe; the
fixed-charge count selects it first.  Keeping the short winding exact while
the other cycle becomes long, we find that both transmission and pair
creation retain the finite winding at leading continuum order.

For the neutral scalar probe, a single insertion takes a right-BPS state
out of the BPS subspace.  On a cycle of length $n$, a fixed covering mode
violates the right-moving BPS bound by $m/n$, so the long-cycle correlator
samples a parametrically dense near-BPS tower of intermediate states.  The
exact finite-circle current block separates a BTZ-like term, a constant
correction proportional to the cycle count $K$, and recurrence images.
In the free fixed-charge ensemble, the short-string-condensation
phase structure consists of the open uncondensed region, the critical line
at its boundary, and the condensed interior.  We find that the cycle count scales
as $K=O_{\Pr}(1)$, $O_{\Pr}(\sqrt N)$, and $O(N)$ in these three regimes,
respectively.  The corresponding chiral BTZ-like window scales as $\log N$,
$\tfrac12\log N$, and $O(1)$.

For the physical orbifold-invariant scalar in the open uncondensed region,
the long-cycle response approaches the extremal-BTZ reference at fixed
time.  The finite-winding correction becomes comparable to the BTZ-like
tail at $t=O(\sqrt N)$, while the exact circle recurs at $t=O(N)$.  These
scales quantify when this probe begins to resolve the microscopic circle
hidden by the strict large-$N$ response, well before the many-body
Heisenberg scale.  Matched BPS states with the same charges, cycle geometry,
mode lattice and recurrence period can nevertheless have different scalar
correlators because their oscillator spectral weights differ.  Thus the
probe resolves the horizon mirage in stages, from the BTZ-like continuum
to the finite circle and then to state-dependent microscopic data, without
turning the leading correlator into a sufficient criterion for a sharp
horizon.

Section~\ref{sec:framework} reviews the orbifold, the continuation tests,
the counting functions, the effective long-string picture and the
finite-circle probe.  Section~\ref{sec:microscopic-fortuity} presents the
individual-state tests.  Section~\ref{sec:long-strings-lifting} derives the
fixed-charge selection law, follows it across the condensation boundary, and
transfers it to the exact and non-graviton BPS sectors.
Section~\ref{sec:twist-main} derives the finite-winding twist kernels,
identifies the time scales at which finite-circle effects resolve the
BTZ-like probe response, and compares matched BPS states beyond their common
cycle geometry.  Section~\ref{sec:discussion-and-open-problems} collects the
resulting picture and the open problems.

\section{The D1--D5 CFT and BPS lifting}
\label{sec:framework}

We use the free-field and cohomological conventions of
Ref.~\cite{ChangLinZhang}; the oscillator and permutation-orbit
normalizations are collected in Appendix~\ref{app:finite-rank-calculations}.
An \emph{oscillator word} is a product of creation modes acting on specified
component-string ground states.

\subsection{Orbifold states and spectral flow}
\label{sec:review-orbifold-bps}
\label{sec:review-BPS-sector}

Consider $N_1$ D1-branes on a circle and $N_5$ D5-branes on that circle
times $T^4$. Their low-energy D1--D5 CFT has
$\mathcal N=(4,4)$ supersymmetry and $c=6N$, with $N=N_1N_5$.
At the symmetric-orbifold point, with $S_N$ the permutation group of the $N$ copies,
\begin{equation}
\begin{aligned}
 \operatorname{Sym}^N(T^4)&=\frac{(T^4)^N}{S_N}, \\
 c&=6N.
\end{aligned}
 \label{eq:orbifold}
\end{equation}
One copy contains four real bosons $X^A$, $A=1,\ldots,4$, four left
fermions and four right fermions. We denote the left fermions by
$\psi^\pm,\bar\psi^\pm$ and the right fermions by
$\widetilde\psi^\pm,\widetilde{\bar\psi}{}^\pm$. The bar on $\bar\psi^\pm$ labels internal polarizations and tilde distinguishes right movers. On one NS copy, bosonic creation modes
have positive integer energies and fermionic creation modes have
positive half-integer energies. Acting with them on the vacuum gives a
Fock basis. We use $\alpha^i,\bar\alpha^i$, $i=1,2$, for the two complex
pairs of left bosonic oscillators and $|1\rangle$ for the one-copy NS
vacuum. For example, $\alpha^i_{-1}|1\rangle$ and
$\psi^+_{-1/2}|1\rangle$ are different one-copy excitations, of left
energies $1$ and $1/2$.

The quotient in Eq.~\eqref{eq:orbifold} has two effects. It retains
permutation-invariant combinations of the ordinary tensor-product
states, and it introduces sectors with permutation-twisted boundary
conditions. A conjugacy class is specified by a partition
$p=(1^{n_1}2^{n_2}\cdots)$ with $\sum_w wn_w=N$; here $n_w$ counts length-$w$ cycles. For a particular
$w$-cycle the permutation part of the boundary condition is
\begin{align}
 X^{[j]}(e^{2\pi i}z)&=X^{[j+1]}(z), \nonumber \\
 \psi^{[j]}(e^{2\pi i}z)&=\psi^{[j+1]}(z),\qquad j\sim j+w,
\end{align}

with the fermionic sign determined by the NS or Ramond spin structure; in the NS sector this combines with the \(w\)-cycle monodromy to give the parity-dependent mode lattice in Eq.~\eqref{eq:setup-fermion-mode-lattices}. A circuit
passes from copy $j$ to copy $j+1$, and $w$ circuits return to the
starting field. Joining those $w$ segments produces a single field on a
circle of circumference $2\pi wR_y$, where $R_y$ is the radius of the
common D1--D5 circle. This is the component string; its
winding $w$ counts joined copies, not oscillator quanta.
To obtain its modes, diagonalize the cyclic permutation. Here $\mathbb Z_w$ is the
cyclic group generated by one-step rotation of the length-$w$ cycle, and
$\ell=0,\ldots,w-1$ labels its character sector. For a bosonic current,

\begin{align}
 \partial X^{(\ell)}(z)&=\frac1{\sqrt w}
 \sum_{j=0}^{w-1}e^{-2\pi i\ell j/w}\partial X^{[j]}(z), \nonumber \\
 \partial X^{(\ell)}(e^{2\pi i}z)
       &=e^{2\pi i\ell/w}\partial X^{(\ell)}(z).
\end{align}

The Laurent powers must reproduce this phase, so

\begin{equation}
 \partial X^{(\ell)}(z)
   =\sum_{r\in\mathbb Z-\ell/w}\alpha_r^{(\ell)}z^{-r-1}.
\end{equation}
Combining all $\ell$ gives energies spaced by $1/w$. In particular,
$\alpha_{-1/3}$ creates energy $1/3$ on a three-cycle, whereas
$\alpha_{-1}$ creates energy $1$ on a singly wound strand. Fermionic modes have
the same fractional spacing with a spin-structure shift. Relative to
the bare NS twisted vacuum, their mode sets are
\begin{equation}
 \mathcal R_w=
 \begin{cases}
  (\mathbb Z+\tfrac12)/w,&w\text{ odd},\\
  \mathbb Z/w,&w\text{ even}.
 \end{cases} \label{eq:setup-fermion-mode-lattices}
\end{equation}

For a fermion in the $\ell$ sector, the allowed modes satisfy

\begin{equation}
 r\in\mathbb Z+\frac12-\frac{\ell}{w},\qquad \ell=0,\ldots,w-1.
\end{equation}
For even $w$, choosing $\ell=w/2$ cancels the NS half-integer shift.
Then $r\in\mathbb Z$, so $r=0$ is allowed despite the underlying NS
spin structure.

A fixed permutation factorizes into independent cycles. Let $\mathscr H_{(w)}$
denote the decorated Hilbert space of one length-$w$ cycle before its cyclic
projection and let $\mathscr H_p$ be the sector associated with the full
partition $p$. Invariance under rotations of each cycle gives a
$\mathbb Z_w$ projection, while exchange of $n_w$ identical cycles gives
the graded symmetric power $S^{n_w}$:
\begin{align}
 \mathscr H_p&\cong\bigotimes_{w\ge1}
 S^{n_w}\!\left(\mathscr H_{(w)}^{\mathbb Z_w}\right), \nonumber \\
 \sum_w wn_w&=N.
\end{align}

The grading imposes the fermionic signs as identical decorated cycles
are exchanged. It is separate from level matching on each individual
cycle \cite{DMVV,DavidMandalWadiaReview,Avery,ChangLinZhang}.
For example, $N=3$ contains

\begin{equation}
 \mathscr H_3=
 S^3(\mathscr H_{(1)})\ \oplus\
 \bigl(\mathscr H_{(2)}^{\mathbb Z_2}\otimes\mathscr H_{(1)}\bigr)
 \ \oplus\ \mathscr H_{(3)}^{\mathbb Z_3}.
\end{equation}
Thus $|u_{(2)}\rangle\otimes|v_{(1)}\rangle$ has shape $(2,1)$,
whereas $|u_{(3)}\rangle$ has shape $(3)$. Specifying a shape still
leaves all allowed oscillator occupations to be specified.

In the oscillator algebra below, $A,B$ and $\alpha,\beta$ are orthonormal
bosonic and fermionic polarization labels, respectively. For the oscillator
norms we use
\begin{equation}
\begin{aligned}
 [\alpha_r^A,\alpha_s^{B\dagger}]
 &=r\,\delta^{AB}\delta_{rs}, \\
 \{\psi_r^\alpha,\psi_s^{\beta\dagger}\}
 &=\delta^{\alpha\beta}\delta_{rs},\qquad r,s>0.
\end{aligned}
 \label{eq:review-oscillator-normalization}
\end{equation}
Thus $n$ identical bosons have squared norm $n!r^n$, while distinct
fermion creators contribute unit norm. Equation~\eqref{eq:appendix-fock-word-norm}
gives the general contraction formula.

For a normalized decorated Fock word $f$ on chosen copies, let
$H\subset S_N$ be its stabilizer, including permutations of graded-identical
decorations. When distinct orbit images are orthogonal, the invariant state
is the signed orbit sum and our convention is
\begin{equation}
\begin{aligned}
 |\{f\}\rangle&=
 \frac1{\sqrt{N!\,|H|}}\sum_{\sigma\in S_N}\sigma|f\rangle, \\
 \langle\{f\}|\{f\}\rangle&=1.
\end{aligned}
 \label{eq:setup-orbit-unit-state}
\end{equation}
Eqs.~\eqref{eq:appendix-orbit-sum-norms} and
\eqref{eq:appendix-fock-word-norm} give the orbit and oscillator norms. Before symmetrization, a local word is specified on
chosen copies; afterwards it represents the physical orbifold state.
We suppress the braces for an already normalized single-cycle orbit.
With the physical cycle sectors in place, we can impose the BPS condition
and translate their states between NS and Ramond conventions.

The left symmetry algebra is generated by the stress tensor $T$, four
supercurrents $G^\pm,G'{}^\pm$, three R-currents $J^a$ ($a=1,2,3$), four bosonic
currents $\alpha^i,\bar\alpha^i$ ($i=1,2$), and four free fermions. The right
algebra has the same generators with tildes. The generators of the
orbifold symmetry are sums over copies: for example,
$J^a_n=\sum_{i=1}^N J^{a,[i]}_n$. Their ordinary NS modes should be
distinguished from the fractional modes on a single component string. These total currents generate the contracted large $\mathcal N=(4,4)$
symmetry algebra of the orbifold \cite{ChangLinZhang}.

Let $L_0$ and $\widetilde L_0$ denote the left- and right-moving
Virasoro zero modes. Their eigenvalues $h$ and $\bar h$ are the
conformal weights. Let $J^3_0$ and $\widetilde J^3_0$ denote the
corresponding R-symmetry Cartan generators, with eigenvalues $j$ and
$\bar j$. In the NS sector, choose the global right-moving supercharges
\begin{align}
 \mathcal Q_0&=\widetilde G^+_{-1/2}, \nonumber \\
 \mathcal Q_0^\dagger&=\widetilde G^-_{1/2},
\end{align}
where $\widetilde G^\pm_r$, $r\in\mathbb Z+\tfrac12$ are the NS modes of the right-moving supercurrents. The small $\mathcal N=4$ superconformal algebra then gives
\begin{align}
 \{\mathcal Q_0,\mathcal Q_0^\dagger\}
 &=2(\widetilde L_0-\widetilde J^3_0). \nonumber \\
 2(\bar h-\bar j)
 &=\langle\psi|\{\mathcal Q_0,\mathcal Q_0^\dagger\}|\psi\rangle
 =\|\mathcal Q_0\psi\|^2+\|\mathcal Q_0^\dagger\psi\|^2\ge0.
\end{align}
The anticommutator holds at zero torus momentum and winding; the last line
is its expectation value in a normalized simultaneous eigenstate. For a state $|\psi\rangle$, $(h,\bar h)$ denote its left- and right-moving
conformal weights and $(j,\bar j)$ the Cartan eigenvalues of the left and
right R-symmetry groups $SU(2)_L$ and $SU(2)_R$, respectively.
A right-BPS state saturates $\bar h=\bar j$, so both norms vanish.
Conversely, if
$\mathcal Q_0|\psi\rangle=\mathcal Q_0^\dagger|\psi\rangle=0$,
then the right-hand side vanishes and the BPS bound is saturated,
$\bar h=\bar j$. We denote the right-BPS sector of the symmetric-orbifold theory by $ V_N$. Once a finite left-charge block
and all connected right cochain degrees are specified, it is a
finite-dimensional space of oscillator states. At a generic torus lattice, we use the zero internal
momentum and winding sector throughout.

The simplest single-cycle states saturate the left BPS bound as well,
so that $h=j$ in addition to $\bar h=\bar j$. Their
bottom component is
\begin{equation}
 |w_{--,--}\rangle,\qquad
 h=j=\bar h=\bar j=\frac{w-1}{2}.
\end{equation}

The two signs before the comma specify the left Clifford component;
the two after it specify the right component. On one cycle, two
normalized right Clifford creators generate the four components with
right weights
\begin{equation}
 \begin{array}{c|cccc}
 \text{component}&--&+-&-+&++\\ \hline
 \bar h=\bar j&(w-1)/2&w/2&w/2&(w+1)/2
 \end{array}.
\end{equation}
These four states are the NS images of the Ramond ground states
generated by right-moving fermion zero modes. The left-moving sector
similarly supplies four states. Their tensor products give sixteen
half-BPS ground-state polarizations. Acting with allowed left oscillators preserves
the right BPS condition, although it generally violates $h=j$.
Thus each right-BPS
basis state is obtained cycle by cycle by acting with a left
fractional-oscillator word on a ground polarization:
\begin{equation}
\begin{aligned}
 |\{u_1,\ldots,u_c\}\rangle,\qquad
 u_i&=A_i^{\rm frac}|w_i;\text{ground polarization}\rangle,\qquad \sum_iw_i=N,
\end{aligned}
\end{equation}
Here $c$ is the number of component strings in the partition and $A_i^{\rm frac}$ denotes an allowed left-moving
fractional-oscillator word, and each cycle satisfies its projection
condition. This is the basis on which the
restricted supercharge becomes a matrix.

Let $h_{\rm NS}$ and $j_{\rm NS}$ denote the left-moving NS conformal weight and the eigenvalue of $J^3_0$, respectively. After spectral flow to the Ramond sector, we denote the left-moving conformal weight by $h_R$ and normalize the left- and right-moving $SU(2)$ Cartan charges as
$J_L=2J^3_0$ and $J_R=2\widetilde J^3_0$. We denote the right-moving Ramond
weight by $\bar h_R$; a right-moving Ramond BPS ground state has
$\bar h_R=N/4$. With our spectral-flow convention,
\begin{align}
 h_R&=h_{\rm NS}-j_{\rm NS}+\frac N4, \nonumber \\
 J_L&=2j_{\rm NS}-N,
\end{align}
and likewise on the right. 
In the D1--D5--P system, the third charge \(P\) is momentum along the common D1--D5 circle; we denote its quantized momentum number by \(N_p\). For a right-BPS Ramond state, \(N_p=h_R-\bar h_R\). Substituting
$N_p=h_R-\bar h_R=h_{\rm NS}-j_{\rm NS}$ gives the dictionary

\begin{align}
 h_{\rm NS}&=N_p+\frac{J_L}{2}+\frac N2, \nonumber \\
 j_{\rm NS}&=\frac{J_L+N}{2}, \nonumber \\
 \bar h_{\rm NS}=\bar j_{\rm NS}&=\frac{J_R+N}{2}. \label{eq:spectral-flow}
\end{align}
For a single length-\(w\) component string, the same spectral-flow formula applies with \(N\) replaced by \(w\), since that component has central charge \(6w\); summing over all component strings restores the total winding \(N\).
For a right Ramond ground on a cycle of winding $w$, rotation by one
copy is a translation through $1/w$ of the long circle. We denote the
component momentum above the Ramond ground by $E_R$; cyclic invariance
requires it to be integral:
\begin{equation}
E_R=h_R-\frac w4=h_{\rm NS}-j_{\rm NS}\in\mathbb Z.
 \label{eq:review-cyclic-level-matching}
\end{equation}
For example, in the Ramond sector a single boson
$\alpha_{-1/2}|2;R\rangle$ has momentum $1/2$ and is removed by the
$\mathbb Z_2$ projection, where $|2;R\rangle$ denotes a length-two Ramond
ground state; two such bosons have momentum $1$ and pass it. This illustrates why a fractional Fock word must be
checked for level matching before it is counted as physical.

\subsection{BPS lifting and continuation}
\label{sec:review-lifting}
\label{sec:review-rank-tower}
The symmetric-orbifold point gives a free description of the D1--D5 CFT.
The interacting theory is reached by an exactly marginal deformation, whose
action on the BPS sector determines which orbifold states survive toward the
supergravity regime.

We first vary the coupling at fixed $N$. Here $S_{\rm orb}$ denotes the
symmetric-orbifold action, $g$ the exactly marginal coupling, and
$\Phi_{(2)}$ the twist-two marginal operator. The deformation is the
R-symmetry-singlet superdescendant of a twist-two chiral primary:
\begin{equation}
\begin{aligned}
 S(g)&=S_{\rm orb}+g\int d^2z\,\Phi_{(2)}(z,\bar z), \\
 (h_\Phi,\bar h_\Phi)&=(1,1).
\end{aligned}
\end{equation}
This is the interaction responsible for BPS lifting away from the free
symmetric-orbifold point.  Accordingly, \(V_N\) is the ambient right-BPS
space of the free theory, not the exact BPS space at finite coupling:
states in \(V_N\) may lift under the deformation.  Conversely, when an exact
BPS subspace is transported regularly back to the orbifold point, its
weak-coupling limit is a subspace of \(V_N\).

The twist-two permutation joins or splits component strings, while the
superdescendant dressing makes the deformation exactly marginal and an
\(SU(2)_L\times SU(2)_R\) singlet.  The twist-two deformation carries a
transposition, which exchanges two copies of the symmetric product.

Consider the transposition $(23)$. If copies $2$ and $3$ initially belong
to different cycles, as in $(12)$, the transposition joins those cycles:
\begin{align}
 (23)(12)&=(132), \nonumber \\
 (2,1)&\longrightarrow(3).
\end{align}
If copies $2$ and $3$ already belong to the same cycle, the same
transposition splits it:
\begin{align}
 (23)(132)&=(12), \nonumber \\
 (3)&\longrightarrow(2,1).
\end{align}
Here permutations are composed from right to left.
On three copies this is $(2,1)\leftrightarrow(3)$. Both sides still
have total winding three: the join/split operation changes a partition
inside one theory, whereas continuation to higher $N$ changes the theory.

The marginal deformation also deforms the supersymmetry generators.
To determine how the orbifold BPS states behave away from the
symmetric-orbifold point, we expand the right-moving
supercharge perturbatively as
\begin{equation}
 \mathcal Q(g)=\mathcal Q_0+g\mathcal Q_1+g^2\mathcal Q_2+O(g^3).
\end{equation}

The subscripts label the order in $g$; $\mathcal Q_0$ is the orbifold
supercharge.
Let $\Pi_{0,N}$ be the orthogonal projector onto $V_N$.
We call

\begin{equation}
 Q_N=\Pi_{0,N}\mathcal Q_1\Pi_{0,N}
\end{equation}

the restricted first-order deformation supercharge: it is the
$O(g)$ correction $\mathcal Q_1$ to the exact right supercharge, with both
its input and output projected to the free-BPS sector $V_N$. The subscript
on $Q_N$ labels the value of $N$. The conformal-perturbation construction preserves
the free BPS sector at each order \cite{ChangLinZhang}. Since $\mathcal Q_0$ annihilates $V_N$, the order-$g^2$ part of
$\mathcal Q(g)^2=0$ gives

\begin{equation}
 Q_N^2
 =\Pi_{0,N}\mathcal Q_1^2\Pi_{0,N}
 =-\Pi_{0,N}\{\mathcal Q_0,\mathcal Q_2\}\Pi_{0,N}
 =0.
\end{equation}

The supersymmetry algebra also determines the leading shift of the
right-moving conformal weight. At order $g^2$, the mixed terms involving
$\mathcal Q_0$ or $\mathcal Q_0^\dagger$ vanish between free BPS states,
leaving
\begin{align}
 \delta\bar h&=g^2\lambda_Q\{Q_N,Q_N^\dagger\}+O(g^3), \nonumber \\
 \delta h&=\delta\bar h,\qquad \lambda_Q>0,
\end{align}
where $\delta\bar h$ denotes the leading shift of the right-moving
conformal weight away from its orbifold value.
The equality of the two shifts expresses conservation of conformal
spin. The positive constant $\lambda_Q$ fixes normalization
\cite{GavaNarain,GuoMathurLevelOne,GuoMathurHigherLevels,ChangLinZhang}.
Within a degenerate free-BPS block, $\delta\bar h$ is the leading
anomalous-dimension operator. Its eigenvectors are the linear combinations
of Fock words with definite leading lifting, and its eigenvalues are their
corresponding shifts. Positivity gives
\begin{equation}
 \langle\psi|\delta\bar h|\psi\rangle
   =g^2\lambda_Q\bigl(\|Q_N\psi\|^2+
                         \|Q_N^\dagger\psi\|^2\bigr)+O(g^3).
\end{equation}

The zero eigenspace of the coefficient of $g^2$ in the lifting operator is therefore characterized exactly by
\begin{align}
 Q_N\psi&=0, \nonumber \\
 Q_N^\dagger\psi&=0. \label{eq:review-harmonic-condition}
\end{align}
The zero-lifting states in Eq.~\eqref{eq:review-harmonic-condition}
can equivalently be described by the cohomology of $Q_N$. Since
$Q_N^2=0$, every exact state $Q_N \chi$ is automatically closed, implying \(\operatorname{im}Q_N\subset\ker Q_N\).
We therefore can define the cohomology:
\begin{align}
 H_{Q_N}(V_N)
=\frac{\ker Q_N}{\operatorname{im}Q_N},\qquad [\psi]=\{\psi+Q_N\chi:\chi\in V_N\}.
\end{align}

Two closed states belong to the same cohomology class if they differ by an exact state \(Q_N\chi\). Exact states are themselves closed, but a nonzero exact state cannot be harmonic. Indeed, if \(v=Q_N\chi\) also satisfied \(Q_N^\dagger v=0\), then
\begin{equation}
 \|v\|^2
=\langle Q_N\chi,v\rangle
=\langle\chi,Q_N^\dagger v\rangle
=0,
\end{equation}
so \(v=0\). Thus each nontrivial cohomology class is represented by a closed state that is not \(Q_N\)-exact.

Conversely, start with any closed state $\psi$. States that differ by
$Q_N\chi$ belong to the same cohomology class, so we may subtract from
$\psi$ its component along $\operatorname{im}Q_N$. The resulting state
is still closed and is orthogonal to every exact
state. Since
\begin{equation}
 \langle Q_N\chi,\psi\rangle
 =\langle\chi,Q_N^\dagger\psi\rangle,
\end{equation}
this orthogonality implies $Q_N^\dagger\psi=0$. Thus every cohomology
class has a harmonic representative, giving
\begin{equation}
 H_{Q_N}(V_N)
 \cong
 \ker Q_N\cap\ker Q_N^\dagger.
\end{equation}
Equivalently, the finite-dimensional space $V_N$ admits the decomposition
\begin{equation}
 V_N
 =\operatorname{im}Q_N
 \oplus
 \bigl(\ker Q_N\cap\ker Q_N^\dagger\bigr)
 \oplus
 \operatorname{im}Q_N^\dagger.
\end{equation}

In the D1--D5 CFT, the first-order deformation supercharge acts by
joining and splitting component strings,
\begin{equation}
 Q_N=Q_{\rm join}+Q_{\rm split}
\end{equation}
The states that remain BPS at first order are represented by
\begin{equation}
 \mathcal H_N:=\ker Q_N\cap\ker Q_N^\dagger
          \cong H_{Q_N}(V_N).
\end{equation}
Because the fixed-charge block $V_N$ is finite dimensional, every class
$[\psi]\in H_{Q_N}(V_N)$ has a unique representative in $\mathcal H_N$.
We therefore use this harmonic representative as the state itself. This
identification gives cohomology the inner product of $\mathcal H_N$; if
$S\subset H_{Q_N}(V_N)$, we realize the quotient
$H_{Q_N}(V_N)/S$ as the orthogonal complement $S^\perp\subset\mathcal H_N$.

In an orthonormal oscillator basis, its matrix elements are
$\langle t_i|Q_N|s_a\rangle$. These matrix elements are computed from
three-point functions with the twist-two superdescendant, using covering
maps to remove the twist-field branch cuts
\cite{ChangLinZhang}. The states in $\mathcal H_N$ are therefore the
orbifold BPS combinations that remain unlifted at leading order in the
deformation. Their continuation
beyond leading order is addressed separately.

The coupling test determines survival in one CFT. We now keep the
orbifold description and ask whether the same class admits a closed ancestor
in a theory with larger $N$. Let
\(F\in\mathcal H_N\) be a BPS representative at \(N\). The most
obvious candidate at a larger \(M>N\) is obtained by adding
\(M-N\) unexcited length-one strands,
\begin{align}
 x_M^{(0)}\ &\propto
\bigl|\{F,1_{--,--}^{\,M-N}\}\bigr\rangle, \nonumber \\
 \pi_{N,M}x_M^{(0)}&=F .
\end{align}

Here $\pi_{N,M}:V_M\to V_N$ is the fixed stringy-exclusion map of
Ref.~\cite{ChangLinZhang}. It removes the added vacuum on all four members
of the right-moving Clifford quartet with their relative normalizations
fixed. More explicitly, let $\eta_D$ and $\bar\eta_D$ be the normalized
diagonal sums of the right-moving zero-mode creators on the retained
component strings. In the diagonal-top component, $\pi_{N,M}$ projects the
removed strand onto its Clifford vacuum and then restores the pair
$\bar\eta_D\eta_D$ on the retained strings; Appendix~\ref{app:F2-heavy-family}
implements this operation explicitly. The second equation fixes the
normalization of the naive extension $x_M^{(0)}$. Although this prescription
is consistent across the Clifford quartet, it need not intertwine the
deformation supercharges: $Q_N\pi_{N,M}$ need not equal
$\pi_{N,M}Q_M$.
Consequently, being BPS at \(N\) does not guarantee that this naive
extension is BPS at \(M\). The added singly wound Ramond vacuum can participate in
the join/split action of \(Q_M\), so one may have
\(Q_Mx_M^{(0)}\neq0\) even though \(Q_NF=0\). We must therefore allow
{additional states at rank \(M\) that may cancel}
its image,

\begin{equation}
 y=x_M^{(0)}+u .
\end{equation}

The correction \(u\) is allowed only if it does not change the
lower-$N$ cohomology class. Equivalently,
\begin{equation}
 \pi_{N,M}y=F+Q_Nz
\end{equation}
for some \(z\in V_N\). The full continuation problem is therefore
\begin{align}
 Q_My&=0, \nonumber \\
 \pi_{N,M}y&=F+Q_Nz . \label{eq:framework-short-affine}
\end{align}
{The first condition requires \(y\) to be $Q_M$-closed at rank $M$;}
the second requires it to reduce to the same cohomology class
\([F]\) at \(N\). If the lower exact space is empty, this simply
means that \(u\in\ker\pi_{N,M}\). A class fails to continue only when no such corrected
closed state \(y\) exists.

\subsection{Black-hole states and fortuity}
\label{sec:review-fortuity-program}

The fortuity program relates the dependence of BPS states on $N$ to the
microscopic origin of black-hole entropy \cite{ChangLinFortuity,ChangLinZhang}.
For D1--D5--P charges $(N;N_p,J_L,J_R)$, the BMPV entropy scale is
\begin{align}
 \nu&=\frac{N_p}{N}, \nonumber \\
 \jmath&=\frac{J_L}{N}, \nonumber \\
 \delta&=\nu-\frac{\jmath^2}{4}>0, \nonumber \\
 S_{\rm BH}&=2\pi N\sqrt{\delta}.
\end{align}

Here $\delta>0$ is the positive-area condition for the BMPV solution
\cite{BMPV}. At fixed densities in its interior, $S_{\rm BH}$ grows linearly
with $N$. For the counting comparison, $\Gamma_N=(N_p,J_L)$ denotes a sequence of
left-charge blocks at rank $N$, with the complete right Ramond multiplets
retained, and $d_{\rm BPS}(\Gamma_N)$ denotes the corresponding BPS
degeneracy. Since the black-hole entropy fixes only the leading
exponential growth of the degeneracy in the large-$N$ limit, subleading
corrections to $\log d_{\rm BPS}$ are allowed provided that they grow more
slowly than $N$. {Thus matching the leading black-hole entropy requires}
\begin{equation}
\log d_{\rm BPS}(\Gamma_N)=S_{\rm BH}+o(N){.}
\end{equation}
This is the charge convention for the modified index reviewed in
Section~\ref{sec:review-counting-index}. We write
$V_{N,\Gamma_N}$ for the corresponding free right-BPS space,
$D_V(\Gamma_N)=\dim V_{N,\Gamma_N}$, and
$D_H(\Gamma_N)=\dim\mathcal H_{N,\Gamma_N}$ for the first-order harmonic
dimension. The same macroscopic charge block can contain protected
supergravity states as well as additional BPS states;
the entropy question concerns their relative multiplicities.

{
A monotone class has a closed representative at $N$ that is the
projection of a closed state at every higher rank. For the fixed maps
$\pi_{N,M}$, we realize the definition of Refs.~\cite{ChangLinZhang,ChangZhang}
directly as harmonic subspaces:
\begin{align}
 \mathcal H_N^{\rm mon}
 &=\mathcal H_N\cap\left(
 \left[\ker Q_N\cap\bigcap_{M>N}
       \operatorname{im}(\pi_{N,M}|_{\ker Q_M})\right]
       +\operatorname{im}Q_N\right), \nonumber\\
 \mathcal H_N^{\rm for}
 &=\mathcal H_N\cap(\mathcal H_N^{\rm mon})^\perp
   \cong\mathcal H_N/\mathcal H_N^{\rm mon}.
 \label{eq:framework-short-monotone}
\end{align}
The intersection requires one closed representative at rank $N$ to admit
a lift at every higher rank; adding $\operatorname{im}Q_N$ and then
intersecting with $\mathcal H_N$ selects its harmonic representative.
Failure of Eq.~\eqref{eq:framework-short-affine} at even one higher rank
therefore gives a nonzero class in the fortuitous quotient. No map on
cohomology induced by $\pi_{N,M}$ is assumed.
} The original holographic proposal
associates monotone states with smooth horizonless descriptions and
fortuitous states with the entropy-dominant black-hole sector
\cite{ChangLinFortuity}. In the D1--D5 theory, explicit low-$N$
supercharge cohomologies and composite constructions make this proposal
accessible to direct calculations \cite{ChangLinZhang,ChangZhang}.

A more refined proposal concerns the distribution of cohomology along the
right R-charge grading. Fixing all other quantum numbers that label the complex, the remaining states are organized by their right R-charge $r$, and $Q_N$ maps the sector of charge $r$ to that of charge $r+1$:
\begin{equation}
\cdots\xrightarrow{Q_N}V_N^{r-1}
\xrightarrow{Q_N}V_N^r
\xrightarrow{Q_N}V_N^{r+1}\xrightarrow{Q_N}\cdots .
\end{equation}
The R-charge-concentration conjecture states that, within a fixed
cochain complex, fortuitous cohomology is supported at a single value of $r$, whereas monotone cohomology may occur at several values.
 Chang and
Zhang find this concentration at $N=2$ and in the longest $N=3$ complex.
For the latter, through $h=4$, all BPS states away from the central spaces of the cochain complex are symmetry descendants of half-BPS states and are
therefore monotone. Thus any fortuitous classes in this range are confined
to the central sector.

This result is important here because it gives evidence that fortuitous
cohomology has a nontrivial collective organization.  The organization it
constrains, however, is in charge and cochain space.  The symmetric-orbifold
CFT contains an independent geometric datum, the cycle partition
\begin{equation}
\begin{aligned}
 p=(1^{n_1}2^{n_2}\cdots), \qquad
 \sum_{w\ge1} w n_w=N,
\end{aligned}
 \label{eq:review-cycle-partition-geometry}
\end{equation}
which specifies how the total winding is distributed among component
strings. A cycle of winding $w=O(N)$ has fractional excitation spacing
$O(1/N)$ and introduces the collective long-string scale relevant for the
interaction and the geometric probe analysis.  The R-charge-concentration
results therefore motivate a distinct question: whether the collective
organization seen in the supercharge complex is accompanied by a
corresponding concentration in the distribution over cycle partitions.

The gravitational comparison is correspondingly sharpened.
 Singly wound Ramond ground states
supply monotone states outside the ordinary multiparticle-supergraviton
spectrum \cite{HughesShigemori}. The generalized-gravity space
$\mathcal G_N$ includes their dressings: it is the span obtained by acting
with total affine modes on multiparticle supergravitons
\cite{ToweringGravitons}. We denote by $\mathcal G_N^{\rm int}$ the part of this construction that
survives in the interacting BPS spectrum. Its proposed identification with
the monotone sector concerns these surviving interacting BPS states,
\begin{equation}
\mathcal G_N^{\rm int}
 \ \stackrel{\rm conjecture}{\longleftrightarrow}\
 \text{interacting monotone BPS sector}.
\end{equation}
{The classification in Eq.~\eqref{eq:framework-short-monotone} always
uses these same maps and the first-order harmonic space. Vacuum removal
need not commute with the deformation supercharge, so its images are not
silently replaced by images of maps on cohomology. The K3 construction
of Ref.~\cite{GiustoInglisRusso} exhibits a modified projection in the
$(h,j)=(1,0)$ sector at first order; it does not establish a compatible
supercharge-intertwining family in every sector. The later monotone projectors refer to
Eq.~\eqref{eq:framework-short-monotone}, and their dimension and cycle
bounds are stated separately as assumptions.}

The supergravity spectrum already has a known entropy deficit. We use
$D_{\rm sugra}^{(\infty)}(h)$ for the unrestricted large-$N$ multiparticle
supergravity degeneracy at left NS weight $h$. Its asymptotic count has
\begin{equation}
\begin{aligned}
 \log D_{\rm sugra}^{(\infty)}(h)&=O(h^{3/4}), \\
 h_N&=N_p+\frac{J_L}{2}+\frac N2=O(N)
\end{aligned}
\end{equation}
\cite{Benjamin:2016pil}. Its generating function provides an upper bound
on the finite-$N$ supergraviton degeneracy. Along the black-hole charge
trajectories this bound is $e^{O(N^{3/4})}$, parametrically smaller than
the $e^{O(N)}$ black-hole degeneracy. The comparison with the full BPS
space must also account for the affine dressings in $\mathcal G_N$ and
for the cycle distribution of the surviving BPS states.

\subsection{State counting and twist coefficients}
\label{sec:review-counting-index}

{This subsection collects the counting functions and
twist-operator matrix elements used below. The unsigned free-orbifold count
resolves winding multiplicities, whereas the modified index is a signed count
invariant under deformation. Their comparison enters the entropy and
gravity-subtraction analysis of Section~\ref{sec:long-strings-lifting}; the
exact twist coefficients enter the interaction calculation of
Section~\ref{sec:twist-main}.}

An unsigned trace has nonnegative coefficients and gives actual
Hilbert-space multiplicities, which can be resolved by cycle winding.
The modified index is a supersymmetric signed trace preserved under the
marginal deformation. The right-moving zero-mode normalization determines
how its absolute value bounds an actual BPS dimension
\cite{MaldacenaMooreStrominger,ChangLinZhang}.

For a single component string, we call $P_+(q,y)$ the unsigned left-moving
oscillator character: it is the generating function over the left-moving
oscillator Fock space, with $q$ marking excitation level and $y$ marking the
left Cartan charge. Its coefficients count oscillator configurations. At
each positive left-moving oscillator level $m$ on the component-string
cover there are four neutral bosonic oscillators and four fermionic
oscillators, two with Cartan charge $+1$ and two with charge $-1$.
The contribution of all oscillator species at this one level is therefore
\begin{equation}
 Z_m(q,y)
 =
 \frac{(1+yq^m)^2(1+y^{-1}q^m)^2}{(1-q^m)^4}.
\end{equation}

Here each bosonic species contributes
$(1-q^m)^{-1}$, while a fermionic species of charge $\pm1$ contributes
$1+y^{\pm1}q^m$.  Multiplying over all positive modes gives

\begin{equation}
 P_+(q,y)
 =
 \prod_{m\ge1}
 \frac{(1+yq^m)^2(1+y^{-1}q^m)^2}{(1-q^m)^4}
 =
 \sum_{m\ge0,\;r\in\mathbb Z}
 A(m,r)q^m y^r .
\end{equation}
Here $q$ records the left excitation level and $y$ records
the left Cartan charge.  The coefficient $A(m,r)$ is the nonnegative number
of left-moving oscillator configurations at level $m$ and charge $r$.

The oscillator character must be multiplied by the Ramond ground-state
multiplicity. Set $\mu:=\log y$ for the Cartan chemical potential. The
sixteen $T^4$ Ramond ground states decompose, in the same charge convention, as
\begin{align}
 16
 &=
 4_{\,r=-1}\oplus 8_{\,r=0}\oplus4_{\,r=+1}, \nonumber \\
 \shortintertext{and hence their unsigned charge character is}
 g_R(y)
 &=
 4y^{-1}+8+4y
 =
 4(y^{-1}+2+y), \nonumber \\
 g_R(e^\mu)&=16\cosh^2\!\frac{\mu}{2}.
\end{align}

A \textit{physical component string} here refers to a length-$w$
cycle after the $\mathbb Z_w$ cyclic projection, equivalently after the
component level-matching condition has been imposed; it does not denote a
separate classical string configuration.  A length-$w$ component string must obey the cyclic level matching in
Eq.~\eqref{eq:review-cyclic-level-matching}. We introduce $\beta>0$ as the auxiliary source conjugate to the physical
left excitation; $\mu$ is the Cartan source defined above. If $L$ denotes the integer total oscillator level on
the component-string cover, level matching requires
\begin{align}
 L&\equiv0\pmod w, \nonumber \\
 \frac1w\sum_{\ell=0}^{w-1}e^{2\pi i\ell L/w}
 &=
 \begin{cases}
 1, & L\equiv0\pmod w,\\
 0, & L\not\equiv0\pmod w,
 \end{cases} \nonumber \\
 Z_w^{\rm phys}(\beta,\mu)
 &=
 \frac{g_R(e^\mu)}{w}
 \sum_{\ell=0}^{w-1}
 P_+\!\left(
 e^{(-\beta+2\pi i\ell)/w},
 e^\mu
 \right). \label{eq:projected-strand-character}
\end{align}
The cyclic phase average is the projector onto
$L\equiv0\pmod w$: it equals one for allowed multiples of $w$ and vanishes
for every other cover level. Applying this projector to the single-cycle
generating function gives the last line.  An allowed cover level can
therefore be written as $L=wE$, where $E=L/w$ is the physical momentum above
the Ramond ground state.  Consequently any fixed-$(E,J)$ single-cycle multiplicity below is evaluated
at cover level $m=wE$.

Thus, for each fixed $w$, $Z_w^{\rm phys}$ is the unsigned generating
function for physical states of a single length-$w$ component string after
the cyclic projection. Keeping the label $w$ explicit separates the contribution of a component
string of each possible winding. The full symmetric-product count is then
assembled by combining the contributions from all allowed windings. This is the appropriate
object for studying cycle statistics, because all coefficients are
nonnegative.
However, it is not protected under the marginal deformation: a state counted
by $Z_w^{\rm phys}$ need not remain BPS away from the symmetric-orbifold
point.  To obtain information about the dimension that must survive the
deformation, we therefore need a protected index.

For protected counting, start instead from the graded Ramond--Ramond trace
of one free $T^4$ seed at the symmetric-orbifold point. The ordinary
elliptic-genus grading contains $(-1)^F$, with $F$ the total fermion number.
When the fugacities are regarded as Jacobi variables we write
$q=e^{2\pi i\tau}$, $y=e^{2\pi i z}$,
$\bar q=e^{-2\pi i\bar\tau}$ and
$\widetilde y=e^{2\pi i\widetilde z}$. We denote the odd Jacobi theta
function by $\vartheta_1$ and the Dedekind eta function by $\eta$. In one
chiral sector the four real bosons contribute $\eta^{-4}$, while the four
real fermions form two complex fermions whose graded Ramond trace contributes
$(\vartheta_1/\eta)^2$. Their product is therefore
$\vartheta_1^2/\eta^6$. Applying the same factorization to the right movers,
the single-copy graded trace is
\begin{align}
 Z_{RR}^{T^4}(q,y;\bar q,\widetilde y)
 &=
 \frac{\vartheta_1(q,y)^2}{\eta(q)^6}
 \frac{\vartheta_1(\bar q,\widetilde y)^2}{\eta(\bar q)^6}, \nonumber \\
 \widetilde y&=e^{2\pi i\widetilde z},
\end{align}

up to the fermion-parity convention used below. The ordinary
elliptic genus sets $\widetilde y=1$, or equivalently
$\widetilde z=0$. Since
\begin{align}
 \vartheta_1(\bar\tau,\widetilde z)
 &=
 2\pi \widetilde z\,\eta(\bar\tau)^3
 +O(\widetilde z^3), \nonumber \\
 \shortintertext{the right-moving factor has a quadratic zero,}
 \frac{\vartheta_1(\bar\tau,\widetilde z)^2}
      {\eta(\bar\tau)^6}
 &=
 (2\pi\widetilde z)^2+O(\widetilde z^4),
\end{align}
and hence vanishes at $\widetilde z=0$.  This quadratic zero is the
partition-function manifestation of the two right-moving fermion zero
modes.

The modified index is defined by extracting the coefficient of this
quadratic zero:
\begin{align}
 \widehat I(q,y)
 :&=
 \frac12
 \left.
 \left(
 \widetilde y\frac{\partial}{\partial\widetilde y}
 \right)^2
 Z_{RR}^{T^4}(q,y;\bar q,\widetilde y)
 \right|_{\widetilde y=1}. \nonumber \\
 \widetilde y\frac{\partial}{\partial\widetilde y}
 &=
 \frac{1}{2\pi i}\frac{\partial}{\partial\widetilde z}, \nonumber \\
 \frac12
 \left.
 \left(
 \widetilde y\frac{\partial}{\partial\widetilde y}
 \right)^2
 (2\pi\widetilde z)^2
 \right|_{\widetilde z=0}
 &=
 -1. \nonumber \\
 \widehat I(q,y)
 &=
 -\frac{\vartheta_1(q,y)^2}{\eta(q)^6}
 =
 \sum_{m,r}\widehat c(m,r)q^my^r .
\end{align}
The two derivatives act on the quadratic zero produced by the two
right-moving fermion zero modes and leave its nonzero coefficient. This is
why $\widehat I(q,y)$ is the \emph{modified} seed index rather than the
ordinary elliptic genus. The hat distinguishes this single-copy seed from
the fixed-$N$ symmetric-product index $I_N$ introduced below. The protection
statement applies to $I_N$: differentiating its graded trace with respect to
the exactly marginal coupling inserts the marginal operator, which is a
supersymmetry descendant and hence $Q$-exact in the protected trace. The
resulting graded trace vanishes, so $\partial_g I_N=0$
\cite{MaldacenaMooreStrominger,ChangLinZhang}.

The two counts are related by their single-cycle seeds.  The unsigned state count is built from the oscillator generating function
$P_+(q,y)$, whereas the protected count is built from the modified-index
seed $\widehat I(q,y)$.  Using the Jacobi product formulas
\begin{align}
 \vartheta_1(q,y)
 &=
 i q^{1/8}y^{-1/2}(1-y)
 \prod_{m\ge1}
 (1-q^m)(1-yq^m)(1-y^{-1}q^m), \nonumber \\
 \eta(q)
 &=
 q^{1/24}\prod_{m\ge1}(1-q^m),
\end{align}

\begin{align}
 \widehat I(q,y)
 &=
 -\frac{\vartheta_1(q,y)^2}{\eta(q)^6} \nonumber \\
 &=
 y^{-1}(1-y)^2
 \prod_{m\ge1}
 \frac{(1-yq^m)^2(1-y^{-1}q^m)^2}
      {(1-q^m)^4} \nonumber \\
 &=
 y^{-1}(1-y)^2 P_+(q,-y). \nonumber \\
 \widehat I(q,-y)
 &=
 -(y^{-1}+2+y)P_+(q,y). \label{eq:three-adjacent-index-identity}
\end{align}
The last line relates the protected seed directly to the unsigned oscillator
character. Thus the protected seed is obtained from the same nonnegative oscillator
multiplicities that enter the unsigned count, multiplied by the
Ramond-ground-state charge polynomial and accompanied by a known
charge-dependent sign.

Comparing the coefficient of $q^m y^r$ in
Eq.~\eqref{eq:three-adjacent-index-identity} gives
\begin{equation}
(-1)^{r+1}\widehat c(m,r)
 =
 A(m,r-1)+2A(m,r)+A(m,r+1)
 \ge 0 .
 \label{eq:positive-seed-identity}
\end{equation}
Thus $|\widehat c(m,r)|=(-1)^{r+1}\widehat c(m,r)$, so the sign of every modified-index coefficient is fixed by its charge,
while its magnitude is a nonnegative combination of the unsigned oscillator
multiplicities in the three neighboring charge sectors.  The weights
$1,2,1$ are precisely the coefficients of the normalized Ramond-ground-state
charge polynomial $y^{-1}+2+y$.  

Multiplying by the overall factor of four
in $g_R(y)=4(y^{-1}+2+y)$ therefore gives $4|\widehat c(m,r)|$ unsigned
single-cycle states at fixed cover level $m$ and left charge $r$.  For each
left polarization the four right Ramond zero-mode states contain two bosonic
and two fermionic states, so these split into $2|\widehat c(m,r)|$ states
of each parity.  This equality is special to the single-cycle seed: it uses
the definite-sign identity above and does not identify a general protected
index with an unsigned Hilbert-space dimension.

We will also need the charge range in which these seed coefficients can be
nonzero.  Since $\widehat I=-\vartheta_1^2/\eta^6$ is an index-one weak
Jacobi form, its elliptic transformation gives
\begin{equation}
 \widehat c(m,r)
 =\widehat c(m+\ell r+\ell^2,r+2\ell),
 \qquad \ell\in\mathbb Z,
 \label{eq:seed-jacobi-elliptic-shift}
\end{equation}
and preserves the discriminant $4m-r^2$.  Choosing $\ell$ so that the
shifted charge is $0$ for even $r$ or $\pm1$ for odd $r$, and using the
absence of negative powers of $q$ in a weak Jacobi form, gives
\begin{equation}
 \widehat c(m,r)\neq0
 \quad\Longrightarrow\quad
 r^2\le4m+1.
 \label{eq:seed-jacobi-support}
\end{equation}

To obtain the protected count of the full
\(N\)-copy theory, the signed seed coefficients \(\widehat c(m,r)\) must
instead be assembled into the symmetric-product index. At fixed rank \(N\),
total momentum \(N_p\), and integer left Cartan charge \(J\), we denote the
resulting coefficient by \(I_N(N_p,J)\). It is given by the divisor sum
\begin{equation}
\begin{aligned}
 I_N(N_p,J)&=\sum_{k\mid g}k\,
 \widehat c\!\left(\frac{NN_p}{k^2},\frac{J}{k}\right), \\
 g&=\gcd(N,N_p,|J|).
\end{aligned}
 \label{eq:charge-plane-DMVV-divisors}
\end{equation}
Here $\gcd(N,N_p,|J|)$ is the greatest common divisor of the
three integers, and the sum runs over its positive divisors $k$. This is the
fixed-charge form of the modified symmetric-product formula
\cite{MaldacenaMooreStrominger,ChangLinZhang}.  Equation~\eqref{eq:divisor-extraction-main} gives the explicit fixed-charge extraction used below.

\label{sec:quartet-dimension}
To convert the signed index into a lower bound on an actual
BPS dimension, we must keep track of the right-moving fermion zero modes.
The two zero modes generate, from each Clifford vacuum, a four-state
multiplet; we call this four-state multiplet a \textit{quartet}. At zero
torus momentum and winding, the two diagonal right fermion zero modes can be
normalized as creation operators $c_1^\dagger,c_2^\dagger$ obeying
\begin{align}
 \{c_a,c_b^\dagger\}&=\delta_{ab}, \nonumber \\
 [J_R,c_a^\dagger]&=c_a^\dagger, \nonumber \\
 \{Q_N,c_a\}&=\{Q_N,c_a^\dagger\}=0.
\end{align}

Their adjoints likewise graded-commute with $Q_N^\dagger$, so both the
cochain and its harmonic subspace decompose into four-state Clifford modules.
If $|\omega\rangle$ is a Clifford vacuum of charge $j_\omega$ and parity
$(-1)^{F_\omega}=\sigma$, the quartet is

\begin{equation}
 \begin{array}{c|cccc}
 \text{state}&|\omega\rangle&c_1^\dagger|\omega\rangle
 &c_2^\dagger|\omega\rangle&c_1^\dagger c_2^\dagger|\omega\rangle\\ \hline
 J_R&j_\omega&j_\omega+1&j_\omega+1&j_\omega+2\\
 (-1)^F&\sigma&-\sigma&-\sigma&\sigma
 \end{array} \label{eq:right-clifford-four-states}
\end{equation}

\begin{align}
 \Tr_{\rm quartet}(-1)^F
 &=
 \sigma-\sigma-\sigma+\sigma
 =0, \nonumber \\
 \Tr_{\rm quartet}(-1)^FJ_R
 &=
 \sigma\bigl[
 j_\omega-2(j_\omega+1)+(j_\omega+2)
 \bigr]
 =0. \nonumber \\
 \frac12\Tr_{\rm quartet}(-1)^FJ_R(J_R-1)
 &=
 \frac{\sigma}{2}\bigl[
 j_\omega(j_\omega-1)
 -2(j_\omega+1)j_\omega
 +(j_\omega+2)(j_\omega+1)
 \bigr] \nonumber \\
 &=\sigma .
\end{align}
For this quartet the constant and linear supertraces vanish. The
quadratic insertion is the first nonzero weighted trace.
Thus the two right-moving fermion zero modes make the ordinary signed
trace and the linear charge-weighted trace vanish, while the quadratic
insertion assigns one signed unit to the complete four-state quartet. This motivates the normalization of the modified index at fixed left
charges $\Gamma$,
\begin{equation}
 I_N(\Gamma)
 =
 \frac12\Tr_{V_{N,\Gamma}}
 (-1)^FJ_R(J_R-1).
\end{equation}

Let the harmonic BPS space at fixed left charges $\Gamma$ contain
$K$ right-moving Clifford quartets.  Label their index signs by
$\sigma_a=\pm1$, $a=1,\ldots,K$.  Since each quartet contains four
physical states but contributes only one signed unit to the modified index,
\begin{align}
 \dim\mathcal H_{N,\Gamma}&=4K, \nonumber \\
 I_N(\Gamma)&=\sum_{a=1}^{K}\sigma_a, \nonumber \\
 |I_N(\Gamma)|&\le K
 =\frac14\dim\mathcal H_{N,\Gamma}, \nonumber \\
 \dim\mathcal H_{N,\Gamma}&\ge 4|I_N(\Gamma)|.
 \label{eq:quartet-dimension-lower-bound}
\end{align}
Thus the modified index gives a rigorous lower bound on the actual
dimension of the BPS harmonic space.

\label{sec:review-exact-twist}
A twist-two insertion joins component strings of winding $M$ and $a$ into
one of winding $L=M+a$. The exact oscillator coefficients describe transmission
of an incoming excitation and pair creation in the outgoing state
\cite{CarsonHamptonMathurTurtonTwist,CarsonHamptonMathurTurton}.

For incoming and outgoing mode numbers $m$ and $k$, define the corresponding dimensionless fractional mode momenta and two auxiliary factors
\begin{align}
 q&=\frac mM, \nonumber \\
 s&=\frac kL, \nonumber \\
 X&=\frac{L^L}{M^M a^a}, \nonumber \\
 \chi_a(s)&=1-e^{-2\pi ias}. \label{eq:giant-dust-definitions}
\end{align}
Here $q$ and $s$ are the fractional mode frequencies; in this subsection
$q$ should not be confused with the counting fugacity used above.  The factor $X$ is a shorthand for the winding-dependent normalization
that arises in the covering-space expression for the twist coefficient, while $\chi_a(s)$ records the mismatch
between the incoming and outgoing mode lattices.
Indeed, since $Ls=k\in\mathbb Z$ and
$M=L-a$,
\begin{align}
 e^{2\pi iMs}
 &=
 e^{2\pi i(L-a)s}
 =
 e^{-2\pi ias}, \nonumber \\
 1-e^{2\pi iMs}&=\chi_a(s).
\end{align}

Let $z_0$ denote the position of the twist insertion on the base plane.
For $q\neq s$, the exact transmission coefficient of a bosonic oscillator
initially living on the length-$M$ component string is
\begin{equation}
 \widetilde f^{B(1)}_{qs}
 =\frac{i}{2\pi L}\,z_0^{s-q}\frac{\chi_a(s)}{s-q}
 X^{s-q}
 \frac{\Gamma(Lq)}{\Gamma(Mq)\Gamma(aq)}
 \frac{\Gamma(Ms)\Gamma(as)}{\Gamma(Ls)}. \label{eq:giant-dust-exact-f}
\end{equation}

The pole at $s=q$ in this representation is removable on a common lattice
mode; evaluating the finite-$M$ coefficient directly gives

\begin{equation}
 \widetilde f^{B(1)}_{qq}=\frac{M}{M+a}. \label{eq:giant-dust-diagonal}
\end{equation}

The exact twist calculation contains two kinds of oscillator data.
First, if an excitation is already present on one of the incoming
component strings, the twist redistributes it among the modes of the
joined string; this is described by the transmission coefficient
$\widetilde f^{B(1)}$ above.  Second, the twist acting on the incoming vacuum itself produces a squeezed
outgoing state. Here $|0\rangle_M$ and $|0\rangle_a$ are the unexcited
component-string vacua of windings $M$ and $a$, $\sigma_2(z_0)$ is the
twist-two insertion at $z_0$, and $C_{M,a}$ is the vacuum normalization:
\begin{equation}
\begin{aligned}
 \sigma_2(z_0)
 |0\rangle_M|0\rangle_a
 &=
 C_{M,a}
 \exp\!\left[
 \sum_{k,k'>0}
 \gamma^B_{kk'}\,
 \alpha_{-k}\alpha_{-k'}
 +\cdots
 \right]
 |0\rangle_L , \\
 L&=M+a ,
\end{aligned}
\end{equation}
where $\gamma^B_{kk'}$ is the bosonic pair-creation coefficient. After
introducing $s=k/L$ and $s'=k'/L$, we write
$\widetilde\gamma^B_{ss'}:=\gamma^B_{kk'}$ for the same coefficient in
fractional-momentum variables.

Carson et al.\ obtained this coefficient exactly for arbitrary finite
incoming windings \cite{CarsonHamptonMathurTurtonTwist}.  Their result is
written in terms of windings $M$ and $N$.  Relabeling their second winding
as $N=a$, setting $L=M+a$, and using
\begin{align}
 s&=\frac{k}{L}, \nonumber \\
 s'&=\frac{k'}{L}, \nonumber \\
 X&=\frac{L^L}{M^M a^a}, \nonumber \\
 \chi_a(s)&=1-e^{-2\pi ias},
\end{align}
their exact result becomes
\begin{equation}
 \widetilde\gamma^B_{ss'}
 =
 \frac{z_0^{s+s'}\chi_a(s)\chi_a(s')}
      {4\pi^2(s+s')}
 \frac{Ma}{L^3}X^{s+s'}
 \frac{\Gamma(Ms)\Gamma(as)}{\Gamma(Ls)}
 \frac{\Gamma(Ms')\Gamma(as')}{\Gamma(Ls')}. \label{eq:giant-dust-exact-gamma}
\end{equation}
The limit $M\to\infty,\quad a=O(1)$
will be imposed only in Section~\ref{sec:twist-main}.

These coefficients describe the action of the twist operator itself.
The corresponding deformation-supercharge matrix elements require the
additional supercurrent insertion and BPS projection reviewed above. The counting functions and the index will be combined with the
higher-$N$ continuation tests and gravity-sector bounds, whereas the twist coefficients
supply the finite-winding interaction data.

\subsection{Long strings and cycle selection}
\label{sec:review-long-string-selection}

It is well known that a long string carries the leading Cardy entropy in the
D1--D5 system. Early D-brane and effective-string analyses showed that a
multiply wound effective string reproduces both the black-hole entropy and
the near-extremal absorption and emission rates
\cite{MaldacenaSusskind,MaldacenaStromingerGreybody,HassanWadiaEffectiveString}.
At the symmetric-orbifold point, a component string of winding $w$ has
central charge $6w$. For a right-BPS component carrying left momentum $L$
and left angular momentum $J$, the leading charged Cardy entropy is
\begin{equation}
 S_w(L,J)=2\pi\sqrt{wL-\frac{J^2}{4}}+o(w),
 \qquad wL-\frac{J^2}{4}>0.
\end{equation}
For one component with $w=N$, $L=N_p$ and $J=J_L$, this gives
\begin{equation}
 S_{\rm BH}=2\pi\sqrt{NN_p-\frac{J_L^2}{4}}
            =2\pi N\sqrt{\delta}.
\end{equation}

The component-string description also permits the conserved charges to be
distributed among several cycles. For a partition $p=(w_1,\ldots,w_c)$,
\begin{equation}
 \sum_iw_i=N,\qquad
 \sum_iL_i=N_p,\qquad
 \sum_iJ_i=J_L.
\end{equation}
Using the charge densities already defined in
Section~\ref{sec:review-fortuity-program}, set
$\nu_i=L_i/w_i$ and $\jmath_i=J_i/w_i$. The leading entropy then obeys
\begin{align}
 S_p
 &=2\pi\sum_iw_i\sqrt{\nu_i-\frac{\jmath_i^2}{4}}+o(N), \nonumber \\
 \frac{S_p}{2\pi N}
 &\le
 \sqrt{\nu-\frac{\jmath^2}{4}}+o(1)
 =\sqrt{\delta}+o(1).
\end{align}
The inequality is the concavity bound at fixed total momentum and spin.
It is saturated when every component carries the same momentum and spin
densities as the full state,
\begin{equation}
 \nu_i=\nu,\qquad \jmath_i=\jmath,
 \qquad\Longleftrightarrow\qquad
 L_i=\nu w_i,\qquad J_i=\jmath w_i.
\end{equation}
For this charge assignment, each component contributes an entropy
proportional to its winding,
\begin{equation}
 S_p
 =2\pi\sqrt{\delta}\sum_i w_i+o(N)
 =2\pi N\sqrt{\delta}+o(N)
 =S_{\rm BH}+o(N).
\end{equation}
For any admissible cycle partition, these conditions fix the
charge carried by each component but not the partition of the total winding
$N$. Every partition that allows the proportional charge assignment has the
same leading Cardy exponent, so the selection mechanism must arise at
subleading order.

BPS typicality has nevertheless been studied directly in the two-charge
Ramond-ground sector.  At vanishing R-charge, the four bosonic and four
fermionic ground-state polarizations give Bose--Einstein and Fermi--Dirac
occupations whose sum at winding $w$ is
\begin{align}
 \langle n_w\rangle
 &=4\left(\frac{1}{e^{\beta w}-1}
          +\frac{1}{e^{\beta w}+1}\right)
 =\frac{8}{\sinh(\beta w)}, \nonumber \\
 N&=\sum_{w\ge1}w\langle n_w\rangle
 \simeq\frac{2\pi^2}{\beta^2}.
 \label{eq:review-two-charge-typical-twists}
\end{align}
Here $n_w$ is the total number of length-$w$ component strings after summing
the ground-state polarizations, and $\beta$ fixes the total winding $N$.
Thus $\beta=O(N^{-1/2})$ and the twists that make a significant contribution
have $w\sim1/\beta=O(\sqrt N)$ \cite{BalasubramanianKrausShigemori}.
For a generic light untwisted operator \(\mathcal O\), this two-charge typical ensemble obeys
\begin{equation}
 \langle\mathcal O(t)\mathcal O(0)\rangle_{\rm typ}
 \simeq G_{M=0\,{\rm BTZ}}(t),
 \qquad t\ll t_c=O(\sqrt N),
\end{equation}
where $G_{M=0\,{\rm BTZ}}$ is the corresponding massless-BTZ two-point
function; the finite-cycle origin of this limit is reviewed in the next
subsection.  At nonzero R-charge, a condensate of length-one strings carries
the charge while the remaining winding follows the approximately neutral
typical distribution.  Thus, specifically in this two-charge BPS ensemble,
typicality does not produce one almost-maximal component string.

For three-charge BPS states, Ref.~\cite{BenaMoulting} identified, at the
symmetric-orbifold point, an additional entropy-dominant configuration
beyond the ordinary BMPV long-string saddle. They called this the
\emph{enigmatic phase}. It consists of one excited long component string of length \(N-\ell\), together with \(\ell\) aligned length-one component strings in the same polarization state. The condensed regime is characterized by \(\ell=O(N)\), so this single aligned species acquires a macroscopic occupation number.

These aligned length-one component strings are Ramond ground states. They carry winding and
angular momentum but no momentum excitations, while all of $N_p$ remains
on the excited long component. Thus
\begin{equation}
\begin{aligned}
 w_{\rm long}&=N-\ell, \\
 N_{p,\rm long}&=N_p, \\
 J_{L,\rm short}&=\ell, \\
 J_{L,\rm long}&=J_L-\ell .
\end{aligned}
\end{equation}
At leading order in the large-$N$ entropy, only the excited long component
contributes extensively. For fixed condensate occupation $\ell$, its Cardy
entropy is therefore
\begin{equation}
 S_{{\rm enigma},\ell}
 =2\pi\sqrt{(N-\ell)N_p-\frac{(J_L-\ell)^2}{4}}
 +o(N).
 \label{eq:review-moulting-long-string-entropy}
\end{equation}
Here $S_{{\rm enigma},\ell}$ denotes the entropy of this long--short
configuration at fixed $\ell$. Maximizing it over the condensate occupation
gives
\begin{align}
 \ell_*&=J_L-2N_p, \nonumber\\
 S_{\rm enigma}
 &=2\pi\sqrt{N_p\bigl(N+N_p-J_L\bigr)},
 \label{eq:review-moulting-extremum}
\end{align}
where $S_{\rm enigma}=S_{{\rm enigma},\ell_*}$ is the entropy of the
entropy-maximizing condensed configuration, $dS_{{\rm enigma},\ell}/d\ell\vert_{\ell=\ell_*}=0$. On the positive-spin branch,
this saddle has a macroscopic condensate when $\ell_*>0$, together with
$N_p>0$ and $N_p+N-J_L>0$; the negative-spin branch follows by charge
conjugation.
Its excess over the BMPV long-string entropy is
\begin{equation}
 \frac{S_{\rm enigma}^2-S_{\rm BMPV}^2}{4\pi^2}
 =\left(\frac{J_L}{2}-N_p\right)^2.
\end{equation}
Thus the transition to the enigmatic phase is a genuine short-string condensation: a finite fraction of the total winding and spin is transferred to aligned length-one Ramond ground states, while the residual long component carries the momentum excitations and the leading entropy.

The cohomology and continuation analysis of
Section~\ref{sec:review-fortuity-program} divides the BPS space according
to higher-$N$ persistence. The cycle geometry of the same orbifold states
is measured by
\begin{equation}
 d=N-w_{\max}.
\end{equation}
The remaining question is how the three-charge BPS sector carrying the black-hole entropy is distributed in cycle depth \(d\), and how this distribution changes after restricting to physically distinct BPS subspaces. In particular, we will compare the exact non-graviton sector and, with additional continuation input, the monotone and fortuitous sectors.

Once a cycle distribution is specified, the winding of each component fixes
the fractional mode spacing and the finite covering circle sampled by an
untwisted operator. The next subsection reviews how those finite circles
appear in probe correlators and how their continuum limit produces the
BTZ-like response.

\subsection{Probe correlators and emergent spacetime}
\label{sec:review-emergent-spacetime}

A long component string provides a natural microscopic setting in which
to ask how an effective spacetime response can emerge.  A component string
of winding $n$ has fractional mode spacing of order $1/n$, so as
$n\to\infty$ its discrete spectrum can approach a continuum and its
light-probe correlators can develop a black-hole-like form.  The important
question is whether such a continuum response is sufficient to diagnose a
semiclassical horizon.  Symmetric-product orbifolds provide a sharp test of
this question: their large-$N$ correlators can reproduce BTZ-like behavior
even though the orbifold point itself is not described by semiclassical
Einstein gravity \cite{BelinBintanjaCastroKnop}.  This is the \textit{mirage}
problem for emergent spacetime.

Finite cycle length, sum over periodic images, and finite-winding recurrences provide additional information beyond the strict large-$n$ correlator.
We first review the two relevant symmetric-product results: large-$N$
thermal universality and the exact twisted-ground-state image formula.
In \cite{BelinBintanjaCastroKnop}, the authors consider a light untwisted single-trace
operator
\begin{equation}
 \mathcal O(z,\bar z)=\frac1{\sqrt N}
       \sum_{i=1}^N\mathcal O^{(i)}_{\rm seed}(z,\bar z),\qquad h_{\mathcal O}=\bar h_{\mathcal O}=O(1).
\end{equation}

Here $\beta$ is the dimensionless inverse temperature on the boundary
cylinder of spatial circumference $2\pi$. For $\beta<2\pi$, its leading
large-$N$ thermal two-point function is

\begin{align}
 \lim_{N\to\infty}
   \langle\mathcal O(z,\bar z)\mathcal O(0)\rangle_\beta =\sum_{j\in\mathbb Z}
 \left[\frac{\pi/\beta}
 {\sinh\!\left(\pi(z+2\pi j)/\beta\right)}\right]^{2h_{\mathcal O}}
 \left[\frac{\pi/\beta}
 {\sinh\!\left(\pi(\bar z+2\pi j)/\beta\right)}\right]^{2\bar h_{\mathcal O}} . \label{eq:review-Castro-thermal}
\end{align}
Here $j$ labels spatial images on the boundary cylinder.  Writing
$z=\phi+i t_E$ with $\phi\sim\phi+2\pi$, the shift
$z\to z+2\pi j$ winds the insertion around the spatial circle rather than
the Euclidean-time circle.  The same image structure appears in the BTZ
boundary correlator because BTZ is obtained by the corresponding angular
identification.  In the symmetric orbifold, a length-$n$ component string permits $n$
relative positions of the two insertions around the cycle.  Summing those
positions produces a finite set of terms.  As $n$ becomes macroscopic, the
fractional spectrum becomes dense and this finite-cycle sum approaches the
infinite BTZ spatial-image sum.

This agreement has a direct implication for the spectral response of a
light probe.  Write the large-$N$ thermal correlator as
\begin{equation}
 G_{\mathcal O}^{(\infty)}(t)
 =
 \int_{\mathbb R}\frac{d\omega}{2\pi}\,
 \rho_{\mathcal O}^{(\infty)}(\omega)e^{-i\omega t},
\end{equation}
where $\rho_{\mathcal O}^{(\infty)}(\omega)$ is the thermal spectral density
of the operator $\mathcal O$.  Its support,
\begin{equation}
 \operatorname{supp}\rho_{\mathcal O}^{(\infty)}
 =
 \overline{
 \left\{
 \omega\in\mathbb R:
 \rho_{\mathcal O}^{(\infty)}(\omega)\neq0
 \right\}},
\end{equation}
is the set of energy differences to which the probe has nonzero spectral
weight.

In a bulk black-hole geometry, the infinite near-horizon redshift stretches the radial tortoise coordinate to infinite depth; on the boundary this is reflected in a continuous large-\(N\) frequency spectrum.
The operator-algebra statements used here concern the algebra generated
by a thermal generalized free field in the strict large-$N$ limit
\cite{LeutheusserLiuCausal,LeutheusserLiuEmergent}. Spectral support alone
is not a general criterion for causal depth. For example, the sufficient
condition in Proposition~2.13 of Ref.~\cite{GesteauLiuStringy} requires
the commutator spectral function to be continuous, to vanish only at the
origin with a continuous nonzero first derivative there, and to decay
no faster than polynomially. With the Wightman convention above, that
function is $(1-e^{-\beta\omega})\rho_{\mathcal O}^{(\infty)}(\omega)$.
Accordingly, the type-III$_1$ and infinite-depth diagnoses below use the
full thermal BTZ generalized-free-field correlator; matching only the
support of $\rho_{\mathcal O}^{(\infty)}$ would not suffice.

The causal depth quantifies the radial extent of the effective bulk region probed by \(\mathcal O\). Finite causal depth corresponds to a finite accessible region, whereas infinite causal depth signals an arbitrarily deep radial region, as expected in the presence of a horizon.

For the untwisted operators, the symmetric-product correlator
is identical at leading large $N$ to the BTZ correlator,
\begin{align}
 G_{\mathcal O,\,{\rm orb}}^{(\infty)}(t)
 &=
 G_{\mathcal O,\,{\rm BTZ}}^{(\infty)}(t), \nonumber \\
 \rho_{\mathcal O,\,{\rm orb}}^{(\infty)}(\omega)
 &=
 \rho_{\mathcal O,\,{\rm BTZ}}^{(\infty)}(\omega).
\end{align}
Each such probe consequently passes the same continuous-spectrum and
infinite-depth tests as it would in a BTZ background.

 The symmetric-product orbifold is a
tensionless string theory with an infinite tower of light higher-spin
degrees of freedom, rather than a CFT in a semiclassical Einstein-gravity
regime.  Nevertheless, every light untwisted probe considered individually
can exhibit the same large-$N$ thermal response as a field propagating in
BTZ. The spectral response therefore captures the effective geometry seen by one probe, but it is not by itself sufficient to establish a universal
semiclassical spacetime with a sharp horizon.

Ref.~\cite{BelinBintanjaCastroKnop} discusses three possible
places where the distinction can reappear.  First, the algebra generated by
all light operators may differ from the algebra of any one probe after the
large stringy spectrum is resummed.  Second, information absent in the
strict $N\to\infty$ limit can enter through subleading corrections. We denote the
finite-$N$ correlator by $G_{\mathcal O,N}$ and its correction relative to the BTZ
reference by $\delta G_{\mathcal O,N}$; schematically
\begin{equation}
\begin{aligned}
G_{\mathcal O,N}(t)
 &=
 G_{\mathcal O,\,{\rm BTZ}}(t)
 +
 \delta G_{\mathcal O,N}(t),
 \\
 \delta G_{\mathcal O,N}(t)&\to0
 \quad
 (N\to\infty\ {\rm at\ fixed}\ t).
\end{aligned}
\end{equation}
Third, different probe operators may acquire different causal depths,
which would be characteristic of a stringy, probe-dependent geometry.

The second possibility is most directly related to the
calculation in this paper. We keep a macroscopic component string at finite
winding $n$ and ask which periodic-image contributions, recurrence scales,
and late-time corrections become invisible in the strict continuum limit.
For a length-$n$ cycle, the fields propagate
on a covering circle of circumference $2\pi n$.  The relative position of the two insertions along the \(n\)-cycle is labeled by \(j=0,1,\ldots,n-1\), and the
orbifold-invariant correlator contains the sum over these $n$ relative
positions.  

  For a twisted ground state containing one cycle of length $n$, let
$|\sigma_n\rangle$ denote that length-$n$ twisted ground state and the
subscript ${\rm c}$ the connected contraction. The connected contribution
of the same light untwisted probe is exactly
\begin{equation}
 \begin{aligned}
 \langle\sigma_n|\mathcal O(z,\bar z)\mathcal O(0)|\sigma_n\rangle_{\rm c}
 &=\frac nN G_n^{(0)}(z,\bar z), \\
 G_n^{(0)}(z,\bar z)
 &=\sum_{j=0}^{n-1}
 \frac{1}{
 [2n\sin((z-2\pi j)/(2n))]^{2h_{\mathcal O}}
 [2n\sin((\bar z-2\pi j)/(2n))]^{2\bar h_{\mathcal O}}}.
\end{aligned}
\end{equation}
For a general cycle partition $p=(1^{n_1}2^{n_2}\cdots)$, the same-cycle
contraction rule gives the exact weighted sum
\begin{equation}
 G_p^{(0)}(z,\bar z)
 =\frac1N\sum_{w\ge1}w n_wG_w^{(0)}(z,\bar z),\qquad \sum_{w\ge1}w n_w=N, \label{eq:review-probe-cycle-weights}
\end{equation}
which is the finite-cycle structure used again in the probe calculation
below.  If $n$ grows with $N$, the finite image sum approaches
\begin{equation}
 G_n^{(0)}(z,\bar z)
 \longrightarrow
 \sum_{j\in\mathbb Z}
 \frac1{(z-2\pi j)^{2h_{\mathcal O}}
         (\bar z-2\pi j)^{2\bar h_{\mathcal O}}},
\end{equation}
which has the image structure of the massless-BTZ limit.  This twisted-ground
state statement is distinct from the massive thermal-BTZ result in
Eq.~\eqref{eq:review-Castro-thermal}.

In Section~\ref{sec:finiteN-probe-response}, we consider a length-\(n\) component string with thermally populated neutral left-moving oscillators, while the right-moving sector is kept in a Ramond ground state. The calculation keeps the exact finite
covering circle for a controlled probe, resolves the image terms invisible at the strict continuum limit, and determines when those finite-winding
terms become important at long times.

The review in Section~\ref{sec:framework} leads to three questions addressed below:
\begin{itemize}

\item \textbf{From R-charge concentration to cycle geometry.}
The R-charge-concentration conjecture suggests that fortuitous cohomology
has a nontrivial organization inside the supercharge complex, but it does
not determine how the same states are organized by cycle partition. We therefore
ask whether higher-$N$ persistence is correlated with the cycle geometry of
individual BPS states. For a partition with component lengths $w_i$, define
\begin{equation}
 d(p)=N-\max_i w_i .
\end{equation}
Does fortuity have a characteristic cycle signature, such as maximal or
parametrically long winding, and conversely can a long component string be
used as an indicator of fortuity? The finite-$N$ tests below compare
higher-$N$ continuation directly with cycle support, including states built
entirely from short component strings.

\item \textbf{Collective cycle geometry of the entropy-carrying BPS sector.}
We next move from individual states to the fixed-charge ensemble relevant
for black-hole entropy. Since the leading Cardy exponent does not distinguish
between admissible cycle partitions, we ask whether the full fixed-charge
count nevertheless selects a universal cycle organization at subleading
order. In particular, does the entropy-carrying sector concentrate on one
component string of winding $N-O_{\Pr}(1)$, and does this collective
long-string behavior persist after restricting first to the exact BPS
sector and then to states outside the generalized-gravity sector?

This also allows the monotone/fortuitous question to be posed in the reverse
direction. If the black-hole-scale BPS degeneracy is carried predominantly
by the fortuitous sector, does the collective fixed-charge measure imply a
corresponding long-string concentration for that sector, even though
fortuity itself is defined through higher-$N$ continuation rather than cycle
geometry?

Finally, we ask how this collective organization changes near the moulting
boundary. Does the finite short-string remainder become a macroscopic
occupation of aligned singly wound strings, producing a transition from an
almost-maximal long string to a long core accompanied by an $O(N)$
short-string condensate? We then ask whether this condensed cycle
organization survives after passing from the free-orbifold count to the
protected BPS spectrum, and, with the required continuation input, to the
monotone and fortuitous sectors separately.

\item \textbf{Finite-winding dynamics and probes of emergent spacetime.}
Once the collective cycle geometry is determined, we ask how it is seen by
interactions and by finite-circle probes. In the long-string regime, the
relevant join/split process is highly asymmetric, with one component of
winding $M\gg1$ interacting with a component of fixed winding $a$. We ask whether the suppression of individual twist amplitudes at large winding is compensated by the growing density of fractional modes, leaving a finite long-string interaction in the continuum limit.

We then ask whether the probe response can resolve the same change of cycle
organization found in the fixed-charge ensemble. In particular, how do the
component-number correction, the departure from the BTZ-like continuum
response, and the long-component recurrence scale change from the
finite-remainder regime, through the critical region, and into the
short-string-condensed phase? This provides a direct test of whether the
finite-circle response can read the order parameter governing the cycle
transition, while remaining sensitive to state-dependent spectral weights
that are not fixed by the cycle geometry alone.

\end{itemize}

\clearpage
\section{Fortuity and cycle geometry}
\label{sec:microscopic-fortuity}

The fortuity program organizes BPS states through their continuation to higher \(N\), whereas the symmetric-orbifold description resolves the same states by component-string cycle structure. Previous work has emphasized the longest D1--D5 cochain complexes and maximally twisted sectors in connection with black-hole microstates \cite{ChangLinFortuity,ChangLinZhang,ChangZhang}. This motivates a direct state-level test of whether fortuity is correlated with maximal or parametrically long winding.

We compare higher-$N$ persistence with component-string geometry in four
examples: the continuation of $F_3$, a same-charge six-cycle pair, an
even-$N$ family supported entirely on length-two components, and a
same-charge $N=4$ pair with reversed cycle ordering. These examples address
whether fortuity requires long winding, whether maximal winding implies
fortuity, and whether cycle length is ordered in any way by higher-$N$
persistence. Throughout, $\mathcal H_N$ denotes the first-order harmonic
space, with continuation defined by Eq.~\eqref{eq:framework-short-affine}.

The normalized states used below are
\begin{align}
 |F_2\rangle
 &=\sqrt{\frac43}\,\psi^-_0
   (\alpha^1_{-1/2})^3|2_{--,--}\rangle, \nonumber \\
 |F_3\rangle
 &=|\{f_3\}\rangle, \nonumber \\
 |f_3\rangle&=\sqrt{\frac{243}{560}}\,
 \psi^-_{-1/2}\psi^-_{-1/6}
 (\alpha^1_{-1/3})^7|3_{--,--}\rangle, \label{eq:F3-corrected} \\
 |F_6\rangle
 &=\frac{6^8}{8!}
 (\alpha^2_{-1/6})^8(\bar\alpha^2_{-1/6})^8
 \psi^-_{-1/6}\bar\psi^-_{-1/6}|6;R_{--}\rangle, \label{eq:F6-maximal-twist} \\
 |M_0\rangle
 &=\frac1{\sqrt{17}}J^-_{-3}|R^{--}_6\rangle
 =\frac1{\sqrt{17}}\sum_{r=1}^{17}
 \psi^-_{-r/6}\bar\psi^-_{-(18-r)/6}|6;R_{--}\rangle. \nonumber
\end{align}
The states \(F_2\) and \(F_3\) are written in NS conventions, while \(F_6\) and \(M_0\) are written in Ramond conventions; \(f_3\) is the oscillator word defining \(F_3\), and Eq.~\eqref{eq:spectral-flow} relates the two descriptions. Bars on the
oscillators in $F_6$ distinguish internal polarizations, not chirality.
The labels \(F_2\), \(F_3\), and \(F_6\) refer to distinct states defined at \(N=2,3,\) and \(6\), respectively. They are not successive continuations of a single state. The known $F_2$ class
\cite{ChangZhang} will also supply the length-two blocks of an even-$N$
family defined in Section~\ref{sec:heavy-fortuitous-family}.

\label{sec:review-fortuity}
Let $\Omega$ be a product of protected chiral primaries and $A$ a word
in total left creation generators. We write $[v]_{Q_N}$ for the
$Q_N$-cohomology class represented by $v$. The reference space for the
individual state comparisons is
\begin{equation}
\mathcal T_N=\operatorname{span}\{[A\Omega]_{Q_N}\}
 \subseteq\mathcal H_N^{\rm mon}.
 \label{eq:intrinsic-tower-definition}
\end{equation}
It is a specified subspace of the larger generalized-gravity space
$\mathcal G_N$ reviewed in Section~\ref{sec:review-fortuity-program}
\cite{ToweringGravitons}. To see its monotonicity, add NS vacuum strands
to obtain $\Omega_M$ and use the same word in total generators at rank $M$:
\begin{align}
 Q_M A_M\Omega_M&=Q_M^\dagger A_M\Omega_M=0, \label{eq:tower-representatives-harmonic} \\
 \pi_{N,M}(A_M\Omega_M)&=A_N\Omega_N,\qquad (M>N). \label{eq:intrinsic-tower-explicit-lift}
\end{align}
The first relation follows from the commutation of \(Q_M\) and \(Q_M^\dagger\) with the total left generators on the free BPS space. Under \(\pi_{N,M}\), all terms exciting the added vacuum strands vanish, leaving \(A_N\Omega_N\).
Appendix~\ref{app:F6-protected-projectors} derives both steps.

For a partition $p=(w_1,\ldots,w_c)$ of $N$, define its cycle depth by
\begin{equation}
 d(p)=N-\max_i w_i,\qquad \sum_iw_i=N.
\end{equation}
Let $P_p$ be the orthogonal projector onto the
orbifold sector with that component-string content.  We also define
\begin{align}
 P_d&=\sum_{p:\,d(p)=d}P_p, \nonumber \\
 P_{\le R}&=\sum_{d=0}^{R}P_d.
\end{align}
Thus $P_{\le R}$ keeps precisely the components for which the longest
string has winding at least $N-R$.  These projectors act on arbitrary
superpositions as well as on states with a definite partition.

Choose a rank-\(M\) source vector \(x_0\) that projects to the prescribed lower-\(N\) class, and let \(x_1,\ldots,x_s\) span the allowed correction directions within the same projection fibre. Then
\begin{equation}
 y=c_0x_0+\sum_{a=1}^s c_ax_a,\qquad c_0\ne0.
\end{equation}
The images $Q_Mx_a$ lie in the adjacent cochain block reached by the
restricted supercharge.  Let $T_\mu$ be covectors on this block.  Their
matrix elements with the source directions define
\begin{align}
 P_{\mu a}&=\langle T_\mu|Q_M|x_a\rangle, \nonumber \\
 Q_My&=0\ \Longrightarrow\ Pc=0. \label{eq:finite-rank-target-matrix}
\end{align}

A closed continuation would require
\begin{equation}
 Pc=c_0P_{\cdot0}+\sum_{a=1}^{s}c_aP_{\cdot a}=0 .
\end{equation}
To show that no choice of the correction coefficients can satisfy this
condition, it is enough to find a covector $\ell$ such that
\begin{equation}
 \ell P_{\cdot a}=0\quad (a=1,\ldots,s),
 \qquad
 \ell P_{\cdot0}\ne0 .
\end{equation}
Indeed, applying $\ell$ to the closure condition gives
\begin{equation}
 0=\ell Pc=c_0\,\ell P_{\cdot0},
\end{equation}
which contradicts $c_0\ne0$.  We refer to such an $\ell$ as a
\emph{dual witness} of non-continuation.
\label{sec:corrected-n3-lift}

\subsection{An explicit all-rank fortuitous class}
\label{sec:F3-all-rank-fortuity}

We now consider the continuation of the harmonic state $F_3$ in
Eq.~\eqref{eq:F3-corrected}, with
$(N;N_p,J_L,J_R)=(3;4,-3,-1)$.  Use the vacuum-removal maps defined above
and restrict to the NS-sector charge block containing $F_3$.  For $M>3$,
let $y_M$ be any candidate with $\pi_{3,M}y_M=c_0F_3+Q_3z$, $c_0\ne0$.
There is an explicit target functional $\ell_M$, supported on the shape
$(4,1^{M-4})$, such that
\begin{equation}
 \widehat{\mathcal A}_M(y_M)
 := {\cal N}_M^{-1}\ell_M Q_My_M
 =1680\sqrt{M-3}\,c_0,\qquad {\cal N}_M\ne0. \label{eq:F3-all-rank-dual-obstruction}
\end{equation}
Here ${\cal N}_M$ is the common nonzero normalization from the
correlator and the supercharge in the residue convention of
Appendix~\ref{app:F3-dual-witness}.
Consequently,
\begin{align}
 \nexists\,(y_M,z):\quad Q_My_M&=0, \nonumber \\
 \pi_{3,M}y_M&=F_3+Q_3z,\qquad M>3.
\end{align}

At $N=4$, fixing the projection to $F_3$ leaves eight independent
correction directions, so every candidate can be written as
\begin{equation}
 y_4=c_0x_0+\sum_{a=1}^{8}c_ax_a,
 \qquad c_0\ne0 .
 \label{eq:F3-main-explicit-fibre}
\end{equation}
The complete source basis is
\begin{center}
\setlength{\tabcolsep}{4pt}
\begin{tabular}{c|l|l}
 & Three-cycle word & Length-one-strand word\\ \hline
$x_0$&$(\alpha^1_{-1/3})^7\psi^-_{-1/6}\bar\psi^+_{-1/6}\psi^+_{-1/6}\psi^-_{-1/2}$&$1$\\
$x_1$&$(\alpha^1_{-1/3})^6\psi^-_{-1/6}\psi^+_{-1/6}$&$\alpha^1_{-1}$\\
$x_2$&$(\alpha^1_{-1/3})^7$&$\psi^-_{-1/2}\psi^+_{-1/2}$\\
$x_3$&$(\alpha^1_{-1/3})^7\bar\psi^-_{-1/6}\psi^-_{-1/6}\psi^+_{-1/6}$&$\psi^+_{-1/2}$\\
$x_4$&$(\alpha^1_{-1/3})^7\psi^-_{-1/6}\bar\psi^+_{-1/6}\psi^+_{-1/6}$&$\psi^-_{-1/2}$\\
$x_5$&$(\alpha^1_{-1/3})^7\psi^-_{-1/2}$&$\psi^+_{-1/2}$\\
$x_6$&$(\alpha^1_{-1/3})^7\psi^+_{-1/2}$&$\psi^-_{-1/2}$\\
$x_7$&$(\alpha^1_{-1/3})^6\alpha^1_{-2/3}\psi^-_{-1/6}$&$\psi^+_{-1/2}$\\
$x_8$&$(\alpha^1_{-1/3})^6\alpha^1_{-2/3}\psi^+_{-1/6}$&$\psi^-_{-1/2}$
\end{tabular}
\end{center}
Here $x_0$ projects to $F_3$, while $x_1,\ldots,x_8$ contain an excitation on the singly wound strand and lie in the projection kernel. The charge and energy census
establishing completeness is given in Appendix~\ref{app:F3-source-census}.

The six cyclically invariant target covectors needed for the obstruction
can be chosen as follows. Entries denote mode numerators in units of $1/4$,
and all bosons are $\alpha^1$.
\begin{center}
\setlength{\tabcolsep}{5pt}
\begin{tabular}{c|c|c|c|c|c}
 & Bosons & $\psi^+$ & $\bar\psi^+$ & $\psi^-$ & $\bar\psi^-$\\\hline
$T_1$&$(1^5,3)$&$(0,1)$&$(0)$&$(1,2)$&$\varnothing$\\
$T_2$&$(1^5,2)$&$(0,1)$&$(0)$&$(1,3)$&$\varnothing$\\
$T_3$&$(1^5,2)$&$(0,1,2)$&$\varnothing$&$(1)$&$(1)$\\
$T_4$&$(1^5,2)$&$(0,1)$&$(1)$&$(1,2)$&$\varnothing$\\
$T_5$&$(1^4,2,3)$&$(0,2)$&$\varnothing$&$(1)$&$\varnothing$\\
$T_6$&$(1^5,2)$&$(1,3)$&$\varnothing$&$(1)$&$\varnothing$
\end{tabular}
\end{center}
In the source and target orderings displayed above, the residue calculation gives
\begin{equation}
 \begingroup\setlength{\arraycolsep}{3pt}
 P=\begin{pmatrix}
 15&0&8&0&20&6&-2&0&20\\
 385&-176&-784&-140&-1540&196&196&-40&-280\\
 0&0&28&350&0&-7&21&10&0\\
 525&0&-56&0&700&-42&14&0&-20\\
 0&0&28&14&0&-7&21&58&0\\
 35&176&784&-1540&-140&-196&-196&280&40
 \end{pmatrix}.
 \endgroup
 \label{eq:F3-residue-matrix}
\end{equation}
The finite residue sums producing these entries are recorded in
Appendix~\ref{app:F3-dual-witness}. The row vector
\begin{equation}
 \ell=(14,1,5,2,-5,1)
\end{equation}
satisfies
\begin{equation}
 \ell P=(1680,0,\ldots,0).
 \label{eq:F3-dual-witness-main}
\end{equation}
Hence a closed candidate would obey $0=\ell Pc=1680c_0$, which is
impossible because $c_0\ne0$.

The same certificate extends to every $M>4$ without enlarging the active
local problem.  The one-transposition precursors of the four-cycle target
with vacuum spectators have shapes
\begin{align}
 &(3,1^{M-3}),\qquad (2,2,1^{M-4}), \nonumber\\
 &(5,1^{M-5}),\qquad (4,2,1^{M-6}),
\end{align}
with negative multiplicities omitted.  Their minimum right chiral weights
are respectively $1,1,2,2$, whereas the source has $\bar h=1$.  The two
split precursors are therefore absent.  Contracting the vacuum spectators
in the $(2,2,1^{M-4})$ channel leaves the empty rank-four $(2,2)$ source
block, while the $(3,1^{M-3})$ channel reduces exactly to the nine columns
above.  Removing one of the $M-3$ identical singly wound Ramond vacua contributes
$\sqrt{M-3}$ to the prescribed source column.  Thus
\begin{equation}
 \ell P_M=1680\sqrt{M-3}\,(1,0,\ldots,0),
\end{equation}
and the same contradiction excludes a closed continuation at every
higher rank.  Thus
\begin{equation}
[F_3]\in\mathcal H_3^{\rm for}.
\end{equation}

\subsection{A matched pair at equal charges}
\label{sec:w6-maximal-twist}
\label{sec:individual-cycle-tests}

We first compare two states with identical charges and identical cycle
partition but opposite continuation behavior.  We show that $F_6$ is
harmonic at $N=6$ yet has no continuation to $N=7$ by evaluating a single
cover-space matrix element, and then contrast it with the all-rank
continuation of $M_0$.

The two states
\begin{align}
 |F_6\rangle
 &=\frac{6^8}{8!}
 (\alpha^2_{-1/6})^8(\bar\alpha^2_{-1/6})^8
 \psi^-_{-1/6}\bar\psi^-_{-1/6}|6;R_{--}\rangle, \nonumber\\
 |M_0\rangle
 &=\frac1{\sqrt{17}}J^-_{-3}|R^{--}_6\rangle
\end{align}
have the same charges,
\begin{equation}
 (N;N_p,J_L,J_R)=(6;3,-3,-1),
\end{equation}
and both consist of a single length-six component string.  Their
continuation properties are nevertheless different.

Here $J_b^3$ is the Cartan generator of $SU(2)_b$ in the internal
$SU(2)_a\times SU(2)_b$, normalized to eigenvalues $\pm1/2$ on a
doublet, as in Appendix~\ref{app:F3-source-census}. Each of the sixteen
bosons in $F_6$ has weight $-1/2$, while its fermions are singlets under
$SU(2)_b$. Thus, on this polarization sector, $-2J_b^3$ counts these
aligned bosonic excitations; it is a charge operator, not a power to which
the state is raised. For the chosen supercharge component,
\begin{equation}
 \begin{aligned}
  2J_b^3 F_6&=-16F_6,\\
  2J_b^3(Q_6F_6)&=-17Q_6F_6,\\
  2J_b^3(Q_6^\dagger F_6)&=-15Q_6^\dagger F_6.
 \end{aligned}
\end{equation}
At rank six the image of a single six-cycle has two cycles, each of
length at most five. The required charges $(N_p,J_L)=(3,-3)$ therefore
exclude both image sectors:
\begin{align}
 2J_b^3=-17&:\qquad N_p^{\min}\ge\frac{17}{5}>3,\nonumber\\
 2J_b^3=-15&:\qquad N_p^{\min}\ge\frac{15}{5}=3.
\end{align}
In the second case the two Ramond ground states give only
$J_L=-2$, so an additional negative left-moving fermion would be required
to make $J_L=-3$.

Hence
\begin{equation}
 Q_6F_6=Q_6^\dagger F_6=0.
 \label{eq:F6-harmonic-state}
\end{equation}

Suppose that the class $[F_6]$ continued to $N=7$.  By
Eq.~\eqref{eq:framework-short-affine}, there would exist an $N=7$ state
$y$ such that
\begin{align}
 Q_7y&=0, \nonumber\\
 \pi_{6,7}y&=F_6+Q_6z
 \label{eq:F6-N7-continuation-condition}
\end{align}
for some $N=6$ state $z$.  Thus $y$ must be $Q_7$-closed and must reduce,
after removal of the added vacuum strand, to the same $Q_6$-cohomology
class as $F_6$.

The direct vacuum extension is
\begin{align}
 S_7&\propto\bigl|\{F_6,1;R_{--}\}\bigr\rangle,
 & (N;N_p,J_L,J_R)_{S_7}&=(7;3,-4,-2), \nonumber\\
 \pi_{6,7}S_7&=F_6.
 \label{eq:F6-N7-source}
\end{align}
The added singly wound strand is a Ramond vacuum, so it changes the total winding
but adds no oscillator excitation.  In particular, $S_7$ retains the
sixteen bosons and the two mode-$1/6$ fermions of $F_6$.

To test $Q_7$-closure it is sufficient to exhibit one state with nonzero
matrix element against $Q_7y$.  We choose
\begin{equation}
 \ket{\Xi_7}=\frac{7^{15/2}}{\sqrt{7!8!}}
 (\alpha^2_{-1/7})^7(\bar\alpha^2_{-1/7})^8
 \psi^-_{-4/7}\psi^-_{-1/7}\bar\psi^-_{-1/7}
 \ket{7;R_{--}}.
 \label{eq:F6-N7-target}
\end{equation}
It lies in the charge sector reached by the chosen supercharge.  In
particular,
\begin{equation}
 N_p(S_7)=\frac{16+1+1}{6}=3,
 \qquad
 N_p(\Xi_7)=\frac{15+4+1+1}{7}=3.
\end{equation}

A source contributing to \(\langle\Xi_7|Q_7|\cdot\rangle\) must contain two
cycles, since the twist-two interaction joins them into the single
seven-cycle.  The only possible partitions are therefore
\((6,1)\), \((5,2)\), and \((4,3)\).  The source sector has internal Cartan
charge \(-16\), so at least sixteen aligned bosonic excitations are required.
Their momentum is minimized by placing all sixteen on the longer cycle.  The
corresponding minimal bosonic momenta are $\frac{16}{6},\ \frac{16}{5},\ \frac{16}{4}$, for \((6,1)\), \((5,2)\), and \((4,3)\), respectively.  Since the available
momentum is \(3\), the latter two partitions are excluded.

For the surviving \((6,1)\) sector, the bosons use \(16/6=8/3\) units of
momentum, leaving \(1/3\).  The fixed charges then require one
\(\psi^-\) and one \(\bar\psi^-\), whose lowest allowed six-cycle modes
contribute $\frac16+\frac16=\frac13$. Thus the remaining momentum is saturated exactly, and Pauli exclusion leaves
no further choice.  The source that projects onto \(F_6\) and can couple to
\(\Xi_7\) is therefore the single direction \(S_7\).

Accordingly, write a putative continuation as
\begin{equation}
 y=cS_7+y_\perp,
 \qquad
 \langle F_6|\pi_{6,7}y_\perp\rangle=0,
 \qquad
 \langle\Xi_7|Q_7y_\perp\rangle=0.
 \label{eq:F6-complete-witness-kernel}
\end{equation}
Taking the inner product of
Eq.~\eqref{eq:F6-N7-continuation-condition} with $\langle F_6|$ and using
$Q_6^\dagger F_6=0$ together with $\pi_{6,7}S_7=F_6$ fixes
\begin{equation}
 c=1.
 \label{eq:F6-source-coefficient}
\end{equation}
It remains to evaluate the single matrix element
$\langle\Xi_7|Q_7S_7\rangle$.  

The relevant term in $Q_7$ contains a twist-two operator that joins the
length-six component of $S_7$ with the singly wound strand into the length-seven
component of $\Xi_7$.  Let \(z=\Gamma(t)\) denote the covering map from the cover coordinate \(t\) to the base-space coordinate \(z\). We evaluate this twist correlator on the covering
surface.  Specializing the standard twist-two covering map \(z=t^M(t-a)^N\) to \(M=6\), \(N=1\), and choosing the twist insertion at \(t=1\), we obtain
\begin{equation}
 z=\Gamma(t)=t^6(7-6t).
\end{equation}
Its derivative is $\Gamma'(t)=42t^5(1-t)$.  Thus $t=1$ is the simple
ramification point associated with the twist-two insertion, while
$t=7/6$ is the second, unramified preimage of $z=0$ in addition to the
order-six preimage at $t=0$.  The relevant preimages are therefore
\begin{equation}
\begin{array}{c|c|c}
\text{base point} & \text{preimage on the cover} & \text{interpretation} \\ \hline
z=0
& t=0\ \text{(multiplicity 6)},\quad t=7/6\ \text{(multiplicity 1)}
& (6,1)\ \text{incoming cycles}
\\
z=1
& t=1\ \text{(multiplicity 2)}
& \text{twist-two insertion}
\\
z=\infty
& t=\infty\ \text{(degree 7)}
& \text{outgoing seven-cycle}
\end{array}.
\end{equation}

We now evaluate the fermionic part directly on the cover. A twist insertion at \(z=z_0\) is lifted to a point \(t=t_0\) satisfying $\Gamma(t_0)=z_0$. The multivalued fermion on the base becomes a single-valued free fermion on the cover. We denote this cover fermion by \(\psi(t)\); since the fermion has conformal weight \(1/2\), it is related to the base-space field by
\begin{align}
\psi(z)=\bigl(\Gamma'(t)\bigr)^{-1/2}\psi(t).
\end{align}

For a
weight-$1/2$ fermion, the standard Lunin--Mathur lift of a fractional
mode is, up to the state-normalization factor,
\begin{equation}
 \oint_{z_0}\frac{dz}{2\pi i}\,
       \psi(z)(z-z_0)^{r-1/2}
 \quad\longrightarrow\quad
 \oint_{t_0}\frac{dt}{2\pi i}\,
       \psi(t)\sqrt{\Gamma'(t)}\,
       \bigl(\Gamma(t)-z_0\bigr)^{r-1/2}.
 \label{eq:F6-main-fractional-lift}
\end{equation}
This is the usual covering-space prescription
\cite{LuninMathurOrbifold,LuninMathurThreePoint,GaberdielGopakumarNairz,ChangLinZhang}.
The remaining powers of $t$ are fixed by the spin fields.  In the
bosonization convention
\begin{equation}
 \psi^+=e^{i\phi_1},\qquad
 \bar\psi^-=-e^{-i\phi_1},\qquad
 \bar\psi^+=e^{i\phi_2},\qquad
 \psi^-=e^{-i\phi_2}{.}
 \label{eq:F6-main-bosonization}
\end{equation}
With this convention, the signs of the bosonized charges are fixed by the
Ramond polarizations of the incoming and outgoing states and by the chosen
component of the supercharge.  The two incoming strands are both in the
\(R_{--}(t)\) Ramond vacuum, while the outgoing state appears as its conjugate
bra.  The four spin-field insertions on the cover are therefore
\begin{align}
R_{--}(0)
&=e^{-\frac{i}{2}(\phi_1+\phi_2)},
\nonumber \\
R_{--}\left(\frac76\right)
&=e^{-\frac{i}{2}(\phi_1+\phi_2)},
\nonumber \\
S_Q(1)
&=e^{+\frac{i}{2}(\phi_1-\phi_2)},
\nonumber \\
R_{--}^{\dagger}(\infty)
&=e^{+\frac{i}{2}(\phi_1+\phi_2)}.
\label{eq:F6-main-spin-fields}
\end{align}
The first two signs thus follow from the \(R_{--}\) polarization, the signs
at infinity are reversed by radial conjugation, and the mixed signs at
\(t=1\) select the supercharge component used in \(Q_7\).

The fermionic correlator separates into the two bosonized species in
Eq.~\eqref{eq:F6-main-bosonization}.  The six-cycle part of $S_7$ carries
one $\bar\psi^-_{-1/6}$ oscillator in the $\phi_1$ sector and one
$\psi^-_{-1/6}$ oscillator in the $\phi_2$ sector.  The second incoming
strand, whose preimage is $t=7/6$, is the added singly wound $R_{--}$ Ramond vacuum:
it contributes the spin field in Eq.~\eqref{eq:F6-main-spin-fields} but no
fermion-mode contour.  On the outgoing side, radial conjugation turns the
$\bar\psi^-_{-1/7}$ mode into a single $\psi^+$ bra insertion in the
$\phi_1$ sector, whereas the two modes $\psi^-_{-4/7}$ and
$\psi^-_{-1/7}$ become two $\bar\psi^+$ bra insertions in the $\phi_2$
sector.  Thus the $\phi_1$ sector contains only one incoming--outgoing
pair and gives a single nonzero contraction, equal to one.  The only
nontrivial fermionic Wick sum is therefore the $\phi_2$ sector, where the
two outgoing $\bar\psi^+$ insertions can contract with two
negative-charge sources: the six-cycle mode $\psi^-_{-1/6}$ and the local
fermion supplied by $Q_7$ at $t=1$.

Substituting $\Gamma(t)=t^6(7-6t)$ into
Eq.~\eqref{eq:F6-main-fractional-lift} and including the spin-field
correlator, the cover contour representing the incoming six-cycle
$\psi^-_{-1/6}$ mode is
\begin{equation}
 \oint_0\frac{du}{2\pi i}\,
 \psi^-(u)\,
 u^{-1}\left(1-\frac{6u}{7}\right)^{-1/6},
 \label{eq:F6-main-incoming-fermion-contour}
\end{equation}
while the two conjugate outgoing bra modes are
\begin{align}
 &\oint_\infty\frac{dt_1}{2\pi i}\,
 \bar\psi^+(t_1)(t_1-1)t_1^2
 \left(1-\frac{7}{6t_1}\right)^{-3/7},
 \label{eq:F6-main-outgoing-47}\\
 &\oint_\infty\frac{dt_2}{2\pi i}\,
 \bar\psi^+(t_2)(t_2-1)t_2^{-1}
 \left(1-\frac{7}{6t_2}\right)^{-6/7}.
 \label{eq:F6-main-outgoing-17}
\end{align}
The local $Q_7$ insertion supplies the remaining $\psi^-(1)$.  Hence the
$\phi_2$ contribution to the left-moving fermionic part of
$\langle\Xi_7|Q_7|S_7\rangle$ is the three-contour integral
\begin{align}
 \mathcal A_{\phi_2}
 ={}&\oint_\infty\frac{dt_1}{2\pi i}
     \oint_\infty\frac{dt_2}{2\pi i}
     \oint_0\frac{du}{2\pi i}\,
 (t_1-1)t_1^2\left(1-\frac{7}{6t_1}\right)^{-3/7}
 \nonumber\\
 &\times
 (t_2-1)t_2^{-1}\left(1-\frac{7}{6t_2}\right)^{-6/7}
 u^{-1}\left(1-\frac{6u}{7}\right)^{-1/6}
 \nonumber\\
 &\times
 \bigl\langle
 \bar\psi^+(t_1)\bar\psi^+(t_2)\psi^-(1)\psi^-(u)
 \bigr\rangle .
 \label{eq:F6-main-integrated-phi2-correlator}
\end{align}
Using
\begin{equation}
 \bar\psi^+(t)\psi^-(u)\sim\frac{1}{t-u},
\end{equation}
Wick's theorem gives
\begin{equation}
 \langle
 \bar\psi^+(t_1)\bar\psi^+(t_2)\psi^-(1)\psi^-(u)
 \rangle
 =\underbrace{\frac{1}{t_1-1}\frac{1}{t_2-u}}
 _{(4/7\ \text{with }Q_7)(1/7\ \text{with incoming }1/6)}
 -\underbrace{\frac{1}{t_1-u}\frac{1}{t_2-1}}
 _{(4/7\ \text{with incoming }1/6)(1/7\ \text{with }Q_7)}.
 \label{eq:F6-main-wick-four-fermion}
\end{equation}
Substituting the first Wick pairing into
Eq.~\eqref{eq:F6-main-integrated-phi2-correlator} factorizes the three
contours as
\begin{align}
 \left[
 \oint_\infty\frac{dt_1}{2\pi i}\,
 t_1^2\left(1-\frac{7}{6t_1}\right)^{-3/7}
 \right]\times
 \left[
 \oint_\infty\frac{dt_2}{2\pi i}
 \oint_0\frac{du}{2\pi i}\,
 \frac{(t_2-1)t_2^{-1}(1-\frac{7}{6t_2})^{-6/7}
       u^{-1}(1-\frac{6u}{7})^{-1/6}}{t_2-u}
 \right]
 =\frac{85}{216}\cdot1.
 \label{eq:F6-main-first-wick-pairing}
\end{align}
Here the pole $1/(t_1-1)$ from the contraction with the local $Q_7$
fermion cancels the explicit $(t_1-1)$ in the $4/7$ outgoing contour.
The second Wick pairing similarly gives
\begin{align}
 \left[
 \oint_\infty\frac{dt_1}{2\pi i}
 \oint_0\frac{du}{2\pi i}\,
 \frac{(t_1-1)t_1^2(1-\frac{7}{6t_1})^{-3/7}
       u^{-1}(1-\frac{6u}{7})^{-1/6}}{t_1-u}
 \right]
\times
 \left[
 \oint_\infty\frac{dt_2}{2\pi i}\,
 t_2^{-1}\left(1-\frac{7}{6t_2}\right)^{-6/7}
 \right]
 =-\frac5{216}\cdot1.
 \label{eq:F6-main-second-wick-pairing}
\end{align}
The minus sign between these two products is the fermionic minus sign in
Eq.~\eqref{eq:F6-main-wick-four-fermion}.  Therefore
\begin{equation}
 \mathcal A_{\phi_2}
 =\left(\frac{85}{216}\right)(1)
  -\left(-\frac5{216}\right)(1)
 =\frac5{12}\ne0.
 \label{eq:F6-main-phi2-amplitude}
\end{equation}
The same Wick antisymmetrization is encoded by the determinant:
\begin{equation}
 \det\!\begin{pmatrix}
 85/216&-5/216\\
 1&1
 \end{pmatrix}
 =\frac5{12},
 \label{eq:F6-main-fermion-determinant}
\end{equation}
where the rows are the outgoing $4/7$ and $1/7$ modes and the columns are
the $Q_7$ fermion and the incoming $1/6$ mode. Appendix~\ref{app:F6-protected-projectors}
performs the individual residues term by term.  

The $\phi_1$ sector gives
the single contraction mentioned above and equals one.  For the bosons, each propagated oscillator
has unit contour residue, while the oscillator assignments give the factor
$(8!)^2$.  The right-moving spin fields contribute
$\sqrt{7}$.  Thus, up to the overall nonzero cover, state-normalization,
and orbit factors, the matrix element contains the nonzero factor
\begin{equation}
 \langle\Xi_7|Q_7S_7\rangle\ne0.
 \label{eq:F6-finite-w-amplitude}
\end{equation}
Using $c=1$ and
Eq.~\eqref{eq:F6-complete-witness-kernel}, any continuation candidate
would satisfy
\begin{equation}
 \langle\Xi_7|Q_7y\rangle
 =\langle\Xi_7|Q_7S_7\rangle\ne0,
\end{equation}
contradicting $Q_7y=0$.  Thus
\begin{equation}
 [F_6]\in\mathcal H_6^{\rm for}.
\end{equation}

The state $M_0$ behaves oppositely because it is a total-current descendant
of a protected seed.  Let $\Omega_M^{\rm NS}$ be the length-six protected
seed with $M-6$ additional NS vacuum strands and define
\begin{equation}
 \widetilde Y_M=J_{-2}^{-,(M)}\Omega_M^{\rm NS}.
\end{equation}
The Ward identities in
Eqs.~\eqref{eq:tower-representatives-harmonic} and
\eqref{eq:intrinsic-tower-explicit-lift} give, in one step,
\begin{equation}
 Q_M\widetilde Y_M=Q_M^\dagger\widetilde Y_M=0,
 \qquad
 \pi_{6,M}\widetilde Y_M=J_{-2}^{-,(6)}\Omega_6^{\rm NS}.
 \label{eq:M0-explicit-every-rank-lift}
\end{equation}
After spectral flow and normalization the state on the right is $M_0$.
Therefore $M_0\in\mathcal H_6^{\rm mon}$.

The two states also differ in their relation to the protected descendant
space: $F_6$ is orthogonal to it, whereas $M_0$ belongs to it.
After spectral flow, a length-six protected seed at the charges of $F_6$
has $h_0=j_0\in\{5/2,3,7/2\}$, so its descendant level obeys
$\ell\le2$. A product of total-mode creation operators at this level
contains at most
$2\ell\le4$ bosonic oscillators, whereas $F_6$ contains sixteen.  Hence
\begin{equation}
 P_{\mathcal T}F_6=0,
 \qquad
 M_0\in\mathcal T_6\subseteq\mathcal H_6^{\rm mon}.
\end{equation}
Both states nevertheless have
\begin{equation}
 w_{\max}=6=N,
 \qquad d=0.
\end{equation}
Thus a maximal component string does not determine whether an individual
BPS class is monotone or fortuitous.

\subsection{Short-string fortuity and reversed cycle ordering}
\label{sec:short-string-cycle-tests}
\label{sec:heavy-fortuitous-family}
\label{sec:review-finite-composite}

The converse failure persists at arbitrarily large even $N$: fortuitous
states need not contain a long component string at all.  On the
$r$th length-two block let $\eta_r$ and $\bar\eta_r$ be the normalized
right-moving fermion zero-mode creators and set
$\bar\eta_D=k^{-1/2}\sum_{r=1}^k\bar\eta_r$.  A normalized representative
built from $k$ copies of the known $F_2$ state is
\begin{equation}
|F_{2k}^{\rm heavy}\rangle
 =\frac1{\sqrt{(2k)!\,2^k k!}}
 \sum_{\sigma\in S_{2k}}\sigma\!\left[
 \bar\eta_D\eta_1\cdots\eta_k
 \prod_{r=1}^{k}|F_2\rangle_r\right].
 \label{eq:heavy-allF-state}
\end{equation}
The graded permutation symmetrizes the $k$ length-two blocks.  The
composite construction of Ref.~\cite{ChangLinZhang} gives
\begin{equation}
\begin{aligned}
 Q_{2k}F_{2k}^{\rm heavy}&=Q_{2k}^\dagger F_{2k}^{\rm heavy}=0,\\
 (N;N_p,J_L,J_R)&=(2k;2k,-2k,1),\\
 w_{\max}&=2,\qquad d=2k-2.
\end{aligned}
 \label{eq:heavy-main-family-ledger}
\end{equation}
Thus these states remain entirely on length-two component strings as
$N=2k$ grows.

The obstruction is already present in the local $N=2\to3$ problem.  At
the charges of $F_2$, the only rank-three source is the vacuum extension
with partition $(2,1)$:
\begin{equation}
 \begin{array}{c|ccc}
  p&(3)&(2,1)&(1^3)\\\hline
  \dim V_p&0&1&0
 \end{array},
 \qquad \dim\ker\pi_{2,3}=0.
 \label{eq:F2-local-source-ledger}
\end{equation}
For the cover $\Gamma(t)=t^2(3-2t)$ and the target
$(\alpha^1_{-1/3})^4|\sigma_3\rangle_L$, the bosonic propagation factor,
outgoing-mode factor, and unpaired-fermion residue are all one.  The four
choices of outgoing boson and the $3!$ propagated assignments therefore
give
\begin{equation}
 \langle T^{(2)}_3|Q_3|x^{(2)}_3\rangle
 =24C_{2\to3}\ne0.
 \label{eq:F2-local-nonzero-amplitude}
\end{equation}
Because there is no correction direction, the projection condition fixes
the source coefficient to one while $Q_3$-closure would require it to
vanish.  Hence $[F_2]$ has no continuation to $N=3$.

For the $2k\to2k+1$ problem, a twist-two action can reach a target of
cycle type $(3,2^{k-1})$ only from
\begin{align}
 (2^k,1),\quad(3,2^{k-2},1^2)&\xrightarrow{\ Q_{\rm join}\ }(3,2^{k-1}), \nonumber\\
 (5,2^{k-2}),\quad(4,3,2^{k-3})&\xrightarrow{\ Q_{\rm split}\ }(3,2^{k-1}).
\end{align}
The required right outer $SU(2)$ charge is $(k-1)/2$.  In this channel a
source with $c$ cycles can carry at most $(c-2)/2$, so one needs
$c\ge k+1$.  The two split channels contain only $k-1$ cycles and are
therefore absent.  In the second join channel the extra length-two factor is
a harmonic spectator, so its contraction with the local supercharge
vanishes.  The only surviving source type is therefore $(2^k,1)$.

Only one target row is needed; a chain-map identity on the whole
rank-$(2k+1)$ complex is unnecessary. Let $s$ denote the added length-one
strand, and let $\chi=\eta F_2$ and $\varphi=\bar\eta\eta F_2$. For each
$i=1,\ldots,k$, define the map to the active length-three subsystem by
\begin{equation}
 C_{k,i}X:=
 \operatorname{proj}_{\bar\eta_{d_{\rm loc}}\eta_{d_{\rm loc}}}
 \left[
   \left(\bigotimes_{r\ne i}\bra{\chi}_r\right)X
 \right],
 \qquad
 d_{\rm loc}=\frac{(\sqrt2,1)}{\sqrt3}.
 \label{eq:main-Cki-definition}
\end{equation}
The bras remove the $k-1$ spectator two-cycles, leaving the $i$th
two-cycle together with $s$. The projector then retains the right-moving
Clifford component containing the normalized diagonal zero-mode pair
$\bar\eta_{d_{\rm loc}}\eta_{d_{\rm loc}}$. Thus $C_{k,i}X$ is a state in
the local $N=3$ subsystem, with no appendix notation left implicit.

Take the local target $T^{(2)}_3$ in the same Clifford component. The
split precursors are absent, the other join precursor is killed by
$\bra\chi Q_2=0$, and the surviving terms have the same spectator sign.
Up to one common nonzero normalization of the symmetrized target, this gives
\begin{equation}
 \langle\{T^{(2)}_3,\chi^{k-1}\}|Q_{2k+1}|X\rangle
 \propto\sum_{i=1}^k
 \langle T^{(2)}_3|Q_3C_{k,i}|X\rangle.
 \label{eq:heavy-target-row-witness}
\end{equation}
For each active pair, the unique local $2\to3$ matrix element is
\begin{equation}
 \langle T^{(2)}_3|Q_3C_{k,i}|X\rangle
 =24C_{2\to3}
 \langle\varphi|\pi_{2,3}C_{k,i}|X\rangle.
\end{equation}
Finally, the exterior-algebra identity
Eq.~\eqref{eq:main-all-k-identity} gives
\begin{equation}
 \sum_{i=1}^k
 \langle\varphi|\pi_{2,3}C_{k,i}|X\rangle
 =\sqrt{k}\,
 \langle F_{2k}^{\rm heavy}|\pi_{2k,2k+1}|X\rangle.
\end{equation}
A continuation would obey $Q_{2k+1}X=0$ and
$\pi_{2k,2k+1}X=F_{2k}^{\rm heavy}+Q_{2k}z$. The first condition makes
the left side of Eq.~\eqref{eq:heavy-target-row-witness} vanish, while
harmonicity and unit normalization give
$\langle F_{2k}^{\rm heavy}|\pi_{2k,2k+1}|X\rangle=1$. The three displays
would then equate zero to the nonzero number
$24C_{2\to3}\sqrt{k}$, a contradiction. Thus
\begin{equation}
 [F_{2k}^{\rm heavy}]\in\mathcal H_{2k}^{\rm for},\qquad k\ge1.
 \label{eq:heavy-fortuity}
\end{equation}

A complementary same-charge $N=4$ pair reverses the cycle-length ordering.
\label{sec:N4-reversed-cycle-pair}
Take the
protected single-cycle seed
\begin{align}
 |\Omega_4\rangle&=|4_{+-,++}\rangle,
 &(h,j;\bar h,\bar j)&=(2,2;5/2,5/2), \nonumber\\
 |M_4\rangle&=\frac1{\sqrt{960}}
 (L_{-1})^2(J_0^-)^2|\Omega_4\rangle.
 \label{eq:M4-explicit-descendant}
\end{align}
The normalization follows from the standard $SU(2)$ and Virasoro norms,
\begin{equation}
 \|(J_0^-)^2\Omega_4\|^2=24,
 \qquad
 \|(L_{-1})^2|h=2\rangle\|^2=40,
 \qquad 24\times40=960.
\end{equation}
The two $J_0^-$ zero modes lower $j$ from $2$ to $0$, while the two
$L_{-1}$ modes raise $h$ from $2$ to $4$.  Thus
\begin{equation}
 (h,j;\bar h,\bar j)=(4,0;5/2,5/2),
 \qquad
 (N;N_p,J_L,J_R)=(4;4,-4,1).
\end{equation}
Because $M_4$ is a total-mode descendant of a protected seed, the same
Ward identities used above give a closed continuation at every $M>4$:
\begin{equation}
 M_4\in\mathcal T_4\subseteq\mathcal H_4^{\rm mon}.
\end{equation}

Setting $k=2$ in Eq.~\eqref{eq:heavy-main-family-ledger} gives a fortuitous
state $F_4^{\rm heavy}$ with exactly the same charges but cycle partition
$(2,2)$.  Hence
\begin{equation}
 \begin{array}{c|c|c|c}
 \text{state}&\text{class}&p&d\\ \hline
 M_4&\text{monotone}&(4)&0\\
 F_4^{\rm heavy}&\text{fortuitous}&(2,2)&2
 \end{array}.
\end{equation}
At fixed charges, the monotone representative therefore has the longer
cycle.

Taken together, these state-level examples show that neither implication
\begin{equation}
 \text{fortuitous}
 \ \Longrightarrow\
 \text{maximal or long component string},
 \qquad
 \text{maximal cycle}
 \ \Longrightarrow\
 \text{fortuitous}
\end{equation}
holds for individual BPS states.  Table~\ref{tab:finite-rank-summary}
summarizes the continuation and cycle-support tests.

\begin{table}[!htbp]
\centering\small
\setlength{\tabcolsep}{3pt}
\begin{tabular}{@{}c|c|c|c|p{0.24\textwidth}|l@{}}
state & $(N;N_p,J_L,J_R)$ & $d$ & $\|P_{\mathcal T}\psi\|$
& continuation to higher $N$ & class \\ \hline
$F_2$ & $(2;2,-2,-1)$ & $0$ & ---
& none at $N=3$ & fortuitous \\
$F_3$ & $(3;4,-3,-1)$ & $0$ & $0$
& none at any $M>3$ & fortuitous \\
$F_6$ & $(6;3,-3,-1)$ & $0$ & $0$
& none at $N=7$ & fortuitous \\
$M_0$ & $(6;3,-3,-1)$ & $0$ & $1$
& exists at every $M>6$ & monotone \\
$M_4$ & $(4;4,-4,1)$ & $0$ & $1$
& exists at every $M>4$ & monotone \\
$F_4^{\rm heavy}$ & $(4;4,-4,1)$ & $2$ & ---
& none at $N=5$ & fortuitous \\
$F_{2k}^{\rm heavy}$, $k\ge3$ & $(2k;2k,-2k,1)$ & $2k-2$ & ---
& none at $N=2k+1$ & fortuitous
\end{tabular}
\caption{Normalized representatives and their first-order continuation
results for the specified continuation maps. Every row has positive BMPV
charge discriminant. Each listed state is supported in a definite cycle
partition, so its depth $d$ has a single value; a general superposition is
described by the projector probabilities defined below.
Projection norm zero means orthogonality to $\mathcal T_N$, and one
means membership. Dashes mark projections not evaluated here.}
\label{tab:finite-rank-summary}
\end{table}

The point of these finite-state examples is only negative: continuation
class does not determine the cycle partition of an individual state.
Section~\ref{sec:long-strings-lifting} therefore changes the question.  We
fix the total charges, count the whole sector, and ask where its dimension
is concentrated on particular cycle partitions.  The resulting large-$N$ statements are
statistical statements about the sector rather than geometric statements
about every representative.

\section{Collective long-string selection}
\label{sec:long-strings-lifting}

Section~\ref{sec:microscopic-fortuity} showed that higher-$N$ continuation
does not determine cycle geometry state by state. Does the BPS sector
carrying the black-hole entropy nevertheless favor a characteristic cycle
partition at large $N$? The relevant observable is $d=N-w_{\max}$, the
winding outside the longest cycle. The leading Cardy entropy alone does
not distinguish partitions: within the Cardy estimate, every partition
attains the same exponent when momentum is proportional to winding.
Their relative multiplicities therefore depend on the subleading factors
in the fixed-charge count.

Section~\ref{sec:sector-ensemble-construction} derives the free
fixed-charge distribution in the uncondensed region, where one cycle carries
all but $O_{\Pr}(1)$ winding. Section~\ref{sec:phase-and-moments} then
analyzes the critical line and the condensed region, where aligned singly
wound Ramond ground states acquire $O(\sqrt N)$ and $O(N)$ occupation,
respectively. Section~\ref{sec:core-counting} uses the index and free count
to constrain exact BPS and non-graviton subspaces, while
Section~\ref{sec:tower-counting} separates the protected-descendant and
monotone/fortuitous questions. Our counting follows the
D1--D5--P partition-function approach of
Refs.~\cite{Shigemori:2019orj,BenaMoulting}; related discussions of long-string
thermodynamics and tensionless-string partition functions appear in
Refs.~\cite{MartinecLongStrings,EberhardtPartitionFunctions}.

We distinguish three regimes of the free fixed-charge count:
\begin{itemize}
\item the uncondensed region, $\nu>|\jmath|/2$;
\item the critical line, $\nu=|\jmath|/2$, with $0<|\jmath|<1$;
\item the condensed region, $\jmath^2/4<\nu<|\jmath|/2$.
\end{itemize}
When we compare with the rotating BMPV index below, we further restrict to
$0<|\jmath|<1$.

\subsection{Free fixed-charge counting and long-string concentration}
\label{sec:sector-ensemble-construction}
\label{sec:projected-concentration}

We construct the free fixed-charge count, resolve it by cycle partition,
and take its large-\(N\) limit. The subleading dependence announced above
enters through the single-cycle multiplicities, which fall as \(w^{-3}\)
once the common Cardy exponential is removed. That decay is what keeps
\(d\) at \(O_{\Pr}(1)\) in the uncondensed region, and it makes the
short-string family \((2^{N/2})\) of
Section~\ref{sec:short-string-cycle-tests} exponentially rare.

Section~\ref{sec:review-counting-index} showed that a physical cycle of
winding \(w\), momentum \(E\), and left charge \(J\) has
\(4|\widehat c(wE,J)|\) states: \(2|\widehat c(wE,J)|\) bosonic and the
same number of fermionic states.  Here \(\widehat c\) enters only through this
single-cycle degeneracy; the function constructed below is an unsigned
state count, not a protected index.

Each physical single-cycle state contributes a standard Fock factor:
\((1-z^wq^Ey^J)^{-1}\) if bosonic, \(1+z^wq^Ey^J\) if fermionic.  The
right-moving Ramond quartet responsible for the factor of four contains two
states of each statistics, so the \(4|\widehat c(wE,J)|\) states at given
\((w,E,J)\) split evenly and the species contributes
\(\bigl[(1+z^wq^Ey^J)/(1-z^wq^Ey^J)\bigr]^{2|\widehat c(wE,J)|}\).
Multiplying over all physical single-cycle species gives the generating function that counts all free right-BPS Fock basis states:
\begin{align}
Z(z,q,y)
&=\prod_{\substack{w\ge1,\ E\ge0\\ J\in\mathbb Z}}
\left(\frac{1+z^wq^Ey^J}{1-z^wq^Ey^J}\right)^{2|\widehat c(wE,J)|},
\nonumber \\
D_V(\Gamma_N)
&=[z^Nq^{N_p}y^{J_L}]Z(z,q,y).
\label{eq:full-positive-fixed-charge-product}
\end{align}
Here \(z\), \(q\), and \(y\) are fugacities for total winding, momentum,
and left charge, respectively. Thus \(D_V(\Gamma_N)\) is the unsigned
number of free-orbifold states with fixed total charges \((N,N_p,J_L)\).

The same Fock-space count can be resolved by cycle partition. For
\(p=(1^{n_1}2^{n_2}\cdots)\), introduce a formal variable \(u_w\) that
records the number of cycles of winding \(w\); extracting the coefficient
of \(\prod_w u_w^{n_w}\) restricts the count to the partition \(p\).
To resolve the right Ramond charge as well, \(\widetilde y\) denotes its
fugacity, \(t\) the total right charge, and \(t_\alpha\) the right charge of
the physical single-cycle state \(\alpha\). Then
\begin{align}
 \dim V_{p,\Gamma;J_R=t}
 &=\left[\prod_wu_w^{n_w}q^{N_p}y^{J_L}\widetilde y^t\right]
 \prod_{\alpha\ {\rm bosonic}}
   (1-u_{w_\alpha}q^{E_\alpha}y^{J_\alpha}\widetilde y^{t_\alpha})^{-1}
 \prod_{\alpha\ {\rm fermionic}}
   (1+u_{w_\alpha}q^{E_\alpha}y^{J_\alpha}\widetilde y^{t_\alpha}),
 \nonumber\\
 \dim V_{p,\Gamma}&=\sum_t\dim V_{p,\Gamma;J_R=t}.
 \label{eq:partition-resolved-positive-count}
\end{align}
Here \(V_{p,\Gamma;J_R=t}\subset V_{N,\Gamma}\) is the free right-BPS
subspace with cycle partition \(p\) and right charge \(t\).  The refinement
locates the states of Section~\ref{sec:individual-cycle-tests} inside the
count: \(F_6\) and \(M_0\) lie in \(p=(6)\), and the family
\(F_{2k}^{\rm heavy}\) of Section~\ref{sec:short-string-cycle-tests} lies in
\(p=(2^{N/2})\).  These states sit at a rank where the long cycle is not yet dominant.  At
$N=6$ with $(N_p,J_L)=(3,-3)$, an exact count gives the single-six-cycle
partition only $5.13\%$ of the fixed-charge dimension, less than the
$14.1\%$ of $(5,1)$ or the $18.5\%$ of $(4,1^2)$.  The concentration
established below is therefore asymptotic: the explicit states of
Section~\ref{sec:individual-cycle-tests} fix the cycle observable, not the
limit.  The eleven-partition census is in
Appendix~\ref{app:finite-N-partition-census}.

Weights of partitions are compared through the cycle projectors \(P_p\) of
Section~\ref{sec:microscopic-fortuity}. Let \(\mathcal S\) be a subspace
with projector \(P_{\mathcal S}\) and dimension \(D_{\mathcal S}\), and let
\(\Psi\) be a normalized state. The weight of a partition in each case is
\begin{align}
 \langle P_p\rangle_{\mathcal S}
 &=\frac{1}{D_{\mathcal S}}\Tr(P_{\mathcal S}P_p), \nonumber \\
 \langle P_p\rangle_{\Psi}
 &=\langle\Psi|P_p|\Psi\rangle=\|P_p\Psi\|^2 .
 \label{eq:cycle-observable-weight}
\end{align}
For
$P_{\mathcal S}=|\Psi\rangle\langle\Psi|$ and $D_{\mathcal S}=1$, the first
line reduces to the second. The same cycle observable therefore applies to
both a normalized state and a finite-dimensional charge block.
Section~\ref{sec:microscopic-fortuity} evaluates it on individual states and
shows that the monotone or fortuitous label does not fix the cycle weight:
the six-cycle pair has the same maximal winding with opposite labels, whereas
the length-two family is fortuitous with $d=N-2=O(N)$. The present section
instead studies the distribution of this observable over a whole charge
block at large $N$.

The winding dependence is contained in the projected single-cycle
partition function \(Z_w^{\rm phys}\) of
Eq.~\eqref{eq:projected-strand-character}. With \(q=e^{-\beta}\) and
\(y=e^\mu\), we obtain the bosonic and fermionic factors in
Eq.~\eqref{eq:full-positive-fixed-charge-product} by taking the logarithm of the Fock-space product. It converts the product over
single-cycle species into a sum over their repetitions. Expanding
\(\log(1+x)\) and \(-\log(1-x)\), with
\(x=z^we^{-\beta E+\mu J}\), cancels all even repetitions and leaves
\begin{equation}
 \log Z(z,e^{-\beta},e^\mu)
 =\sum_{w\ge1}\sum_{\substack{k\ge1\\k\ {\rm odd}}}
   \frac{z^{kw}}{k}Z_w^{\rm phys}(k\beta,k\mu).
 \label{eq:positive-product-from-character}
\end{equation}
For the identity term in the cyclic projection, the unsigned oscillator
character of Section~\ref{sec:review-counting-index} is evaluated at
\(q=e^{-\beta/w}\).  Its logarithm separates into fermionic and bosonic
sums, treated by Euler--Maclaurin and by the Dedekind product in
Appendix~\ref{app:fixed-charge-technical-bounds}.  The result is
\begin{equation}
 P_+(e^{-\beta/w},e^\mu)
 \sim
 \frac{\exp[w(\pi^2+\mu^2)/\beta]}
      {4\cosh^2(\mu/2)}
 \left(\frac{\beta}{2\pi w}\right)^2 .
\end{equation}

In the cyclic projection, the identity term gives the leading large-$w$
contribution, while the nonidentity roots are exponentially suppressed.
Using \(g_R(e^\mu)=16\cosh^2(\mu/2)\),
\begin{equation}
 Z_w^{\rm phys}(\beta,\mu)
 \sim
 \frac{g_R(e^\mu)}{w}
 P_+(e^{-\beta/w},e^\mu)
 \sim
 \frac{\beta^2}{\pi^2w^3}
 \exp\!\left[\frac{w(\pi^2+\mu^2)}{\beta}\right].
 \label{eq:projected-canonical-asymptotic}
\end{equation}
The Ramond ground-state factor cancels the denominator in
the first line. The explicit $1/w$ from the cyclic average combines with the
$w^{-2}$ oscillator prefactor to give
$Z_w^{\rm phys}\sim w^{-3}e^{w(\pi^2+\mu^2)/\beta}$. The uniform suppression
of the remaining terms in the cyclic average is proved in
Appendix~\ref{app:fixed-charge-technical-bounds}: these are precisely the
$\ell=1,\ldots,w-1$ summands in
Eq.~\eqref{eq:projected-strand-character}, whereas the displayed asymptotic
comes from the identity summand $\ell=0$.

At fixed momentum \(E\) and left charge \(J\), coefficient extraction
reverses the fugacity weight \(e^{-\beta E+\mu J}\), giving the factor
\(e^{\beta E-\mu J}\).  The exponent controlling the coefficient is therefore
\begin{equation}
 w\frac{\pi^2+\mu^2}{\beta}
 +\beta E-\mu J .
\end{equation}
Extremizing in \(\mu\) and \(\beta\) gives \(2w\mu/\beta=J\) and
\(w(\pi^2+\mu^2)/\beta^2=E\).
For a cycle carrying the macroscopic charge densities
\(E/w=\nu\) and \(J/w=\jmath\), the saddle conditions give
\begin{align}
 \mu_*&=\frac{\beta_*\jmath}{2},
 \qquad
 \frac{\pi^2+\mu_*^2}{\beta_*^2}=\nu,
 \nonumber\\
 \beta_*&=\frac{\pi}{\sqrt{\delta}}.
 \label{eq:wedge-critical-parameters}
\end{align}
The second saddle equation also gives
\(\left(\pi^2+\mu_*^2\right)/\beta_*=\beta_*\nu\), so
\begin{equation}
 Z_w^{\rm phys}(\beta_*,\mu_*)
 \sim
 \frac{e^{\beta_*\nu w}}{\delta w^3}.
\end{equation}
Removing this common exponential, define
\(a_w:=e^{-\beta_*\nu w}Z_w^{\rm phys}(\beta_*,\mu_*)\). Then
\begin{equation}
 a_w\sim\frac{1}{\delta w^3}.
 \label{eq:critical-projected-weight}
\end{equation}

The exponential factor of each cycle is \(e^{\beta_*\nu w}\); its product
over any partition is therefore \(e^{\beta_*\nu N}\). In contrast, the
prefactors depend on the individual windings. The oscillator spectrum contributes \(w^{-2}\) and the cyclic projection
contributes \(w^{-1}\), so
\(a_w\sim(\delta w^3)^{-1}\).  In particular,
\(\sum_w w a_w<\infty\), which is the convergence condition underlying the
finite-winding remainder derived below.

For a physical cycle of winding $w$ and momentum $E$, the cover level is
$m=wE$.  Hence Eq.~\eqref{eq:seed-jacobi-support} gives
$J^2\le4wE+1$.  For $w\ge1$ and $E\ge0$, this further implies
$|J|\le\sqrt{4wE+1}\le2E+w$.  Using
$\mu_*=\beta_*\jmath/2$, we obtain
\begin{align}
 \beta_*\nu w+\beta_*E-\mu_*J
 &\ge\beta_*\left[\left(\nu-\frac{|\jmath|}{2}\right)w
                  +(1-|\jmath|)E\right], \nonumber \\
 e^{-\beta_*\nu w-\beta_*E+\mu_*J}
 &\le e^{-\beta_*(\nu-|\jmath|/2)w}.
 \label{eq:no-pole-chamber}
\end{align}
For $|\jmath|<1$ and $\nu>|\jmath|/2$, the first line is strictly
positive for every physical species. Hence every bosonic occupation
weight is strictly smaller than one and no bosonic occupation factor is
singular at the saddle. Each aligned, singly wound bosonic Ramond ground
state has $E=0$ and $J=\operatorname{sgn}\jmath$. Its occupation factor
becomes singular at $\nu=|\jmath|/2$. 

In the uncondensed region, the cycles outside the longest one are
therefore counted by the same Fock generating function
\(Z(ze^{-\beta_*\nu},e^{-\beta_*},e^{\mu_*})\), evaluated at the
fixed-charge saddle after removing the common factor
\(e^{\beta_*\nu w}\).  {The paired Bose--Fermi product has nonnegative logarithmic coefficients:
only odd repetitions survive in Eq.~\eqref{eq:positive-product-from-character}.
For a repeated term of total winding $r=kw$ with odd $k\ge3$,
Eq.~\eqref{eq:no-pole-chamber} supplies an extra factor bounded by
$e^{-cr}$ for some $c>0$; repeated occupations therefore do not change the
power-law tail set by the first repetition $k=1$. Appendix~
\ref{app:fixed-charge-technical-bounds} then proves directly that the
single-cycle law $a_r\sim(\delta r^3)^{-1}$ gives the same $r^{-3}$ tail
after exponentiating the Fock logarithm. The $k=1$ term is present for every
winding $w\ge1$, because a length-$w$ cycle always has Ramond ground states.
The asymptotic below therefore holds for every sufficiently large integer
$r$, rather than only along a sublattice such as the even integers. Together
with the finite first moment $\sum_r r a_r<\infty$, this places the Fock
product in the convergent weighted-partition regime
\cite{StuflerConvergent}. It yields}
\begin{equation}
 [z^r]Z(ze^{-\beta_*\nu},e^{-\beta_*},e^{\mu_*})
 \sim\frac{Z(e^{-\beta_*\nu},e^{-\beta_*},e^{\mu_*})}{\delta r^3}.
 \label{eq:remainder-subexponential}
\end{equation}
The zeroth and first winding sums of this remainder distribution therefore
converge.  At fixed total winding \(N\), this is the mechanism that favors
one macroscopic cycle while the other cycles carry only a finite total
winding.  Equivalently, the free count lies in the convergent regime of
weighted partitions \cite{StuflerConvergent,StuflerUnlabelled}.

This winding argument does not yet impose the exact total momentum and
spin.  The quantity \(Z_w^{\rm phys}(\beta_*,\mu_*)\) sums over all
momentum and spin carried by a length-\(w\) cycle, whereas in the
fixed-charge sector the long cycle must carry whatever charges remain after
the finite remainder is removed.  We therefore write
\begin{equation}
 E=\nu w+u,
 \qquad
 J=\jmath w+v,
\end{equation}
where $(E,J)$ lies on the integer charge lattice and $u=E-\nu w$,
$v=J-\jmath w$ remain bounded for each fixed remainder state. Since
$E$ and $J$ are integers, arbitrary $w$ can produce non-integer $(E,J)$, making
$(\nu w+u,\jmath w+v)$ an unallowed charge pair. Every limit below therefore
uses only those values of $w$ for which it is allowed. There is no further
even--odd restriction: Ramond ground states realize the charge difference
$(\Delta E,\Delta J)=(0,1)$, and a neutral boson at cover level $w$
realizes $(1,0)$. These two differences generate the full lattice
$\mathbb Z^2$.

To isolate one definite pair \((E,J)\), normalize the charge-resolved
single-cycle weights and take their Fourier transform,
\begin{equation}
 \Psi_w(t,\phi)
 :=\frac{1}{Z_w^{\rm phys}(\beta_*,\mu_*)}
 \sum_{E,J}4|\widehat c(wE,J)|
 e^{-\beta_*E+\mu_*J}e^{i(tE+\phi J)},
 \qquad \Psi_w(0,0)=1.
 \label{eq:single-cycle-charge-characteristic}
\end{equation}
Fourier inversion then extracts the fraction of
\(Z_w^{\rm phys}\) carried by the required momentum and spin.  Near the
fixed-charge saddle the distribution is a nondegenerate two-dimensional
Gaussian.  Denoting its covariance matrix by \(\Sigma_w\), Appendix~
\ref{app:fixed-charge-technical-bounds} gives
\(\Sigma_w=w\Sigma_*+O(1)\) with
\(\det\Sigma_*=4\pi^2/\beta_*^4\).  Hence, for fixed \(u\) and \(v\),
\begin{align}
&\frac{4|\widehat c(w(\nu w+u),\jmath w+v)|
 e^{-\beta_*(\nu w+u)+\mu_*(\jmath w+v)}}
 {Z_w^{\rm phys}(\beta_*,\mu_*)}
\nonumber\\
&\quad=
 \frac{1}{(2\pi)^2}
 \int_{-\pi}^{\pi}dt\,d\phi\,
 e^{-i[(\nu w+u)t+(\jmath w+v)\phi]}
 \Psi_w(t,\phi)
\nonumber\\
&\quad=
 \frac{1+o(1)}{(2\pi)^2}
 \int_{\mathbb R^2}dt\,d\phi\,
 \exp\!\left[
 -\frac12
 \begin{pmatrix}t&\phi\end{pmatrix}
 \Sigma_w
 \begin{pmatrix}t\\ \phi\end{pmatrix}
 -i(ut+v\phi)
 \right]
\nonumber\\
&\quad=
 \frac{1+o(1)}
 {2\pi\sqrt{\det\Sigma_w}}\,
 \exp\!\left[
 -\frac12
 \begin{pmatrix}u&v\end{pmatrix}
 \Sigma_w^{-1}
 \begin{pmatrix}u\\ v\end{pmatrix}
 \right]
\nonumber\\
&\quad=
 \frac{\beta_*^2}{4\pi^2w}\,[1+o(1)]
 =
 \frac{1+o(1)}{4\delta w}.
 \label{eq:main-one-string-local-limit}
\end{align}
Because \(u\) and \(v\) are fixed while
\(\Sigma_w^{-1}=O(w^{-1})\), the quadratic offset in the Gaussian is
\(O(w^{-1})\) and is absorbed into \(1+o(1)\).  Thus fixing momentum and
spin contributes one additional factor \(w^{-1}\).  Appendix~\ref{app:fixed-charge-technical-bounds} proves exponential Fourier
suppression away from the origin and a uniform $C/w$ bound for every
charge coefficient. The latter also justifies summing over the unbounded
energy of the finite-winding remainder.

The additional \(w^{-1}\) factor now converts the long-cycle weight into
an absolute fixed-charge multiplicity.  For \(r=d<N/2\), there is a unique
cycle of winding \(N-r\), while all remaining cycles carry total winding
\(r\).  Summing over the states of this finite remainder and using the saddle relation
\(\beta_*\nu N+\beta_*N_p-\mu_*J_L=S_{\rm BH}\) give
\begin{equation}
 \Tr_VP_r
 =\frac{e^{S_{\rm BH}}}{4\delta^2N^4}
 [z^r]Z(ze^{-\beta_*\nu},e^{-\beta_*},e^{\mu_*})[1+o(1)],
 \qquad r\ \text{fixed}.
 \label{eq:main-fixed-r-weight}
\end{equation}
The \(N^{-4}\) prefactor combines the \(N^{-3}\) projected single-cycle
weight with the additional \(N^{-1}\) from fixing the momentum and spin of
the long cycle.

Equation~\eqref{eq:main-fixed-r-weight} gives the multiplicity for each
fixed finite \(r\).  To turn this into a statement about the whole
fixed-charge block, we need both the total weight of the tail \(r>R\) and
the total normalization:
\begin{align}
 e^{-S_{\rm BH}}\sum_{r>R}\Tr_VP_r
 &\le C_1N^{-4}(1+R)^{-2}+C_2N^{-6},
 \label{eq:main-uniform-weight-tail}\\
 D_V(\Gamma_N)
 &=\frac{e^{S_{\rm BH}}}{4\delta^2N^4}
 Z(e^{-\beta_*\nu},e^{-\beta_*},e^{\mu_*})[1+o(1)].
 \label{eq:main-full-fixed-charge-asymptotic}
\end{align}
In the first line, \(e^{-S_{\rm BH}}\) only removes the common leading
black-hole degeneracy.  The \((1+R)^{-2}\) term comes from the \(r^{-3}\)
tail of configurations with one cycle longer than \(N/2\), while
configurations with no such cycle require at least two large cycles and
are \(O(N^{-6})\).  The second line is obtained by summing the fixed-\(r\)
weights, so \(D_V(\Gamma_N)\) is precisely their total normalization.
The uniform estimates are given in
Appendix~\ref{app:fixed-charge-technical-bounds}.

Dividing the tail estimate by the total fixed-charge dimension gives
\begin{align}
 \frac{\sum_{r>R}\Tr_VP_r}{D_V(\Gamma_N)}
 &\le
 \frac{
 e^{S_{\rm BH}}
 \left[
 C_1N^{-4}(1+R)^{-2}+C_2N^{-6}
 \right]
 }{
 \dfrac{e^{S_{\rm BH}}}{4\delta^2N^4}
 Z(e^{-\beta_*\nu},e^{-\beta_*},e^{\mu_*})[1+o(1)]
 }
 \nonumber\\
 &=
 \frac{4\delta^2}
 {Z(e^{-\beta_*\nu},e^{-\beta_*},e^{\mu_*})}
 \left[
 \frac{C_1}{(1+R)^2}
 +\frac{C_2}{N^2}
 \right]
 [1+o(1)]
 \nonumber\\
 &\le
 \frac{4\delta^2(C_1+C_2)}
 {Z(e^{-\beta_*\nu},e^{-\beta_*},e^{\mu_*})}
 \frac{1+o(1)}{(1+R)^2}
 \nonumber\\
 &\le
 \frac{C}{(1+R)^2},
 \qquad N\ge N_0.
 \label{eq:uniform-finiteN-depth-tail}
\end{align}
The common factor \(e^{S_{\rm BH}}\) and the overall \(N^{-4}\)
normalization have therefore cancelled.  The remaining bound is directly
the fraction of the fixed-charge dimension with more than \(R\) units of
winding outside the longest cycle.

The bound shows that the total winding carried outside the longest cycle
does not grow with \(N\).  More explicitly, for any desired accuracy
\(\epsilon>0\), one can choose a fixed number \(R\), independent of \(N\),
such that, for sufficiently large \(N\), more than a fraction
\(1-\epsilon\) of the fixed-charge states satisfy
\(N-w_{\max}\le R\).
Equivalently,
\begin{equation}
 w_{\max}=N-O_{\Pr}(1).
\end{equation}

For each fixed remainder \(r\), the normalized weight has the limit
\begin{equation}
 \frac{\Tr_VP_r}{D_V}
 \longrightarrow
 \frac{[z^r]Z(ze^{-\beta_*\nu},e^{-\beta_*},e^{\mu_*})}
 {Z(e^{-\beta_*\nu},e^{-\beta_*},e^{\mu_*})}.
 \label{eq:projected-remainder-law}
\end{equation}
The same \(r^{-3}\) tail also determines the moments of the finite
remainder:
\begin{align}
 \frac{[z^r]Z(ze^{-\beta_*\nu},e^{-\beta_*},e^{\mu_*})}
 {Z(e^{-\beta_*\nu},e^{-\beta_*},e^{\mu_*})}
 &\sim\frac1{\delta r^3}, \nonumber \\
 \frac{\sum_{r>R}[z^r]Z(ze^{-\beta_*\nu},e^{-\beta_*},e^{\mu_*})}
 {Z(e^{-\beta_*\nu},e^{-\beta_*},e^{\mu_*})}
 &\sim\frac1{2\delta R^2},
 \label{eq:projected-tail-and-moments} \\
 \frac{\sum_r r\,\Tr_VP_r}{D_V}
 &\longrightarrow
 \left.\partial_z\log Z(ze^{-\beta_*\nu},e^{-\beta_*},e^{\mu_*})\right|_{z=1}<\infty,
 \label{eq:mean-depth-convergence} \\
 \frac{\sum_r(N-r)\Tr_VP_r}{ND_V}
 &=1-\frac1N
 \left.\partial_z\log Z(ze^{-\beta_*\nu},e^{-\beta_*},e^{\mu_*})\right|_{z=1}
 +o(N^{-1}).
 \label{eq:mean-longest-cycle-concentration}
\end{align}
Because the remainder distribution falls as \(r^{-3}\), its mean winding
is finite:
\(\sum_r r\,P(r)\sim\sum_r r^{-2}<\infty\).
Its second moment instead behaves as
\(\sum_r r^2P(r)\sim\sum_r r^{-1}\), and therefore diverges
logarithmically.  The uniform tail bound above ensures that no additional
weight escapes to large \(r\) as \(N\) grows, so the finite mean survives
the large-\(N\) limit.

\label{sec:two-cycle-weight}
The natural generalization of the explicit short-string family is the
entire sector with no dominant long component,
\(w_{\max}\le N/2\).  We bound this sector by a generating function whose
coefficients are manifestly positive.  For each positive single-species
activity \(x<1\),
\begin{equation}
 1+x\le \frac{1}{1-x}=1+x+x^2+\cdots.
\end{equation}
Replacing a fermionic factor \(1+x\) by
\((1-x)^{-1}=1+x+x^2+\cdots\) only adds nonnegative multiple-occupation
terms.  Therefore the resulting bosonic product gives an upper bound on
the number of states at every fixed total winding.

At the fixed-charge saddle, write the logarithm of this bosonic upper
bound as
\begin{equation}
 \log Z_{\rm bos}(z)=\sum_{w\ge1}b_w z^w,
 \qquad 0\le b_w\le \frac{C}{w^3}.
 \label{eq:no-giant-log-envelope}
\end{equation}
The \(w^{-3}\) bound is the uniform version of
Eq.~\eqref{eq:critical-projected-weight}: the single-cycle term has this
power law, while the repeated terms in the logarithmic Fock expansion are
exponentially suppressed by Eq.~\eqref{eq:no-pole-chamber}; the finitely
many small windings are absorbed into \(C\).

{
For configurations with physical cycle lengths at most $N/2$, use
the truncated logarithmic envelope
\begin{equation}
 Z_{\rm bos}^{\le N/2}(z)
 =\exp\!\left(\sum_{w\le N/2}b_wz^w\right).
\end{equation}
The truncation removes every logarithmic term of degree greater than $N/2$.
This includes some repeated occupations of permitted physical cycles, so we
must bound the error introduced by the truncation. Every such removed term
has repetition number at least two; Eq.~\eqref{eq:no-pole-chamber}, summed
over the possible divisors of its degree $r$, bounds its coefficient by
$Ce^{-cr}$ for some $c>0$. In the exponential expansion, a monomial of total
degree $N$ can contain at most one removed term, because two degrees greater
than $N/2$ would already sum to more than $N$. The rest of the monomial comes
from the retained factor, whose nonnegative coefficients are bounded by
those of $Z_{\rm bos}$. Since $Z_{\rm bos}(1)<\infty$ follows from
$\sum_w b_w<\infty$, the coefficient error caused by the truncation obeys
\begin{equation*}
 C\sum_{N/2<r\le N}e^{-cr}[z^{N-r}]Z_{\rm bos}(z)
 \le Ce^{-cN/2}Z_{\rm bos}(1).
\end{equation*}
After decreasing $c$ if necessary, it is therefore enough to bound the
coefficient of the truncated envelope with an additive $O(e^{-cN})$
error:

}
\begin{equation}
 [z^N]Z_{\rm bos}^{\le N/2}(z)
 =
 \sum_{k\ge2}\frac1{k!}
 \sum_{\substack{w_1+\cdots+w_k=N\\1\le w_i\le N/2}}
 \prod_{i=1}^k b_{w_i}.
 \label{eq:no-giant-k-cycle-expansion}
\end{equation}
For a contribution with $k$ logarithmic parts,
\begin{equation}
 w_1+\cdots+w_k=N,
 \qquad
 w_i\le\frac N2.
\end{equation}
At least two of these parts must satisfy
\(w_i\ge N/(2k)\).  Otherwise, even if the largest part had its maximal
allowed winding \(N/2\), the total winding would obey
\begin{align}
 \sum_{i=1}^k w_i
 &<
 \frac N2+(k-1)\frac{N}{2k}
 \nonumber\\
 &<N,
\end{align}
contradicting the fixed total winding.

Choose two such parts, $w_a$ and $w_b$.  Since
\(b_w\le Cw^{-3}\),
\begin{align}
 b_{w_a}b_{w_b}
 \le
 \frac{C^2}{w_a^3w_b^3}
\le
 C^2\left(\frac{2k}{N}\right)^6.
 \label{eq:no-giant-two-large-parts}
\end{align}
Each term of the truncated envelope therefore contains two
parts whose combined weight supplies an $N^{-6}$ suppression.

If the remaining $k-2$ parts carry total winding
\(r=N-w_a-w_b\), then \(w_a+w_b=N-r\).
Since both \(w_a\) and \(w_b\) are at most \(N/2\), one must have
\[
 \frac N2-r\le w_a\le\frac N2.
\]
The integer $w_a$ has at most $r+1$ allowed values, and
$w_b$ is then fixed by $w_b=N-r-w_a$. The remaining sums are finite because
\(\sum_w b_w<\infty\) and \(\sum_w wb_w<\infty\).  Including the at most
{$\binom{k}{2}\le k^2/2$ choices for which two of the $k$
parts are $w_a$ and $w_b$. Combining this factor with the $k^6N^{-6}$ bound
in Eq.~\eqref{eq:no-giant-two-large-parts} gives the $k^8N^{-6}$ factor below:}
\begin{align}
 [z^N]Z_{\rm bos}^{\le N/2}(z)
 &\le
 CN^{-6}\sum_{k\ge2}\frac{k^8C^k}{k!}
 +O(e^{-cN})
 \nonumber\\
 &=O(N^{-6}).
 \label{eq:no-giant-envelope-main}
\end{align}

The bosonic product overcounts the original positive Fock coefficients,
and the fixed-charge sector is one positive charge sector at the saddle.
This gives:
\begin{equation}
 e^{-S_{\rm BH}}
 \sum_{p:\,w_{\max}(p)\le N/2}\dim V_{p,\Gamma_N}
\le
 [z^N]Z_{\rm bos}^{\le N/2}(z)+O(e^{-cN})
 =O(N^{-6}).
 \label{eq:no-giant-fixed-charge-bound}
\end{equation}
Using
\(D_V(\Gamma_N)\sim e^{S_{\rm BH}}N^{-4}\), we finally obtain
\begin{equation}
 \frac{
 \displaystyle\sum_{p:\,w_{\max}(p)\le N/2}
 \dim V_{p,\Gamma_N}}
 {D_V(\Gamma_N)}
 =
 O(N^{-2}).
 \label{eq:no-giant-sector-suppression}
\end{equation}
The whole sector in which no single cycle carries more than
half of the total winding occupies a vanishing fraction of the fixed-charge
block.
The family \(F_{2k}^{\rm heavy}\), with \(w_{\max}=2\), is an explicit
state-level example lying inside this statistically suppressed sector.

This concentration law holds in the open uncondensed region.
At $\nu=|\jmath|/2$, the aligned winding-one Bose weight reaches one, so
its Fock factor develops a pole and the argument changes.

\subsection{Critical line and short-string condensation}
\label{sec:phase-and-moments}
At \(\nu=|\jmath|/2\), the effective weight of an aligned winding-one
Ramond ground state reaches one:
\begin{equation}
 e^{-\beta_*(\nu-|\jmath|/2)}=1.
\end{equation}
Its bosonic Fock factor therefore becomes singular at
\(\nu=|\jmath|/2\), the boundary of the uncondensed region. The \(O_{\Pr}(1)\) law for
\(d\) therefore changes: on the critical line these states have
\(O_{\Pr}(\sqrt N)\) occupation, while in the condensed region their
occupation is extensive and carries the excess winding and spin.  The
fixed-charge coefficients determine the corresponding occupation
laws and the change of the leading entropy.

At $\nu=|\jmath|/2$, with $0<|\jmath|<1$, the two aligned singly wound
bosonic Ramond ground states have a divergent mean occupation. These are
the first species whose Bose factors become singular. Their contribution
and its first two occupation cumulants are
\begin{align}
 \left(1-ze^{-\beta_*(\nu-|\jmath|/2)}\right)^{-2}
 &=\sum_{m\ge0}(m+1)e^{-m\beta_*(\nu-|\jmath|/2)}z^m, \nonumber \\
 \left.z\partial_z\log
 \left(1-ze^{-\beta_*(\nu-|\jmath|/2)}\right)^{-2}\right|_{z=1}
 &=\frac{2}{e^{\beta_*(\nu-|\jmath|/2)}-1}, \nonumber \\
 \left.(z\partial_z)^2\log
 \left(1-ze^{-\beta_*(\nu-|\jmath|/2)}\right)^{-2}\right|_{z=1}
 &=\frac{2e^{\beta_*(\nu-|\jmath|/2)}}
 {(e^{\beta_*(\nu-|\jmath|/2)}-1)^2}.
 \label{eq:moulting-singleton-pole}
\end{align}
Here $N\to\infty$ is
taken before $\nu\to|\jmath|/2^+$. Their mean occupation diverges as
\(2/[\beta_*(\nu-|\jmath|/2)]\) as
\(\nu\to|\jmath|/2^+\).  By contrast, after excluding these two aligned
winding-one bosons, the expected total winding carried by all remaining
single-cycle species stays finite.

We now keep $0<|\jmath|<1$ and resolve these two singular species
explicitly in the fixed-charge count, as in
Ref.~\cite{Shigemori:2019orj}.  For \(0<|\jmath|<1\), the singular modes are the two bosonic
Ramond-ground-state polarizations with
\((w,E,J)=(1,0,\operatorname{sgn}J_L)\).
On the \(J_L>0\) branch, their combined Fock factor is therefore
\((1-zy)^{-2}\).  Although each state has winding one, their total
occupation can scale with $N$ and hence carry a macroscopic amount of
winding and spin.

Let $m$ be the total occupation of the two species.  There are $m+1$
ways to distribute $m$ identical excitations between them.  They carry
winding $m$, momentum zero, and left charge $m$, so the remaining Fock
space must carry $(N-m,N_p,J_L-m)$.  Removing the two singular factors
from $Z$ therefore gives
\begin{align}
 D_V(\Gamma_N)
 =\sum_{m=0}^{N} (m+1)[z^{N-m}q^{N_p}y^{J_L-m}](1-zy)^2Z(z,q,y){.}
 \label{eq:singleton-resolved-coefficient}
\end{align}
For $J_L<0$, the same argument applies with $y$ replaced by $y^{-1}$.

For a short-string condensate of total occupation \(m\), the aligned
winding-one Ramond ground states carry
\begin{equation}
 w_{\rm short}=m,\qquad
 N_{p,\rm short}=0,\qquad
 J_{L,\rm short}=\operatorname{sgn}(J_L)m.
\end{equation}
The remaining excited component therefore carries
\begin{equation}
 w_{\rm long}=N-m,\qquad
 N_{p,\rm long}=N_p,\qquad
 J_{L,\rm long}=\operatorname{sgn}(J_L)(|J_L|-m).
\end{equation}
Thus increasing \(m\) transfers winding and spin into the aligned
short-string sector while leaving all momentum on the excited component.
Using Eq.~\eqref{eq:review-moulting-long-string-entropy}, the entropy at
fixed \(m\) is
\begin{align}
 S(m)^2
 &=
 4\pi^2\left[
 (N-m)N_p-\frac{(|J_L|-m)^2}{4}
 \right]
 \nonumber\\
 &=
 4\pi^2N_p(N+N_p-|J_L|)
 -\pi^2(m-|J_L|+2N_p)^2.
 \label{eq:condensed-saddle-square}
\end{align}

The position of the maximum is fixed by
\begin{equation}
 \nu-\frac{|\jmath|}{2}
 =
 \frac{2N_p-|J_L|}{2N}.
\end{equation}
The three cases are therefore
\begin{align}
 |J_L|<2N_p
 \quad\left(\nu>\frac{|\jmath|}{2}\right):
 &\qquad
 m_*=0,\qquad
 S_{\max}=S_{\rm BH}
 =2\pi\sqrt{NN_p-\frac{J_L^2}{4}},
 \nonumber\\
 |J_L|=2N_p
 \quad\left(\nu=\frac{|\jmath|}{2}\right):
 &\qquad
 m_*=0,\qquad
 S_{\max}=S_{\rm BH}=S_{\rm enigma},
 \nonumber\\
 |J_L|>2N_p
 \quad\left(\nu<\frac{|\jmath|}{2}\right):
 &\qquad
 m_*=|J_L|-2N_p,\qquad
 S_{\max}=S_{\rm enigma}
 =2\pi\sqrt{N_p(N+N_p-|J_L|)}.
 \label{eq:critical-condensed-saddles}
\end{align}
Thus the uncondensed region has its maximum at the endpoint \(m=0\).
The critical line is the point at which this endpoint becomes stationary,
and in the condensed region the maximum moves to the interior value
\(m_*=|J_L|-2N_p\), reproducing the enigmatic saddle reviewed in
Eq.~\eqref{eq:review-moulting-extremum}.

The quadratic entropy takes a different form on the critical line
and in the condensed interior:
\begin{align}
 |J_L|=2N_p:\qquad
 S(m)
 &=
 S_{\rm BH}
 -\frac{\pi^2m^2}{2S_{\rm BH}}
 +O\!\left(\frac{m^4}{N^3}\right),
 \qquad m=O(\sqrt N),
 \nonumber\\
 |J_L|>2N_p:\qquad
 S(m)
 &=
 S_{\rm enigma}
 -\frac{\pi^2(m-m_*)^2}{2S_{\rm enigma}}
 +O\!\left(\frac{(m-m_*)^4}{N^3}\right),
 \qquad m-m_*=O(\sqrt N).
 \label{eq:condensed-saddle-curvature}
\end{align}

At the condensed saddle \(m_*=|J_L|-2N_p\), the residual excited
component carries
\begin{align}
 w_{\rm long}
 &=N-m_*
 =N-|J_L|+2N_p,
 \nonumber\\
 N_{p,\rm long}
 &=N_p,
 \nonumber\\
 J_{L,\rm long}
 &=\operatorname{sgn}(J_L)(|J_L|-m_*)
 =\operatorname{sgn}(J_L)2N_p,
 \qquad
 |J_{L,\rm long}|=2N_p.
 \label{eq:condensed-critical-core-charges}
\end{align}
Its charges therefore lie on the critical line, and
\begin{equation}
 S_{\rm BH}\!\left(
 N-|J_L|+2N_p,\,
 N_p,\,
 \operatorname{sgn}(J_L)2N_p
 \right)
 =
 S_{\rm enigma}.
 \label{eq:condensed-critical-core}
\end{equation}
The residual excited component carries the leading entropy, while the
aligned winding-one ground states carry the excess winding and spin.

To obtain the fixed-\(m\) counting law, we must first verify that the two
singular bosonic modes isolated above are the only modes whose occupation
can become macroscopic.  At the condensed saddle, the residual excited
component lies on the critical line.  Evaluating the remaining Fock
product at the saddle associated with this residual component gives
\begin{NoHyper}
\begin{equation}
 \beta=\frac{\pi(N-|J_L|+2N_p)}
 {\sqrt{N_p(N+N_p-|J_L|)}},
 \qquad
 \mu=\frac{\pi\,\operatorname{sgn}(J_L)N_p}
 {\sqrt{N_p(N+N_p-|J_L|)}},
 \qquad
 \frac{\pi^2+\mu^2}{\beta}=|\mu|.
 \nonumber
\end{equation}
\end{NoHyper}

At this saddle, a bosonic single-cycle species \((w,E,J)\) has Fock
factor
\begin{equation}
 \left(1-z^w e^{-|\mu|w-\beta E+\mu J}\right)^{-1}.
\end{equation}
It can become singular as \(z\to1\) only if its effective Boltzmann cost
\begin{equation}
 C(w,E,J)=|\mu|w+\beta E-\mu J
\end{equation}
vanishes.  On the positive-spin branch, \(\mu>0\).  Since
Eq.~\eqref{eq:seed-jacobi-support} implies \(|J|\le 2E+w\),
\begin{align}
 C(w,E,J)
 &\ge \mu w+\beta E-\mu(2E+w)
 \nonumber\\
 &=(\beta-2\mu)E.
\end{align}
The residual excited component lies on the critical line, so
\(\mu=\beta\jmath_{\rm c}/2\) with \(0<\jmath_{\rm c}<1\), and hence
\(\beta-2\mu>0\).  Therefore every species with \(E>0\) has strictly
positive cost.  For \(E=0\), Eq.~\eqref{eq:seed-jacobi-support} gives
\(|J|\le1\), and
\begin{equation}
 C(w,0,J)=\mu(w-J)\ge\mu(w-1).
\end{equation}
Thus \(C=0\) is possible only for \(w=1\) and \(J=1\), namely the two
singular bosonic Ramond-ground-state polarizations isolated above.  The
negative-spin branch follows by charge conjugation.

The same bound also controls the large-winding tail.  Using
\(J^2\le4wE+1\) together with
\((\pi^2+\mu^2)/\beta=|\mu|\) gives
\begin{equation}
 |\mu|w+\beta E-\mu J
 \ge
 \frac{\pi^2}{\beta}w-\frac{\beta}{4w},
\end{equation}
which grows linearly with \(w\).  Hence, after removing the double Bose
pole of the two singular modes, the remaining Fock product and its first
winding and energy moments are finite.  In the shifted product
\(Z(ze^{-|\mu|},e^{-\beta},e^\mu)\), the two singular modes contribute
exactly \((1-z)^{-2}\).  Therefore
\begin{NoHyper}
\begin{equation}
 0<\lim_{z\to1^-}(1-z)^2
 Z(ze^{-|\mu|},e^{-\beta},e^\mu)<\infty
 \nonumber
\end{equation}
\end{NoHyper}
is the finite contribution of all remaining species at the fixed critical saddle $(\beta,\mu)$.

We can now convert the entropy profile into the fixed-\(m\) multiplicity.
For a given condensate occupation \(m\), three factors contribute: the
multiplicity \(m+1\) of the two singular bosonic polarizations, the
fixed-charge degeneracy of the residual excited component, and the finite
Fock factor above.  The \(N^{-4}\) fixed-charge prefactor derived in
Eq.~\eqref{eq:main-fixed-r-weight} is evaluated using the charges of the
residual component.  At the condensed saddle, the combination appearing in that prefactor becomes
\begin{align}
 4\left(
 \frac{N_p}{w_{\rm long}}
 -\frac{N_p^2}{w_{\rm long}^2}
 \right)^2w_{\rm long}^4
 &=
 4N_p^2(N+N_p-|J_L|)^2
 \nonumber\\
 &=
 4\nu^2(1+\nu-|\jmath|)^2N^4.
\end{align}

For \(m-m_*=O(\sqrt N)\), with
\(m_*=|J_L|-2N_p\), the entropy expansion gives
\begin{equation}
 e^{S(m)}
 =
 e^{S_{\rm enigma}}
 \exp\!\left[
 -\frac{\pi^2(m-m_*)^2}{2S_{\rm enigma}}
 \right]
 [1+o(1)].
\end{equation}
Combining the three factors gives
\begin{align}
 D_V^{(m)}(\Gamma_N)
 &=
 (m+1)
 [z^{N-m}q^{N_p}y^{J_L-m}]
 (1-zy)^2Z(z,q,y)
 \nonumber\\
 &=
 (m+1)
 \frac{e^{S(m)}}
 {4\nu^2(1+\nu-|\jmath|)^2N^4}
 \left[
 \lim_{z\to1^-}(1-z)^2
 Z(ze^{-|\mu|},e^{-\beta},e^\mu)
 \right]
 [1+o(1)]
 \nonumber\\
 &=
 \frac{e^{S_{\rm enigma}}}
 {4\nu^2(1+\nu-|\jmath|)^2N^4}
 \left[
 \lim_{z\to1^-}(1-z)^2
 Z(ze^{-|\mu|},e^{-\beta},e^\mu)
 \right]
 \nonumber\\
 &\qquad\times
 (m+1)
 \exp\!\left[
 -\frac{\pi^2(m-|J_L|+2N_p)^2}
 {2S_{\rm enigma}}
 \right]
 [1+o(1)],
 \label{eq:condensate-fixed-charge-weight}
\end{align}
where we have used the large-\(N\) asymptotic of the coefficient
extraction, whose leading exponential is \(e^{S(m)}\):
\begin{equation}
 \log\!\left(
 [z^{N-m}q^{N_p}y^{J_L-m}]
 (1-zy)^2Z(z,q,y)
 \right)
 =
 S(m)-4\log N+O(1),
 \qquad
 m-m_*=O(\sqrt N).
\end{equation}
This estimate is uniform when
\((m-m_*)/\sqrt N\) is bounded.  Hence, the
leading \(m\)-dependence of \(D_V^{(m)}\) is entirely contained in the
multiplicity \(m+1\) and in the Gaussian variation of \(S(m)\).

The ground-state occupations enhance the total multiplicity by different
powers of $N$ on the critical line and in the condensed region:
\begin{align}
 \sum_{m\ge0}(m+1)
 \exp\!\left[-\frac{\pi^2(m-|J_L|+2N_p)^2}{2S_{\rm enigma}}\right]
 &\sim\begin{cases}
 S_{\rm enigma}/\pi^2,&|J_L|=2N_p,\\
 (|J_L|-2N_p)\sqrt{2S_{\rm enigma}/\pi},&|J_L|>2N_p,
 \end{cases} \nonumber \\
 D_V(\Gamma_N)
 &\sim\frac{e^{S_{\rm enigma}}}{4\nu^2(1+\nu-|\jmath|)^2N^4}
 \left[\lim_{z\to1^-}(1-z)^2Z(ze^{-|\mu|},e^{-\beta},e^\mu)\right]\nonumber\\
 &\quad\times\begin{cases}
 S_{\rm enigma}/\pi^2,&\nu=|\jmath|/2,\\
 (|J_L|-2N_p)\sqrt{2S_{\rm enigma}/\pi},&\nu<|\jmath|/2.
 \end{cases}
 \label{eq:critical-condensed-total-count}
\end{align}
On the critical line $|J_L|=2N_p$, one has $m_*=0$, so the
saddle is at the lower boundary. The occupation sum is then proportional
to $S_{\rm enigma}=\Theta(N)$ and changes the $N^{-4}$ prefactor to
$N^{-3}$. In the condensed interior, $m_*=|J_L|-2N_p=\Theta(N)$.
The multiplicity $m+1\sim m_*$ supplies a factor $N$ and the Gaussian
width supplies a factor $\sqrt N$, giving
$N^{-4}N\sqrt N=N^{-5/2}$.
Consequently,
\begin{equation}
 \log D_V=
 \begin{cases}
 S_{\rm BH}-3\log N+O(1),
 & \nu=|\jmath|/2,\\
 S_{\rm enigma}-\frac52\log N+O(1),
 & \nu<|\jmath|/2.
 \end{cases}
\end{equation}

Dividing the fixed-\(m\) count by the total multiplicity gives
\begin{equation}
 \frac{D_V^{(m)}(\Gamma_N)}{D_V(\Gamma_N)}
 =
 \frac{
 (m+1)\exp[-\pi^2(m-m_*)^2/(2S_{\rm enigma})]
 }{
 \displaystyle
 \sum_{q\ge0}(q+1)
 \exp[-\pi^2(q-m_*)^2/(2S_{\rm enigma})]
 }
 [1+o(1)]{.}
 \label{eq:condensate-local-probability}
\end{equation}
On the critical line, \(m_*=0\), and the scaled occupation follows the
Rayleigh law.  In the condensed region,
\(m_*=|J_L|-2N_p\), and the fluctuations about \(m_*\) are Gaussian.
The regular Fock factor has a finite first winding moment, so all cycles
outside the residual excited component and the condensate carry only
\(O_{\Pr}(1)\) total winding.  Thus \(d=N-w_{\max}\) is determined by the
condensate occupation up to \(O_{\Pr}(1)\) corrections:
\begin{equation}
 \boxed{
 \begin{array}{ll}
 d=O_{\Pr}(1),
 &\nu>|\jmath|/2,\\[1mm]
 d=O_{\Pr}(\sqrt N),
 &\nu=|\jmath|/2,\\[1mm]
 d=(|\jmath|-2\nu)N+O_{\Pr}(\sqrt N),
 &\nu<|\jmath|/2.
 \end{array}}
 \label{eq:three-region-free-cycle-law}
\end{equation}

The leading entropy has the BMPV value in the uncondensed region and
the larger value $S_{\rm enigma}$ in the condensed region:
\begin{align}
 \frac{S_{\rm free}}{2\pi N}
 &=\begin{cases}
 \sqrt{\nu-\jmath^2/4},&\nu\ge|\jmath|/2,\\[1mm]
 \sqrt{\nu(1+\nu-|\jmath|)},&\nu\le|\jmath|/2,
 \end{cases}
 \label{eq:two-phase-free-entropy} \\
 S_{\rm enigma}^2-S_{\rm BH}^2&=\pi^2(|J_L|-2N_p)^2,
 \label{eq:condensed-entropy-gap} \\
 \left.\partial_\nu S_{\rm enigma}\right|_{\nu=|\jmath|/2}
 &=\left.\partial_\nu S_{\rm BH}\right|_{\nu=|\jmath|/2}
 =\frac{\pi N}{\sqrt{|\jmath|/2-\jmath^2/4}}, \nonumber \\
 \left.(\partial_\nu^2S_{\rm enigma}-\partial_\nu^2S_{\rm BH})
 \right|_{\nu=|\jmath|/2}
 &=\frac{2\pi N}{\sqrt{|\jmath|/2-\jmath^2/4}}.
 \label{eq:condensed-second-derivative-jump}
\end{align}
At $\nu=|\jmath|/2$, the entropy and its first $\nu$ derivative are
continuous, but its second derivative jumps. The condensate fraction
vanishes linearly on approaching this line from the condensed region.

The preceding analysis assumed $0<|\jmath|<1$.  The endpoint
\(\jmath=0\) is different.  In this case
\(\mu_*=0\), \(\delta=\nu\), and
\begin{equation}
 \beta_*\nu=\pi\sqrt{\delta}.
\end{equation}
A winding-\(w\) Ramond ground state therefore has effective weight
\(e^{-\pi\sqrt{\delta}\,w}\).  There are eight bosonic and eight
fermionic Ramond ground states at each winding, whose mean occupation
numbers are respectively
\(1/(e^{\pi\sqrt{\delta}\,w}-1)\) and
\(1/(e^{\pi\sqrt{\delta}\,w}+1)\).  Their contribution to the mean total
winding is therefore
\begin{equation}
 8\sum_{w\ge1}w\left(
 \frac{1}{e^{\pi\sqrt{\delta}\,w}-1}
 +
 \frac{1}{e^{\pi\sqrt{\delta}\,w}+1}
 \right).
\end{equation}
For \(\delta\to0\), Euler--Maclaurin gives
\begin{align}
 \sum_{w\ge1}\frac{w}{e^{\pi\sqrt{\delta}\,w}-1}
 &=\frac{1}{6\delta}+O(\delta^{-1/2}),
 \nonumber\\
 \sum_{w\ge1}\frac{w}{e^{\pi\sqrt{\delta}\,w}+1}
 &=\frac{1}{12\delta}+O(\delta^{-1/2}),
 \nonumber\\
 8\sum_{w\ge1}w\left(
 \frac{1}{e^{\pi\sqrt{\delta}\,w}-1}
 +
 \frac{1}{e^{\pi\sqrt{\delta}\,w}+1}
 \right)
 &=\frac{2}{\delta}+O(\delta^{-1/2}).
 \label{eq:jzero-ground-winding-main}
\end{align}
Thus, unlike the generic branch where only the two aligned winding-one
bosons become singular first, at \(\jmath=0\) Ramond ground states of all
windings collectively produce a divergent mean remainder winding.

\subsection{Exact BPS counting and the non-graviton sector}
\label{sec:core-counting}

Sections~\ref{sec:sector-ensemble-construction}
and~\ref{sec:phase-and-moments} determine the cycle distribution of the
full free fixed-charge space, including its change across the short-string
condensation transition.  We now ask how much of this cycle organization
survives after restricting to physically distinct BPS subspaces.  We first
consider the exact BPS space and then remove the generalized-gravity sector
to isolate an entropy-carrying non-graviton subspace.

The protected index fixes the deformation-invariant entropy scale against
which lifting can be measured.  It is built from the same seed coefficients
as the unsigned count, with the grading fixed in
Section~\ref{sec:review-counting-index}, and for fixed \(\delta>0\) its
fixed-charge divisor formula is dominated by the \(k=1\) term.  This gives
\begin{align}
 4|I_N|
 &=\frac{e^{S_{\rm BH}}}{4\delta^2N^4}[1+o(1)],
 \nonumber\\
 \log|I_N|
 &=S_{\rm BH}-4\log S_{\rm BH}+O_{\nu,\jmath}(1).
 \label{eq:CFT-index-invariant-log}
\end{align}
The estimate holds also when \(\gcd(N,N_p,|J_L|)>1\), where subleading
divisor terms are present, and assumes nothing about their signs
(Appendix~\ref{app:fixed-charge-technical-bounds}).  Since each signed unit
of the index requires a complete four-state right-moving Clifford quartet,
the quartet bound of Section~\ref{sec:review-counting-index} turns this
protected count into a bound on an actual dimension,
\begin{equation}
 D_H(\Gamma_N)\ge4|I_N|.
\end{equation}

The logarithmic term already has a direct bulk interpretation.  For a
nonzero fixed rotation ratio, Eq.~\eqref{eq:CFT-index-invariant-log}
agrees with Sen's rotating \(T^5\) BMPV index through one loop:
\begin{equation}
 \left.
 \log|\widetilde I^{(6)}_{\rm BMPV}|
 \right|_{\text{through one loop}}
 =
 S_{\rm BH}-4\log S_{\rm BH}+O(1).
 \label{eq:BMPV-one-loop-match}
\end{equation}
The charge dictionary is
\begin{equation}
 Q_1Q_5=N,\qquad
 n=N_p,\qquad
 J=J_L=2J_{3L},\qquad
 S_{\rm BH}
 =
 2\pi\sqrt{Q_1Q_5n-J^2/4}.
 \label{eq:our-to-Sen-charge-ray}
\end{equation}
Sen's \(T^5\) result is \(-6\log\Lambda\) for
\(Q_1,Q_5,n\sim\Lambda\) and \(J\sim\Lambda^{3/2}\) at fixed nonzero
rotation ratio \cite{SenLogBMPV}; since
\(S_{\rm BH}\propto\Lambda^{3/2}\), this is
\(-4\log S_{\rm BH}\).  The difference between the full sixth helicity trace and the
center-of-mass-factored quadratic insertion used here stays finite at the
rotating saddle, so it changes only the \(O(1)\) term; the factor is
evaluated in Appendix~\ref{app:fixed-charge-technical-bounds}.  The Cartan-fixed and
Casimir-fixed indices likewise have the same logarithmic coefficient for
nonzero macroscopic rotation; the nonrotating singlet is exceptional
because the difference factor vanishes at the saddle
\cite{SenLogBMPV}.

The ratio of the free count to the protected lower bound measures how
much lifting is compatible with the index. Its large-$N$ behavior differs
in the three regions:
\begin{align}
 \frac{D_V}{4|I_N|}
 &\longrightarrow
 Z(e^{-\beta_*\nu},e^{-\beta_*},e^{\mu_*}),
 &&\nu>\frac{|\jmath|}{2},
 \nonumber\\
 \frac{D_V}{4|I_N|}
 &=\Theta(N),
 &&\nu=\frac{|\jmath|}{2},
 \nonumber\\
 \log\frac{D_V}{4|I_N|}
 &=S_{\rm enigma}-S_{\rm BH}+o(N),
 &&\frac{\jmath^2}{4}<\nu<\frac{|\jmath|}{2}.
 \label{eq:free-protected-three-region}
\end{align}
Here \(\Theta(N)\) denotes linear growth in \(N\), so the free count
exceeds the protected scale by one power of \(N\).  In the open
uncondensed region the protected count remains a finite fraction of the
free dimension, leaving no exponential room for lifting.  On the critical
line a polynomial mismatch opens, while in the condensed interior the
mismatch is exponential.  The phase boundary therefore exhibits the point at
which the free spectrum is no longer rigidly tied to the protected scale.

The condensed-side mismatch can also be seen directly in the signed seed.
Applying the same \(m\)-resolved decomposition as in
Eq.~\eqref{eq:singleton-resolved-coefficient}, on the positive-spin branch
one finds
\begin{align}
 \sum_{m=0}^{N}(m+1)
 \widehat c((N-m)N_p,J_L-m)
 &=
 [q^{NN_p}y^{J_L}]
 \frac{\widehat I(q,y)}{(1-yq^{N_p})^2},
 \nonumber\\
 \frac{\widehat I(q,y)}{(1-yq^{N_p})^2}
 &=
 y^{-1}(1-y)^2
 \frac{(1-y^{-1}q^{N_p})^2}{(1-q^{N_p})^4}
 \prod_{\substack{r\ge1\\r\ne N_p}}
 \frac{(1-yq^r)^2(1-y^{-1}q^r)^2}{(1-q^r)^4}.
 \label{eq:condensed-pole-zero-cancellation}
\end{align}
The two aligned bosonic occupations produce
\((1-yq^{N_p})^{-2}\), whereas the modified-index seed contains the exact
Jacobi factor \((1-yq^{N_p})^2\).  These factors cancel, so the
macroscopic short-string occupation that enhances the unsigned count has
no corresponding pole enhancement in the protected index.  Charge
conjugation gives the negative-spin branch.  Ref.~\cite{BenaMoulting}
finds the analogous short-string cancellation using a theta sum.

We now transfer the long-string geometry from the free block to the exact
BPS space.  Along a regular deformation path, an exact BPS state that
survives to the interacting theory has a limiting representative inside the
free right-BPS block, so the free dimension is a ceiling while the index is
a deformation-invariant floor.  Let \(S^0_{N,\Gamma_N}\) be such
a limiting exact-BPS subspace and
\(D_{\rm gen}=\dim S^0_{N,\Gamma_N}\).  Then
\begin{equation}
 4|I_N|\le D_{\rm gen}\le D_V.
 \label{eq:exact-BPS-index-squeeze}
\end{equation}
For differentiable representatives, the projected first-order equations
of Section~\ref{sec:review-lifting} also imply
\(S^0\subseteq\mathcal H_N\).

Dimension alone does not determine cycle geometry, because a BPS state may
superpose several cycle partitions.  The relevant physical constraint is positivity.
A subspace carrying the protected entropy scale cannot hide in a region to
which the entire free block assigns asymptotically little weight.  For any set \(E\) of cycle partitions and any subspace
\(S\subseteq V_{N,\Gamma_N}\), the orthogonal cycle projectors therefore
obey
\begin{NoHyper}
\begin{equation}
 0\le
 \Tr\!\left[\left(\sum_{p\in E}P_p\right)P_S\right]
 \le
 \Tr_V\!\left(\sum_{p\in E}P_p\right).
 \label{eq:cycle-set-positivity}
\end{equation}
\end{NoHyper}
The required projectors are constructed in
Appendix~\ref{app:projector-constructions}.  In the uncondensed region,
Eqs.~\eqref{eq:uniform-finiteN-depth-tail}
and~\eqref{eq:CFT-index-invariant-log} give
\begin{equation}
 \limsup_{N\to\infty}
 \frac{\Tr_V(1-P_{\le R})}{4|I_N|}
 \le\frac{C}{(1+R)^2},
 \label{eq:positive-tail-envelope}
\end{equation}
where \(C\) is independent of \(N\) on a compact set of charge ratios.
Together with the first line of Eq.~\eqref{eq:free-protected-three-region}
and the squeeze \eqref{eq:exact-BPS-index-squeeze}, this yields
\begin{align}
 \liminf_{N\to\infty}\frac{D_{\rm gen}}{D_V}
 &\ge
 \frac1{Z(e^{-\beta_*\nu},e^{-\beta_*},e^{\mu_*})},
 \nonumber\\
 \log D_{\rm gen}
 &=S_{\rm BH}-4\log S_{\rm BH}+O_{\nu,\jmath}(1),
 \nonumber\\
 \limsup_{N\to\infty}
 \frac{\Tr[(1-P_{\le R})P_{S^0}]}{D_{\rm gen}}
 &\le\frac{C}{(1+R)^2}.
 \label{eq:exact-BPS-uncondensed-transfer}
\end{align}
Thus we do not claim that the exact-BPS subspace has the same detailed
cycle distribution as the free ensemble.  What transfers is only the
long-string tail bound: its normalized cycle weight lies, with probability
approaching one as \(R\to\infty\), in sectors with
\(w_{\max}\ge N-R\).

The decisive question is whether this entropy and long-string
concentration are carried merely by states already accounted for by
supergravity.  We therefore remove the generalized-gravity sector
\(\mathcal G_N\) reviewed in Section~\ref{sec:review-fortuity-program}.
It is generated from multiparticle supergraviton states by total affine
creation modes, including the singly wound Ramond-ground-state dressings.
For the argument below, only an upper bound on its fixed-charge dimension
\(D_G=\dim\mathcal G_{N,\Gamma_N}\) is needed.

The finite supergravity field content on the near-horizon geometry behaves
thermodynamically as a gas rather than as a black hole: its entropy at
energy \(h\) grows as \(h^{3/4}\), not linearly in \(h\).  Since the
black-hole charge ray has \(h_N=O(N)\), this makes the supergravity sector
subexponential compared with the \(e^{O(N)}\) black-hole degeneracy.  The
standard partition-function estimate makes this precise.  The nonzero-energy
contribution to the unsigned supergravity partition function of
Ref.~\cite{Benjamin:2016pil} is proportional to
\(\prod_{r\ge1}[(1+q^{r/2})/(1-q^{r/2})]^{8r^2+4}\).  At \(q=e^{-t}\) its logarithm is \(8\pi^4/(3t^3)+O(t^{-1})\), and since the
coefficients are nonnegative a Legendre transform
(Appendix~\ref{app:fixed-charge-technical-bounds}) gives
\begin{equation}
 \log D_{\rm sugra}^{(N)}(h)
 \le
 \frac{4\pi\,2^{3/4}}{3}h^{3/4}
 +O(h^{1/4}+\log N).
 \label{eq:sugra-finiteN-envelope}
\end{equation}
Let \(s\) denote the conformal level added by the total-affine
(singleton) creation modes.  A generalized-gravity state of total level
\(h_N\) is obtained by dressing an underlying multiparticle supergraviton
state of level \(h_N-s\) with affine excitations of total level \(s\).
There are eight bosonic and eight fermionic left-moving affine-generator
species.  Dropping the fermionic exclusion rule and all charge restrictions
bounds the number of affine dressings at level \(s\) by
\(\exp[O(\sqrt{s})]\).  Hence
\begin{align}
 D_G(\Gamma_N)
 &\le
 \sum_{\substack{0\le s\le h_N\\2s\in\mathbb Z}}
 D_{\rm sugra}^{(N)}(h_N-s)e^{O(\sqrt s)},
 \nonumber\\
 \log D_G(\Gamma_N)
 &=O(N^{3/4})=o(N).
 \label{eq:generalized-gravity-subexponential}
\end{align}
This overcounts the generalized-gravity space. Imposing the
fixed-charge conditions, quotienting null states, and identifying
linearly dependent descendants can only decrease the actual dimension.

Since this overcounted generalized-gravity sector has only
\(\log D_G=O(N^{3/4})\), it is exponentially negligible on the
black-hole scale. Removing it from the exact BPS space changes neither the leading dimension
nor the long-string concentration. Projecting \(S^0\) onto \(\mathcal G_N\) can remove at most
\(D_G\) independent directions.  Hence
\begin{equation}
 D_{\rm gen}-D_G
 \le
 \dim(S^0\cap\mathcal G_N^\perp)
 \le D_{\rm gen}.
\end{equation}
Since \(D_G/D_{\rm gen}\) is exponentially small, the non-graviton
subspace retains the full black-hole entropy,
\begin{equation}
 \boxed{
 \begin{aligned}
 \log\dim(S^0\cap\mathcal G_N^\perp)
 &=S_{\rm BH}-4\log S_{\rm BH}+O(1),\\
 \limsup_{N\to\infty}
 \frac{\Tr[(1-P_{\le R})P_{S^0\cap\mathcal G_N^\perp}]}
 {\dim(S^0\cap\mathcal G_N^\perp)}
 &\le\frac{C}{(1+R)^2}.
 \end{aligned}}
 \label{eq:non-graviton-sector-law}
\end{equation}
Thus an entropy-carrying exact non-graviton subspace remains concentrated
at \(w_{\max}=N-O_{\Pr}(1)\).  

The preceding trace bound is an average statement over the non-graviton
subspace.  It can be strengthened to a state-wise statement.  Let
\(S=S^0\cap\mathcal G_N^\perp\) and diagonalize
\(P_SP_{\le R}P_S\).  An eigenvalue close to one means that the
corresponding state has almost all of its norm in the long-string region.
For an eigenvector \(u_i\), the quantity \(1-\lambda_i\) is precisely
its weight outside the long-string region.  The trace bound controls the
sum of these weights over the whole subspace, so only a small fraction of
independent directions can have a substantial short-string component. After discarding these directions, one retains a subspace with
the same black-hole entropy scale in which every unit vector satisfies
\begin{equation}
 \langle\psi|P_{\le R}|\psi\rangle\ge1-\epsilon .
 \label{eq:non-graviton-core-entropy}
\end{equation}
The elementary spectral estimate is given in
Appendix~\ref{app:projector-constructions}.

This does not restore a state-by-state geometric criterion of the kind
ruled out in Section~\ref{sec:microscopic-fortuity}.  Rather, it shows
that although individual BPS states need not obey a universal cycle rule,
the entropy-carrying sector contains a black-hole-sized subspace whose
vectors are uniformly concentrated on long-string configurations.

The transfer of long-string concentration relies on the protected lower
bound having the same asymptotic scale as the full free count, and therefore
holds only in the open uncondensed region.  On the critical line and in the
condensed interior, the index still guarantees exponentially many exact
non-graviton states, but no longer determines where their cycle weight is
concentrated.  From Eqs.~\eqref{eq:exact-BPS-index-squeeze}
and~\eqref{eq:generalized-gravity-subexponential},
\begin{align}
 \log D_{\rm gen}
 &\ge S_{\rm BH}+o(N),
 \nonumber\\
 \log\dim(S^0\cap\mathcal G_N^\perp)
 &\ge S_{\rm BH}+o(N).
 \label{eq:condensed-nongrav-lower-bound}
\end{align}
Thus exponentially many exact non-graviton states still survive in both
regions, but their cycle distribution is no longer fixed by the free
count.  The Rayleigh law on the critical line and the Gaussian law in the
condensed interior remain the universal large-\(N\) scaling laws of the
free fixed-charge ensemble derived in
Section~\ref{sec:phase-and-moments}.  What fails is their transfer to the
exact or non-graviton BPS sectors.

\subsection{Protected descendants and the monotone/fortuitous split}
\label{sec:tower-counting}
\label{sec:tower-giant-exclusion}

This subsection asks how much of the non-graviton long-string
concentration of Section~\ref{sec:core-counting} can be restated in the
monotone and fortuitous language. We distinguish a proved tower bound,
a conditional dimension statement, and a further condition on cycle
weights. The explicit protected descendant tower is a directly controlled
monotone subsector whose normalized weight at fixed cycle depth tends to
zero. If the full monotone space is
subexponential, the fortuitous sector inherits both the black-hole entropy
and the long-string concentration.  Separating the two cycle distributions needs a further assumption on monotone
representatives.

The first step is an upper bound on the dimension of the tower
\(\mathcal T_N\) of Eq.~\eqref{eq:intrinsic-tower-definition}.
For this purpose we deliberately overcount.  The eight bosonic and eight
fermionic affine-generator species are treated as sixteen unrestricted
bosonic species.
Let \(p_{16}(n)\) denote the resulting sixteen-colour partition number,
defined by
\begin{equation}
 \sum_{n\ge0}p_{16}(n)x^n
 =
 \prod_{r\ge1}(1-x^r)^{-16}.
 \label{eq:p16-generating-function}
\end{equation}
For \(x=e^{-t}\), its canonical growth obeys
\begin{align}
 \log\sum_{n\ge0}p_{16}(n)e^{-tn}
 &=
 16\sum_{k\ge1}\frac{1}{k(e^{tk}-1)}
 \le
 \frac{8\pi^2}{3t},
 \nonumber\\
 p_{16}(n)e^{-tn}
 &\le
 \sum_{m\ge0}p_{16}(m)e^{-tm},
 \nonumber\\
 \log p_{16}(n)
 &\le
 \inf_{t>0}
 \left(
 nt+\frac{8\pi^2}{3t}
 \right)
 =
 2\pi\sqrt{\frac{8n}{3}}.
 \label{eq:p16-upper-bound}
\end{align}

A state in \(\mathcal T_N\) is specified by two pieces: a product
\(\Omega\) of protected single-cycle seeds and a word \(A\) of total-affine
creation operators.  We bound these two choices separately.  Since the
cycle windings in \(\Omega\) sum to \(N\), treating all protected seed
species as unrestricted bosonic colours gives
\begin{equation}
 \#\{\Omega\}\le p_{16}(N).
\end{equation}
If the affine word adds conformal level \(\ell\), then
\(0\le\ell\le h_N\).  After doubling the half-integer grading and again
treating all sixteen affine-generator species as bosonic colours,
\begin{equation}
 \#\{A\}\le p_{16}(2h_N).
\end{equation}
With the finite zero-mode
multiplicity this gives
\begin{equation}
 \log\dim\mathcal T_{N,\Gamma_N}
 \le2\pi\sqrt{\frac83}(\sqrt N+\sqrt{2h_N})+O(\log N)
 =O(\sqrt N).
 \label{eq:tower-dimension-upper}
\end{equation}
The estimate deliberately ignores Bose--Fermi restrictions and null
relations.

To show that the protected tower avoids the fixed-depth long-string
region, an upper bound on that region is not enough: we also need a lower
bound on the total size of the tower.  On the even-\(N\) charge ray
\(\Gamma_N=(N,-N)\), we construct an explicit family of linearly
independent tower states whose multiplicity already grows as
\(e^{O(\sqrt N)}\):
\begin{align}
 \dim\mathcal T_{N,(N,-N)}
 &\ge [x^N]\prod_{w\ge1}\frac{(1+x^w)^4}{(1-x^w)^4}
      \;[x^{N/2}]\prod_{r\ge1}(1-x^r)^{-4}, \nonumber \\
 \log\dim\mathcal T_{N,(N,-N)}
 &\ge2\pi\left(1+\frac1{\sqrt3}\right)\sqrt N-O(\log N).
 \label{eq:tower-independent-lower-exponent}
\end{align}
The coefficient product above counts distinct choices of protected seed
states and bosonic-current occupation numbers. To use this count as a
lower bound on the dimension of the tower, we must verify that these
choices give linearly independent states. 

The protected seeds have \(h_0=j_0=N/2\), whereas the target charge
sector has \(h=N\) and \(j=0\).  We first raise the conformal weight
without changing the spin, so the neutral total currents must contribute
exactly \(N/2\) units of level. Write such a descendant as
\begin{equation}
 A_{\boldsymbol k}
 =
 \prod_{m,\mu}
 \left(\alpha_{-m}^{\mu,({\rm T})}\right)^{k_{m,\mu}},
 \qquad
 \sum_{m,\mu}m k_{m,\mu}=\frac N2 .
\end{equation}
The total bosonic currents satisfy
\begin{equation}
 \left[
 \alpha_m^{\mu,({\rm T})},
 \alpha_{-n}^{\nu,({\rm T})}
 \right]
 =
 Nm\,\delta_{mn}\delta^{\mu\nu}.
\end{equation}
Repeated use of this algebra gives
\begin{equation}
 \langle\Omega_a|
 A_{\boldsymbol k}^\dagger A_{\boldsymbol l}
 |\Omega_b\rangle
 =
 \delta_{ab}\delta_{\boldsymbol k,\boldsymbol l}
 \prod_{m,\mu}
 k_{m,\mu}!(Nm)^{k_{m,\mu}}.
 \label{eq:tower-Heisenberg-Gram}
\end{equation}
Thus the Gram matrix is diagonal with strictly positive diagonal entries,
so all states counted above are linearly independent. Finally,
$(J_0^-)^{N/2}$ lowers the spin from $j=N/2$ to the required $j=0$.
The $SU(2)$ algebra multiplies every norm by the same nonzero factor $N!$,
and therefore preserves linear independence.

We next bound \(\Tr(P_{\mathcal T}P_{\le R})\).
Recall that
\(d=N-w_{\max}\), so \(d\le R\) means that one cycle carries at least
\(N-R\) units of winding, while all remaining cycles carry at most \(R\)
units in total.  For fixed \(R\) and \(N>2R\), the long cycle is unique
and the possible protected seed configurations outside it form a finite
set independent of \(N\).  Their number is therefore \(O_R(1)\).

It remains to count the total-current descendants that can be built on
these seeds.  For an upper bound we ignore charge constraints and null
relations.  Introducing \(x\) to count twice the total conformal level,
the eight bosonic and eight fermionic total creation modes give
\begin{equation}
 \prod_{m\ge1}(1-x^{2m})^{-8}
 \prod_{m\ge0}(1+x^{2m+1})^8 .
 \label{eq:tower-fixed-depth-descendant-product}
\end{equation}
Setting \(x=e^{-t}\), its small-\(t\) growth is
\begin{equation}
 \log\prod_{m\ge1}(1-e^{-2tm})^{-8}
 +\log\prod_{m\ge0}(1+e^{-t(2m+1)})^8
 =
 \frac{\pi^2}{t}+O(\log(1/t)).
 \label{eq:tower-fixed-depth-descendant-asymptotic}
\end{equation}
Writing
\(Z_{\rm desc}(x)=\sum_{n\ge0}a_nx^n\) with \(a_n\ge0\), positivity gives
\begin{equation}
 a_N e^{-tN}
 \le
 Z_{\rm desc}(e^{-t}),
\end{equation}
and hence
\begin{align}
 \log a_N
 &\le
 Nt+\frac{\pi^2}{t}+O(\log(1/t)). \nonumber
\\ 
 [x^N]Z_{\rm desc}(x)
 &\le
 \exp\!\left(
 Nt+\frac{\pi^2}{t}+O(\log(1/t))
 \right).
\end{align}
The exponent is minimized at \(t=\pi/\sqrt N\), yielding
\begin{equation}
 [x^N]Z_{\rm desc}(x)
 \le
 e^{2\pi\sqrt N+O(\log N)}.
\end{equation}

On the common even-$N$ ray, the tower and the exact non-graviton space
therefore satisfy opposite normalized trace limits:
\begin{equation}
 \boxed{
 \begin{aligned}
 \frac{\Tr(P_{\mathcal T}P_{\le R})}{\dim\mathcal T_{N,(N,-N)}}
 &\longrightarrow0
 &&\text{for every fixed }R,\\
 \lim_{R\to\infty}\limsup_{N\to\infty}
 \frac{\Tr[P_{S^0\cap\mathcal G_N^\perp}(1-P_{\le R})]}
 {\dim(S^0\cap\mathcal G_N^\perp)}
 &=0.&&
 \end{aligned}}
 \label{eq:two-sided-reference-core-separation}
\end{equation}
The two limits separate the cycle behavior of the two
subspaces. For every fixed $R$, the fraction of the explicit descendant
tower $\mathcal T_N$ in the region $d=N-w_{\max}\le R$ vanishes as
$N\to\infty$. This proves only that $d$ is not $O(1)$ in this tower; it does
not distinguish growth such as $\sqrt N$, $N^\alpha$, or $N$. For
$S^0\cap\mathcal G_N^\perp$, the weight at
$d>R$ instead vanishes after taking $N$ large and then $R$ large. The exact
non-graviton subspace is therefore concentrated at
$w_{\max}=N-O_{\Pr}(1)$, in contrast to the explicit tower.

Extending the comparison to the full monotone sector requires
an additional argument.
A subexponential bound on the monotone dimension implies that fortuitous
states account for all but an exponentially small fraction of the
harmonic BPS space in the uncondensed region. Their normalized weight is
then concentrated at finite $d$. This does not determine where the
monotone states lie. A different cycle distribution for the monotone
sector requires an additional bound on $\Tr(P_{\rm mon}P_{\le R})$ for
fixed $R$.

Let $P_{\rm mon}$ be the orthogonal projector onto
$\mathcal H^{\rm mon}_{N,\Gamma_N}$ as defined in
Eq.~\eqref{eq:framework-short-monotone}, using the same $\pi_{N,M}$ as
in the finite-rank tests. Let $P_{\rm for}$ project onto its orthogonal
complement inside $\mathcal H_{N,\Gamma_N}$. The explicit continuations
in Eq.~\eqref{eq:intrinsic-tower-explicit-lift} give
$\mathcal T_{N,\Gamma_N}\subseteq\operatorname{Ran}P_{\rm mon}$, and
$\rank P_{\rm for}=D_H-\rank P_{\rm mon}$.

First assume only that the monotone space is subexponential on the
charge ray:
\begin{equation}
 \log\rank P_{\rm mon}=o(N).
 \label{eq:stable-dimension-hypothesis}
\end{equation}
In the uncondensed region, $4|I_N|\le D_H\le D_V$ fixes the harmonic
dimension on the same scale as the exact BPS dimension. Under
Eq.~\eqref{eq:stable-dimension-hypothesis}, the monotone fraction is
exponentially small. Its orthogonal complement has the same leading
multiplicity and long-string concentration:
\begin{align}
 \frac{\rank P_{\rm mon}}{D_H}
 &\le e^{-S_{\rm BH}+o(N)}, \nonumber \\
 \log\rank P_{\rm for}
 &=S_{\rm BH}-4\log S_{\rm BH}+O(1), \nonumber \\
 \limsup_{N\to\infty}
 \frac{\Tr[(1-P_{\le R})P_{\rm for}]}{\rank P_{\rm for}}
 &\le\frac{C}{(1+R)^2}.
 \label{eq:BH-wall-theorem}
\end{align}
Hence $w_{\max}=N-O_{\Pr}(1)$ in the fortuitous sector, even though
individual short-string fortuitous families remain possible.

Trace additivity also gives, for any set $E$ of cycle partitions,
\begin{equation}
 \sup_E\left|
 \frac{\Tr[(\sum_{p\in E}P_p)(P_{\rm mon}+P_{\rm for})]}{D_H}
 -\frac{\Tr[(\sum_{p\in E}P_p)P_{\rm for}]}{\rank P_{\rm for}}
 \right|
 \le\frac{\rank P_{\rm mon}}{D_H}
 \le e^{-S_{\rm BH}+o(N)}.
 \label{eq:harmonic-fortuitous-measure-comparison}
\end{equation}
Thus removing the monotone space changes the normalized harmonic
weights assigned to cycle partitions only by an exponentially small amount.

For differentiable regular transport, $S^0\subseteq\mathcal H_N$. The
exact BPS states orthogonal to both the generalized-gravity and monotone
spaces have dimension at least
\begin{equation}
 \dim\!\left[S^0\cap\mathcal G_N^\perp
 \cap(\operatorname{Ran}P_{\rm mon})^\perp\right]
 \ge D_{\rm gen}-D_G-\rank P_{\rm mon}.
 \label{eq:exact-nongrav-fortuitous-dimension}
\end{equation}
In the open uncondensed region this retains a fraction
$1-e^{-S_{\rm BH}+o(N)}$ of the dimension of the limiting exact BPS subspace, together with
the $-4\log S_{\rm BH}$ correction and the same long-string tail bound.
Here fortuity is the first-order continuation classification. An all-orders
identification additionally requires non-renormalization and norm-preserving
identifications that intertwine the continuation maps at each rank
\cite{ChangLinZhang}.

The third step separates the two cycle distributions and requires an
additional assumption about the harmonic monotone representatives. Keep
the already established inclusion
\begin{equation}
 \mathcal T_{N,\Gamma_N}\subseteq\operatorname{Ran}P_{\rm mon},
 \label{eq:geometric-monotone-hypothesis}
\end{equation}
and assume directly the fixed-depth bound
\begin{equation}
 \Tr(P_{\rm mon}P_{\le R})
 \le e^{2\pi\sqrt N+O_R(\log N)}
 \quad\text{on }\Gamma_N=(N,-N).
 \label{eq:gravity-creation-finite-depth-bound}
\end{equation}
This is a condition on cycle weights, not a consequence of the
subexponential dimension bound alone. It avoids assuming an unspecified
class of creation operators for all monotone states.

Combining this with the tower lower bound and
Eq.~\eqref{eq:BH-wall-theorem} yields:
\begin{equation}
 \boxed{
 \begin{aligned}
 \frac{\Tr(P_{\rm mon}P_{\le R})}{\rank P_{\rm mon}}
 &\le e^{-2\pi\sqrt{N/3}+O_R(\log N)}\longrightarrow0,\\
 \limsup_{N\to\infty}
 \frac{\Tr[P_{\rm for}(1-P_{\le R})]}{\rank P_{\rm for}}
 &\le\frac{C}{(1+R)^2}.
 \end{aligned}}
 \label{eq:conditional-monotone-fortuitous-separation}
\end{equation}
Under the dimension assumption
\eqref{eq:stable-dimension-hypothesis} and the additional trace assumption
\eqref{eq:gravity-creation-finite-depth-bound}, the two normalized cycle
distributions separate after taking $N$ large and then $R$ large. The trace
assumption is stronger than a gravity/monotone identification at the level
of the interacting spectrum \cite{ToweringGravitons}: it also constrains
harmonic representatives in the specified orbifold continuation convention
\cite{GiustoInglisRusso}.

On the critical line and in the condensed region, the fortuitous
multiplicity is still exponential under
Eq.~\eqref{eq:stable-dimension-hypothesis}:
\begin{equation}
 \log\rank P_{\rm for}\ge S_{\rm BH}+o(N).
 \label{eq:condensed-fortuitous-lower-bound}
\end{equation}
The same dimension argument applies to the exact non-graviton intersection
when differentiable transport is available. However, the comparison with
the free coefficient loses a power of $N$ at criticality and an
exponential factor in the condensed interior, so it does not determine the
condensate fraction of these BPS subspaces. The three-region law
\eqref{eq:three-region-free-cycle-law} supplies the free-theory reference distribution of $d$
for Section~\ref{sec:twist-main}; the exact-BPS result established here
is restricted to the open uncondensed region and the stated
continuation assumptions.

\section{Twist kernels and finite-winding probes}
\label{sec:twist-main}

Section~\ref{sec:review-emergent-spacetime} posed the mirage
problem: a light untwisted probe in a symmetric-product orbifold can reproduce
the BTZ response at large $N$ even though the orbifold point is not a
semiclassical Einstein-gravity regime \cite{BelinBintanjaCastroKnop}.
The cycle distribution derived in
Section~\ref{sec:long-strings-lifting} allows us to ask which microscopic
data remain visible before the strict continuum limit. A BTZ-like leading
term will refer only to the response of the chosen probe, not to an
independent criterion for a sharp bulk horizon.

The twist operator changes the cycle partition by joining a
long cycle to a finite one \cite{CarsonHamptonMathurTurtonTwist}, whereas an
untwisted current probe preserves the partition and measures propagation on
the corresponding finite circles \cite{BelinBintanjaCastroKnop}. Together
with matched-state comparisons, these observables separate the finite
winding retained by joining kernels, the response of a finite circle, and
the oscillator data not fixed by the cycle geometry.

The cycle distribution does not determine either the oscillator occupations
or the full deformation-supercharge matrix elements.  In the uncondensed
region, Section~\ref{sec:long-strings-lifting} establishes long-string
concentration for the exact BPS sector under the conditions stated there.
On the critical line and in the condensed region, the explicit cycle laws
used below remain those of the free fixed-charge ensemble.  The occupation
of the two aligned singly wound Ramond ground states is
$O_{\Pr}(\sqrt N)$ at criticality and
$(|\jmath|-2\nu)N+O_{\Pr}(\sqrt N)$ in the condensed interior.  We do not
assume the same detailed occupation law for the exact BPS sector in those
regions.  The stronger monotone and fortuitous statements remain subject
to the assumptions of Section~\ref{sec:tower-counting}.

\subsection{Asymmetric twist kernel and its continuum limit}
\label{sec:twist-expansion}

Section~\ref{sec:long-strings-lifting} shows that a typical state in the
uncondensed region contains one cycle whose winding differs from $N$ by only
a finite amount, together with cycles carrying finite total winding.  These
finite cycles are thermodynamically subleading in the counting of
Section~\ref{sec:long-strings-lifting}, but they need not be dynamically
invisible.  We therefore ask whether their winding $a$ remains visible when
a twist-two operator joins a long cycle of winding $M$ to a finite cycle,
\begin{equation}
 (M)+(a)\longrightarrow L=M+a,
\end{equation}
and the limit $M\to\infty$ is taken with $a$ fixed.

Section~\ref{sec:review-exact-twist} reviewed the two exact oscillator
coefficients associated with this joining process.  The coefficient
$\widetilde f^{B(1)}_{qs}$ is the finite-$M$ transmission amplitude for a
bosonic excitation with incoming fractional frequency $q=m/M$ to emerge on
the joined cycle with outgoing fractional frequency $s=k/L$.  The
coefficient $\widetilde\gamma^B_{ss'}$ instead multiplies the pair of
outgoing modes with $s=k/L$ and $s'=k'/L$ in the squeezed state produced
when the twist acts on the incoming vacuum.  Transmission and pair creation
therefore test separately whether the finite winding $a$ survives in the
continuum action of the twist.  The exact finite-winding formulas are
Eqs.~\eqref{eq:giant-dust-exact-f} and
\eqref{eq:giant-dust-exact-gamma}
\cite{CarsonHamptonMathurTurtonTwist,CarsonHamptonMathurTurton}.  This
subsection concerns the twist operator itself at the orbifold point.  The
supercurrent insertion and BPS projection required for the full lifting
operator are recalled at the end.

We first consider transmission.  At finite $M$,
$\widetilde f^{B(1)}_{qs}$ is a discrete amplitude between one incoming
mode and one outgoing mode.  Each such coefficient is $O(L^{-1})$, while
the number of incoming modes per unit fractional frequency is $O(L)$.
The density-rescaled limit therefore defines the continuum transmission
kernel
\begin{equation}
 \lim_{M\to\infty}L\widetilde f^{B(1)}_{qs}
 =\mathcal F_a(q,s),
 \qquad
 \mathcal F_a(q,s)=\frac{i}{2\pi}z_0^{s-q}
 \frac{1-e^{-2\pi ias}}{s-q}
 \left(\frac ea\right)^{a(s-q)}q^{aq}s^{-as}
 \frac{\Gamma(as)}{\Gamma(aq)}.
 \label{eq:continuum-transmission-kernel}
\end{equation}
Thus $\widetilde f^{B(1)}_{qs}$ is the exact discrete transmission
amplitude, whereas $\mathcal F_a(q,s)$ is the kernel that remains after the
growing mode density is included.  The limit is uniform when $q$ and $s$
remain in any fixed finite interval $\epsilon\le q,s\le\Lambda$ with
$\epsilon>0$, provided they stay away from $q=s$.  The restriction keeps
the fractional frequencies bounded away from zero while $M$ grows.  When
an incoming and outgoing lattice mode coincide, the exact coefficient is
instead Eq.~\eqref{eq:giant-dust-diagonal}.

The continuum limit relevant for an operator is the limit of the complete
mode sum.  When $q$ stays away from $s$, the usual Riemann sum gives, for a
smooth function $g$,
\begin{equation}
 \frac1L\sum_m g(m/M)
 =\frac ML\left[\frac1M\sum_m g(m/M)\right]
 \longrightarrow\int dq\,g(q).
\end{equation}
The same replacement is not uniform when the incoming and outgoing
fractional frequencies approach one another because
$\mathcal F_a(q,s)$ contains the Cauchy factor $(s-q)^{-1}$.  The region
$q\simeq s$ must therefore be treated separately.  This is not a distinct
physical process.  It is required because replacing the discrete sum by an
integral there would miss the finite offset between the two mode lattices.

To separate the Cauchy factor from the part that is regular at $q=s$, define
\begin{equation}
 H_a(q,s):=(s-q)\mathcal F_a(q,s),
 \qquad
 H_a(s,s)=\frac{i}{2\pi}\chi_a(s),
 \qquad
 \chi_a(s)=1-e^{-2\pi ias}.
 \label{eq:Ha-definition}
\end{equation}
The function $H_a$ contains the smooth numerator of the transmission
kernel.  Subtracting its value at equal incoming and outgoing frequencies
removes the apparent singularity,
\begin{equation}
 \frac{H_a(q,s)\psi(q)}{s-q}
 =\frac{H_a(q,s)\psi(q)-H_a(s,s)\psi(s)}{s-q}
 +\frac{H_a(s,s)\psi(s)}{s-q}.
 \label{eq:twist-pole-subtraction}
\end{equation}
The first numerator vanishes at $q=s$ and therefore has an ordinary
continuum limit.  The second term keeps track of the relative displacement
of the two discrete lattices.  Since $q=m/M$, $s=k/L$ and $L=M+a$,
\begin{align}
 M(s-q)
 &=Ms-m \nonumber\\
 &=(L-a)s-m \nonumber\\
 &=k-m-as
 =j-as,
 \qquad j:=k-m.
 \label{eq:shifted-lattice}
\end{align}
Hence the singular part is controlled by a lattice shifted by the finite
amount $as$ even when both spacings go to zero.

At leading order the mode sum can now be split before taking the limit.
Writing $q_m=m/M$,
\begin{align}
 \frac1L\sum_m
 \frac{H_a(q_m,s)\psi(q_m)}{s-q_m}
 &=\frac1L\sum_m
 \frac{H_a(q_m,s)\psi(q_m)-H_a(s,s)\psi(s)}{s-q_m}
 \nonumber\\
 &\quad+
 \frac ML H_a(s,s)\psi(s)
 \sum_m\frac1{k-m-as}.
 \label{eq:twist-mode-sum-split}
\end{align}
The first sum is regular and becomes the continuum principal-value term
when the subtracted constant is restored.  In a symmetric window around
$q=s$, the second sum tends to the shifted principal-value lattice sum.
The identities
\begin{align}
 \operatorname{PV}\sum_{j\in\mathbb Z}\frac1{j-as}
 &=-\pi\cot(\pi as), \nonumber\\
 H_a(s,s)\operatorname{PV}\sum_{j\in\mathbb Z}\frac1{j-as}
 &=-\frac i2\chi_a(s)\cot(\pi as)
 =\frac{1+e^{-2\pi ias}}2
 \label{eq:shifted-lattice-contact}
\end{align}
then give the continuum transmission operator
\begin{equation}
 \boxed{\begin{aligned}
 \sum_m\widetilde f^{B(1)}_{m/M,s}\psi(m/M)
 &\longrightarrow
 \frac{1+e^{-2\pi ias}}2\psi(s)\\
 &\quad+\operatorname{PV}\!\int dq\,\mathcal F_a(q,s)\psi(q).
 \end{aligned}}
 \label{eq:giant-dust-continuum-limit}
\end{equation}
The first term is local in fractional frequency and records the displaced
mode lattice near $q=s$.  The second term mixes different incoming and
outgoing frequencies.  At a common lattice mode the local coefficient is
one by continuity.  Both pieces retain $a$, so the continuum transmission
operator still distinguishes which finite cycle joined the long one.

Having obtained the operator limit, we next determine the finite-$M$ error
of the transmission kernel.  At finite $M$,
$\widetilde f^{B(1)}_{qs}$ is the exact discrete amplitude, while
$\mathcal F_a(q,s)/L$ is its leading continuum approximation for fixed
$q$ and $s$ away from $q=s$.  Their ratio
\begin{equation}
 R_M(q,s;a)
 :=\frac{L\widetilde f^{B(1)}_{qs}}{\mathcal F_a(q,s)}
 =X^{s-q}\left(\frac ea\right)^{-a(s-q)}q^{-aq}s^{as}
 \frac{\Gamma(Lq)}{\Gamma(Mq)}
 \frac{\Gamma(Ms)}{\Gamma(Ls)}
 \label{eq:RM-definition}
\end{equation}
measures the multiplicative finite-$M$ correction and satisfies $R_M\to1$.
The large-$M$ expansions entering this ratio are
\begin{align}
 \log X&=a+a\log(M/a)+\frac{a^2}{2M}-\frac{a^3}{6M^2}+O(M^{-3}),\nonumber\\
 \log\frac{\Gamma(z+u)}{\Gamma(z)}
 &=u\log z+\frac{u(u-1)}{2z}
 -\frac{u(2u^2-3u+1)}{12z^2}+O(z^{-3}).
 \label{eq:twist-Stirling-main}
\end{align}
In the Gamma ratios, use $(z,u)=(Mq,aq)$ and $(Ms,as)$.  Expanding
$\log R_M$ gives
\begin{align}
 \log R_M(q,s;a)
 &=\frac1M\left[
 \frac{a^2(s-q)}2+\frac{a(aq-1)}2-\frac{a(as-1)}2
 \right]\nonumber\\
 &\quad+\frac1{M^2}\left[
 -\frac{a^3(s-q)}6
 -\frac{a(2a^2q^2-3aq+1)}{12q}
 +\frac{a(2a^2s^2-3as+1)}{12s}
 \right]+O(M^{-3})\nonumber\\
 &=\frac{a}{12M^2}\left(\frac1s-\frac1q\right)+O(M^{-3}).
 \label{eq:twist-cancellation-main}
\end{align}
Therefore
\begin{equation}
 \widetilde f^{B(1)}_{qs}
 =\frac{\mathcal F_a(q,s)}{L}
 \left[1+\frac{a}{12M^2}\left(\frac1s-\frac1q\right)+O(M^{-3})\right].
 \label{eq:main-twist-result}
\end{equation}
The relative error therefore begins at $M^{-2}$.  The $a$ dependence in
$\mathcal F_a$ is already part of the leading continuum transmission
kernel rather than a finite-$M$ correction.  The estimate is uniform on
any fixed finite momentum interval bounded away from zero.

We now turn to the second oscillator coefficient reviewed in
Section~\ref{sec:review-exact-twist}.  The exact quantity
$\widetilde\gamma^B_{ss'}$ in
Eq.~\eqref{eq:giant-dust-exact-gamma} is the coefficient of the outgoing
pair with fractional frequencies $s=k/L$ and $s'=k'/L$ in the squeezed
state created when the twist acts on the incoming vacuum.  An individual
pair coefficient is $O(L^{-2})$, while the two outgoing mode sums contain
$O(L^2)$ possible pairs.  The appropriate continuum object is therefore
the density-rescaled limit $L^2\widetilde\gamma^B_{ss'}$.

The finite-winding formula of
Eq.~\eqref{eq:giant-dust-exact-gamma} can be written in a form that makes
this limit explicit,
\begin{align}
 L^2\widetilde\gamma^B_{ss'}
 &=\frac{z_0^{s+s'}\chi_a(s)\chi_a(s')}{4\pi^2(s+s')}
 \frac{Ma}{L}
 \prod_{x\in\{s,s'\}}
 \left[
 X^x\frac{\Gamma(Mx)\Gamma(ax)}{\Gamma(Lx)}
 \right],
 \nonumber\\
 \frac{Ma}{L}&\longrightarrow a,
 \qquad
 X^x\frac{\Gamma(Mx)}{\Gamma(Lx)}
 \longrightarrow e^{ax}(ax)^{-ax}.
 \label{eq:pair-continuum-derivation}
\end{align}
It follows that
\begin{equation}
 \lim_{M\to\infty}L^2\widetilde\gamma^B_{ss'}
 =\mathcal G_a(s,s'),
 \qquad
 \mathcal G_a(s,s')=
 \frac{a}{4\pi^2}z_0^{s+s'}
 \frac{\chi_a(s)\chi_a(s')}{s+s'}
 e^{a(s+s')}(as)^{-as}(as')^{-as'}\Gamma(as)\Gamma(as').
 \label{eq:continuum-pair-kernel}
\end{equation}
Thus $\mathcal G_a$ is not an independent ansatz.  It is the continuum
pair-creation kernel obtained directly from the exact finite-winding
coefficient $\widetilde\gamma^B$.

For a smooth test function $u$ supported in a fixed finite momentum window
bounded away from zero,
\begin{equation}
 \sum_{k,k'}\widetilde\gamma^B_{k/L,k'/L}
       u(k/L,k'/L)
 \longrightarrow
 \iint ds\,ds'\,\mathcal G_a(s,s')u(s,s').
 \label{eq:pair-density-compensation}
\end{equation}
The $L^{-2}$ suppression of one pair is exactly compensated by the
$L^2$ density of outgoing pairs.  Pair creation therefore survives as a
finite continuum contribution.  Since $\mathcal G_a$ also depends
explicitly on $a$, the finite cycle is remembered both when a pre-existing
excitation is transmitted and when the twist creates a pair from the
vacuum.

Although the finite cycles outside the longest cycle carry
only subleading total winding in the uncondensed ensemble, their winding
remains visible at leading order in the continuum twist-two joining process:
both $\mathcal F_a(q,s)$ and $\mathcal G_a(s,s')$ retain $a$. These kernels
describe a specified joining $(M)+(a)\to(M+a)$; averaging them over the cycle
distribution is a separate dynamical problem.

\subsection{Finite-circle current blocks}
\label{sec:finiteN-probe-response}

{
We use the neutral untwisted scalar reviewed in
Section~\ref{sec:review-emergent-spacetime}.  Its sum over copies is the
standard invariant untwisted probe
\cite{BelinBintanjaCastroKnop}.
\begin{equation}
\begin{aligned}
 \mathcal O^{AB}(u,\bar u)&=\frac1{\sqrt N}
 \sum_{i=1}^N J^{A,[i]}(u)\widetilde J^{B,[i]}(\bar u), \\
 J^A&=i\partial X^A.
\end{aligned}
 \label{eq:probe-definition}
\end{equation}
Here $u,\bar u$ are cylinder coordinates, $A,B$ label the internal
$T^4$ bosonic polarizations, and $\widetilde J^B=i\bar\partial X^B$.  Let
$P_{\rm BPS}$ denote the orthogonal projector onto the right-BPS
external-state subspace used here.  A single insertion is not an operator
acting within this subspace.  The current $\widetilde J^B$ creates a right-moving bosonic
excitation above the Ramond ground, so
\begin{equation}
 P_{\rm BPS}\,\mathcal O^{AB}\,P_{\rm BPS}=0.
 \label{eq:probe-BPS-compression-zero}
\end{equation}
Consequently the two-point function is mediated entirely by states outside
the BPS subspace.  Inserting the resolution of the identity gives
\begin{align}
 P_{\rm BPS}\mathcal O^{AB}(\tau)\mathcal O^{AB}(0)P_{\rm BPS}
 &=P_{\rm BPS}\mathcal O^{AB}(\tau)
   \bigl[P_{\rm BPS}+(1-P_{\rm BPS})\bigr]
   \mathcal O^{AB}(0)P_{\rm BPS}
   \nonumber\\
 &=P_{\rm BPS}\mathcal O^{AB}(\tau)P_{\rm BPS}
   \mathcal O^{AB}(0)P_{\rm BPS}
   +P_{\rm BPS}\mathcal O^{AB}(\tau)(1-P_{\rm BPS})
   \mathcal O^{AB}(0)P_{\rm BPS}
   \nonumber\\
 &=P_{\rm BPS}\mathcal O^{AB}(\tau)(1-P_{\rm BPS})
   \mathcal O^{AB}(0)P_{\rm BPS},
 \label{eq:probe-nonbps-intermediate}
\end{align}
where the first term on the second line vanishes by
Eq.~\eqref{eq:probe-BPS-compression-zero}.  Thus the correlator does not
measure an operator spectrum internal to the protected Hilbert space.  It
measures the excitations that the probe can create around a BPS microstate.  On a length-$n$ cycle, a right-moving
current mode of fixed covering level $m$ raises the energy above the
Ramond ground by $m/n$ without changing the R charge.  The intermediate
state is therefore non-BPS at finite $n$ but becomes parametrically
near-BPS for fixed $m$ as $n\to\infty$.  This is conceptually parallel,
without implying a mode-by-mode identification, to low-temperature
AdS$_3$ supergravity analyses in which fluctuation eigenvalues approach
zero at extremality \cite{BacCastroJainNearBPS}.  In the present CFT, the
dense near-BPS tower is the microscopic spectrum sampled by the long-cycle
probe.  The untwisted insertions preserve the cycle partition.  We set the
common circle radius to one and the internal bosonic zero modes to zero
throughout.

The cycle length fixes the available frequencies, but their spectral
weights depend on oscillator occupations.  Let $\tau$ denote the complex
Euclidean time separation on the covering cylinder, with Lorentzian
continuation $\tau=\epsilon+it$ and $\epsilon>0$.  For a normalized
number state on a cycle of winding $n$, $N_m^A$ denotes the number of
left-moving neutral bosonic quanta of polarization $A$ at covering level
$m$, whose physical frequency is $m/n$.  The two possible oscillator
contractions contribute $N_m^A+1$ and $N_m^A$; each current supplies a
factor $1/n$ when written on the length-$n$ cover.  The exact left-moving
block is
\begin{equation}
C^L_{n,\{N_m^A\}}(\tau)
 =\frac1{n^2}\sum_{m\ge1}m\left[
 (N_m^A+1)e^{-m\tau/n}+N_m^Ae^{m\tau/n}\right].
 \label{eq:probe-number-state-block}
\end{equation}
Equation~\eqref{eq:probe-number-state-block} is the exact pure-state
matrix element at the symmetric-orbifold point.  To define the canonical
reference state associated with the long-string saddle, consider one
bosonic mode of energy $m/n$.  A state containing $r$ quanta carries the
Boltzmann weight $e^{-\beta_*mr/n}$, so
\begin{equation}
 \overline N_m
 =\frac{\sum_{r\ge0}r e^{-\beta_*mr/n}}
        {\sum_{r\ge0}e^{-\beta_*mr/n}}
 =\frac1{e^{\beta_*m/n}-1},
 \qquad \beta_*=\frac{\pi}{\sqrt\delta}.
 \label{eq:probe-bose-occupation}
\end{equation}
Replacing $N_m^A$ by $\overline N_m$ therefore defines a canonical
reference correlator.  It is neither a fixed-charge projection of the
oscillator state nor the correlator of every BPS state on the same cycle.
The right movers remain in a Ramond ground state.  Keeping $n$ finite
distinguishes the dense near-BPS spectrum from its strict continuum limit
and from the recurrences of the finite circle.  This is the microscopic
version of the mirage test reviewed in
Section~\ref{sec:review-emergent-spacetime}.  The $n\to\infty$ response can
look BTZ-like, while finite winding can retain information absent from that
leading limit \cite{BelinBintanjaCastroKnop}.

We first evaluate the reference correlator in the Euclidean thermal strip
$0<\operatorname{Re}\tau<\beta_*$, where the geometric expansions below
converge.  Lorentzian time is introduced afterward through
$\tau=\epsilon+it$.  Substituting Eq.~\eqref{eq:probe-bose-occupation}
in Eq.~\eqref{eq:probe-number-state-block} gives
\begin{align}
 C^L_{n,\beta_*}(\tau)
 &=\frac1{n^2}\sum_{m\ge1}m\left[
 \frac{e^{-m\tau/n}}{1-e^{-\beta_*m/n}}
 +\frac{e^{m\tau/n}}{e^{\beta_*m/n}-1}\right] \nonumber \\
 &=\frac1{4n^2}\sum_{k\in\mathbb Z}
 \operatorname{csch}^2\!\left(\frac{\tau+k\beta_*}{2n}\right). \label{eq:probe-finite-n-block}
\end{align}

To obtain the second line, expand the Bose denominators in powers of
$e^{-\beta_*m/n}$ and use
$\sum_{m\ge1}m e^{-mx}=1/[4\sinh^2(x/2)]$ on each term.
This expresses the canonical block as thermal images of the vacuum block.
The spatial period of the covering circle follows from the
partial-fraction identity

\begin{equation}
 \frac{1}{4n^2}\operatorname{csch}^2\!\left(\frac{x}{2n}\right)
 =\sum_{\ell\in\mathbb Z}\frac{1}{(x+2\pi i n\ell)^2}. \label{eq:probe-csch-partial-fraction}
\end{equation}
The two periods are therefore $\beta_*$ and $2\pi in$ in complex
Euclidean time.  In the limit of infinite winding, the thermal block is
$C^L_{\infty,\beta_*}$ below.  At finite $n$, its covering-circle images alone
are not the whole answer.  The exact finite-circle block is
\begin{equation}
 \boxed{\begin{aligned}
 C^L_{n,\beta_*}(\tau)
 &=\sum_{\ell\in\mathbb Z}
 C^L_{\infty,\beta_*}(\tau+2\pi i n\ell)-\frac1{n\beta_*}, \\
 C^L_{\infty,\beta_*}(\tau)
 &=\left(\frac{\pi}{\beta_*}\right)^2
 \csc^2\!\left(\frac{\pi\tau}{\beta_*}\right).
\end{aligned}} \label{eq:probe-btz-and-first-finite-n}
\end{equation}

The difference between the finite-circle block and the image sum is
entire and doubly periodic, hence constant.  Their averages over the
imaginary period are
\begin{align}
 \frac1{2\pi n}\int_0^{2\pi n}C^L_{n,\beta_*}(\epsilon+it)\,dt&=0,\nonumber\\
 \int_{-\infty}^{\infty}C^L_{\infty,\beta_*}(\epsilon+it)\,dt
 &=\frac{2\pi}{\beta_*},
\end{align}
which fixes the constant to $-1/(n\beta_*)$.
Equation~\eqref{eq:probe-btz-and-first-finite-n} therefore separates the
strict BTZ-like block from two finite-circle effects.  The constant
$-1/(n\beta_*)$ measures a finite-size correction, while the terms with
$\ell\ne0$ remember the covering-circle period.  After
$\tau=\epsilon+it$, reindexing the image sum gives
\begin{equation}
 C^L_{n,\beta_*}\bigl(\epsilon+i(t+2\pi n)\bigr)
 =C^L_{n,\beta_*}(\epsilon+it),
 \label{eq:probe-chiral-recurrence}
\end{equation}
so the exact chiral block recurs with period $2\pi n$.  At fixed time only
the BTZ-like term survives as $n\to\infty$.  The chiral probe therefore
sees the dense near-BPS tower as a continuum at leading order, while finite
winding leaves both a constant correction and an exact recurrence.  This is
precisely the type of subleading information left open by the mirage
analysis of symmetric-product universality
\cite{BelinBintanjaCastroKnop}.  Leading agreement with BTZ is a statement
about the response of this probe, not by itself an identification of the
state with a semiclassical black-hole geometry.

For $\tau=\epsilon+it$ with fixed $0<\epsilon<\beta_*$, the
$\ell=0$ BTZ-like chiral term decays as $e^{-2\pi t/\beta_*}$ while the
finite-circle correction has magnitude $1/(n\beta_*)$.  We denote by
$t_{\rm chiral}$ the Lorentzian crossover time at which these two
contributions become comparable.  Thus
\begin{equation}
 t_{\rm chiral}
 =\frac{\beta_*}{2\pi}\log n+O_{\beta_*,\epsilon}(1).
 \label{eq:probe-logarithmic-crossover}
\end{equation}
For a single long cycle in the uncondensed region,
$n=N-O_{\Pr}(1)$, this becomes
\begin{equation}
 t_{\rm chiral}
 =\frac{\beta_*}{2\pi}\log N+O_{\Pr}(1).
 \label{eq:probe-long-cycle-chiral-scale}
\end{equation}
For $t\ll t_{\rm chiral}$ the left chiral block cannot resolve the
finite circle at leading relative accuracy and therefore looks BTZ-like.
For $t\gtrsim t_{\rm chiral}$ the finite-size term is no longer negligible.
This crossover measures how long the chiral block of a
finite-circle system approximates the BTZ response; it is not a horizon
lifetime.

The same correction also reconnects the probe directly to the change in
the number of cycles across the three regions of
Section~\ref{sec:long-strings-lifting}.  For a partition
$p=(n_1,\ldots,n_K)$ with $\sum_a n_a=N$, averaging the chiral block over
copies gives
\begin{equation}
 C_p^L(z)=\frac1N\sum_{a=1}^K n_a C^L_{n_a,\beta}(z)
 =C^L_{\infty,\beta}(z)-\frac{K}{N\beta}
 +\frac1N\sum_{a=1}^K n_a\sum_{\ell\ne0}
 C^L_{\infty,\beta}(z+2\pi i n_a\ell).
 \label{eq:probe-component-counter}
\end{equation}
Thus the constant finite-circle correction counts cycles.  Its size is
$K/N$, while the image terms retain the individual cycle lengths.  In the
free fixed-charge ensemble, Section~\ref{sec:phase-and-moments} gives
\begin{equation}
 \frac{K}{N}=
 \begin{cases}
 O_{\Pr}(N^{-1}),&\nu>|\jmath|/2,\\[1mm]
 O_{\Pr}(N^{-1/2}),&\nu=|\jmath|/2,\\[1mm]
 (|\jmath|-2\nu)+O_{\Pr}(N^{-1/2}),&\nu<|\jmath|/2.
 \end{cases}
 \label{eq:probe-phase-component-count}
\end{equation}
For the condensed reference block the excited long cycle has inverse
temperature
$\beta_{\rm c}=\pi(1-|\jmath|+2\nu)/\sqrt{\nu(1+\nu-|\jmath|)}$,
which agrees with $\beta_*$ on the critical line.  Provided the nonzero
images are still negligible, comparison with the exponentially decaying
chiral BTZ-like term gives
\begin{equation}
 t_{\rm chiral}^{(p)}=
 \begin{cases}
 \dfrac{\beta_*}{2\pi}\log N+O_{\Pr}(1),&\nu>|\jmath|/2,\\[2mm]
 \dfrac{\beta_*}{4\pi}\log N+O_{\Pr}(1),&\nu=|\jmath|/2,\\[2mm]
 O_{\Pr}(1),&\nu<|\jmath|/2.
 \end{cases}
 \label{eq:probe-phase-crossover}
\end{equation}
The chiral BTZ-like window therefore grows as $\log N$ in
the uncondensed region, as $\tfrac12\log N$ on the critical line, and remains
$O(1)$ in the condensed region. The latter two are free-ensemble reference
statements; the exact-BPS cycle concentration established in
Section~\ref{sec:long-strings-lifting} is used below only in the open
uncondensed region.

\subsection{Orbifold-invariant scalar and the horizon mirage}
\label{sec:orbifold-scalar-recurrence}

The chiral block of
Section~\ref{sec:finiteN-probe-response} is not yet the physical mirage test.
The scalar in Eq.~\eqref{eq:probe-definition} includes both chiralities and
the sum over relative insertion positions around the cycle. We determine how
long its strict extremal-BTZ approximation persists before the probe resolves
the underlying finite circle.

The right mover remains in a Ramond ground state, so on a length-$n$ cycle
\begin{align}
 C^R_n(\tau)&=\frac1{4n^2\sinh^2(\tau/(2n))}, \nonumber \\
 C^R_n(\epsilon+it)
 &=-\frac1{[2n\sin((t-i\epsilon)/(2n))]^2}.
\end{align}
The second insertion can lie at any of the $n$ relative copy positions.
The orbifold-invariant scalar contribution is therefore
\begin{equation}
 G_n(t)=\sum_{j=0}^{n-1}
 C^L_{n,\beta_*}\bigl(\epsilon+i(t+2\pi j)\bigr)
 C^R_n\bigl(\epsilon+i(t-2\pi j)\bigr).
 \label{eq:probe-physical-images}
\end{equation}
Both insertions contract within the same cycle
\cite{BelinBintanjaCastroKnop}.  Using
Eq.~\eqref{eq:probe-btz-and-first-finite-n}, the relative-position sum can
be unfolded to
\begin{equation}
\begin{split}
 G_n(t)={}&\sum_{j\in\mathbb Z}
 C^L_{\infty,\beta_*}\bigl(\epsilon+i(t+2\pi j)\bigr)
 C^R_n\bigl(\epsilon+i(t-2\pi j)\bigr)\\
 &-\frac{C^R_1(\epsilon+it)}{n\beta_*}.
\end{split}
 \label{eq:probe-physical-unfolding}
\end{equation}
At fixed time, $C^R_n(\tau)\to\tau^{-2}$ as $n\to\infty$.  Replacing the
right block by this limit defines the extremal-BTZ reference scalar
$G_{\rm eBTZ}^{\rm img}(t)$, and
\begin{equation}
 G_n(t)=G_{\rm eBTZ}^{\rm img}(t)
 -\frac{C^R_1(\epsilon+it)}{n\beta_*}+O(n^{-2}).
 \label{eq:probe-physical-correction}
\end{equation}
Thus the strict long-cycle limit reproduces the BTZ-like response of the
chosen probe, exactly the situation underlying the horizon mirage of
Ref.~\cite{BelinBintanjaCastroKnop}.  The finite circle first appears in an
explicit $1/n$ correction that is invisible at fixed time in the strict
$n\to\infty$ limit.

The time at which this correction becomes resolvable is different from the
logarithmic scale of the left chiral block.  For $t=2\pi m+u$ with fixed
phase $u$, the extremal-BTZ image sum has the late-time form
\begin{equation}
 G_{\rm eBTZ}^{\rm img}(t)
 =-\frac1{4t^2}\sum_{j\in\mathbb Z}
 C^L_{\infty,\beta_*}(\epsilon+i(u+2\pi j))+O(t^{-3}).
 \label{eq:probe-physical-fixed-phase-tail}
\end{equation}
At a regulated phase for which the leading image sum is nonzero, the two
competing contributions therefore scale as
\begin{align}
 |G_{\rm eBTZ}^{\rm img}(t)|&\sim\frac{c(u,\epsilon,\beta_*)}{t^2},
 \nonumber\\
 \left|\frac{C_1^R(\epsilon+it)}{n\beta_*}\right|
 &\sim\frac{c_R(u,\epsilon,\beta_*)}{n},
 \label{eq:probe-scalar-competing-scales}
\end{align}
with nonzero $O(1)$ coefficients at fixed phase.  The physical scalar
begins to resolve the finite-circle correction when these magnitudes become
comparable, namely when $t^{-2}\sim n^{-1}$.  Hence
\begin{equation}
 t_{\rm scalar}=O(\sqrt n).
 \label{eq:probe-scalar-resolution-scale}
\end{equation}
For $t\gtrsim t_{\rm scalar}$ the extremal-BTZ approximation
to this scalar is no longer parametrically accurate. The scale refers to the
chosen probe rather than to a universal breakdown of semiclassical
spacetime.

The finite circle becomes still more explicit through exact recurrence.
Both current blocks have period $2\pi n$ after Lorentzian continuation,
so
\begin{equation}
 C^{L,R}_n(t+2\pi n)=C^{L,R}_n(t),
 \qquad
 G_n(t+2\pi n)=G_n(t).
 \label{eq:probe-exact-recurrence}
\end{equation}
For a partition containing one cycle of winding $N-d$ and shorter cycles
with total winding $d$, write
\begin{equation}
 G_{\{C_r\}}(t)=\frac{N-d}{N}G_{N-d}(t)
 +\frac1N\sum_{r\ge1}rC_rG_r(t),
 \qquad \sum_r rC_r=d.
\end{equation}
If the regulated blocks satisfy $|G_r|\le B$, the recurrence of the longest
cycle at $T_g=2\pi(N-d)$ gives
\begin{equation}
 |G_{\{C_r\}}(t+T_g)-G_{\{C_r\}}(t)|
 \le \frac{2Bd}{N}.
 \label{eq:probe-long-cycle-recurrence-bound}
\end{equation}
Hence a partition with $d=O(1)$ has an $O(1/N)$-accurate recurrence at a
time of order $N$.  Section~\ref{sec:long-strings-lifting} shows that this
is precisely the cycle geometry selected in the open uncondensed region,
with the qualifications on exact BPS subspaces stated there.

In the open uncondensed region, the three probe scales are therefore
\begin{equation}
 \boxed{
 t_{\rm chiral}\sim\log N,
 \qquad
 t_{\rm scalar}\sim\sqrt N,
 \qquad
 T_{\rm circle}\sim N.}
 \label{eq:probe-mirage-time-hierarchy}
\end{equation}
The physical scalar resolves finite winding at
$O(\sqrt N)$, parametrically later than the chiral building block but before
the $O(N)$ recurrence of the covering circle. These polynomial scales are
distinct from the exponentially large many-body scale required to resolve
black-hole level spacings \cite{BarbonRabinovici}. Their separation refines
the mirage problem by identifying information discarded by the strict
large-$N$ thermal correlator \cite{BelinBintanjaCastroKnop}.

The preceding time scales answer when the probe resolves the finite circle.
At a known winding $n$, the left-current block also determines the entire
occupation profile of its chosen polarization.  To make this statement
precise, let $|\psi\rangle$ have fixed finite left conformal weight on that
cycle.  Bilinears with unequal mode numbers, or with two creation or two
annihilation operators, change this weight and have vanishing expectation
values.
Equation~\eqref{eq:probe-number-state-block} therefore extends to such a
state with $N_m^A$ replaced by $\langle N_m^A\rangle_\psi$, where $N_m^A$
inside the expectation value is the number operator.  Subtracting the
vacuum block $C^L_{n,\rm vac}(\tau)=[4n^2\sinh^2(\tau/(2n))]^{-1}$ gives
\begin{equation}
 C^L_{n,\psi}(\tau)-C^L_{n,\rm vac}(\tau)
 =\frac{2}{n^2}\sum_{m\ge1}m\langle N_m^A\rangle_\psi
   \cosh\!\left(\frac{m\tau}{n}\right).
 \label{eq:probe-occupation-profile}
\end{equation}
At fixed finite energy this sum has finite support.  The positive-frequency
Fourier coefficient over one period recovers each occupation expectation,
\begin{equation}
 \langle N_m^A\rangle_\psi
 =\frac{n e^{-m\epsilon/n}}{2\pi m}
 \int_0^{2\pi n}\!dt\,e^{-imt/n}
 \left[C^L_{n,\psi}(\epsilon+it)-C^L_{n,\rm vac}(\epsilon+it)\right].
 \label{eq:probe-occupation-inversion}
\end{equation}
Thus, at fixed $n$, two such left-current blocks agree exactly if and only
if their occupation expectations agree for every mode of the measured
polarization.  This is reconstruction from the full block, not from its
BTZ-like approximation or from a sum over unresolved cycle lengths.

The pair of Section~\ref{sec:individual-cycle-tests} is a concrete example
in which a difference of occupations also survives the physical scalar's
relative-position sum.  The states $F_6$ and $M_0$ have the same charges
$(N;N_p,J_L)=(6;3,-3)$ and the same single-cycle partition $p=(6)$.
Their fractional mode lattice $\Delta\omega=1/6$ and recurrence period
$12\pi$ therefore agree.  Any remaining difference comes from the state
on that circle rather than from its geometry.  No canonical average enters
this comparison.

Choose the left current polarization $A=\alpha^2$ or $A=\bar\alpha^2$ in
the complex internal basis of Eq.~\eqref{eq:F6-maximal-twist}.  The explicit
states of Section~\ref{sec:individual-cycle-tests} have
\begin{equation}
 N_1^A(F_6)=8,
 \qquad
 N_1^A(M_0)=0,
 \label{eq:matched-pair-occupation-difference}
\end{equation}
because $F_6$ contains eight bosonic quanta of that polarization at covering
level $m=1$, whereas $M_0$ is built from fermionic current descendants and
contains none.  Since both states have $n=6$, the exact number-state block
then gives
\begin{align}
 C^L_{F_6}(\tau)-C^L_{M_0}(\tau)
 &=\frac{8}{6^2}\left(e^{-\tau/6}+e^{\tau/6}\right)
 \nonumber\\
 &=\frac49\cosh\frac{\tau}{6}.
 \label{eq:matched-pair-chiral-difference}
\end{align}
The two states have the same right Ramond ground.  Summing over the six
relative copy positions produces the physical scalar difference
\begin{align}
 G_{F_6}^{\alpha^2 B}(t)-G_{M_0}^{\alpha^2 B}(t)
 &=\sum_{j=0}^{5}\frac49
 \cosh\!\left(\frac{\tau+2\pi ij}{6}\right)
 C_6^R(\tau-2\pi ij) \nonumber \\
 &=\boxed{\frac{\cosh(2\tau/3)+5\cosh(\tau/3)}
 {54\sinh^2(\tau/2)}}.
 \label{eq:matched-pair-full-difference}
\end{align}
Thus the $F_6/M_0$ difference measures a change in the $m=1$ occupation,
not a change in the available frequencies.  It exhibits the occupation
resolution of Eq.~\eqref{eq:probe-occupation-inversion} within two states
whose geometric data have been matched.

The same formula also fixes what these blocks do not measure.  Number
operators are diagonal in the oscillator number basis, so the block
retains occupation expectations but not relative phases between number
configurations.  The existing harmonic states give an explicit example.
Since $M_0$ has no neutral bosonic excitations,
$N_m^A|M_0\rangle=0$ and $\langle F_6|N_m^A|M_0\rangle=0$ for every $m,A$.
Consequently,
\begin{align}
 |\Psi_\pm\rangle
 &=\frac{|F_6\rangle\pm|M_0\rangle}{\sqrt2},
 \nonumber\\
 \langle N_m^A\rangle_{\Psi_+}
 &=\langle N_m^A\rangle_{\Psi_-}
 =\frac12\langle N_m^A\rangle_{F_6},
 \nonumber\\
 C^L_{6,\Psi_+}(\tau)
 &=C^L_{6,\Psi_-}(\tau)
 =\frac12\left[C^L_{F_6}(\tau)+C^L_{M_0}(\tau)\right].
 \label{eq:probe-phase-degeneracy}
\end{align}
Both combinations are orthogonal harmonic representatives for the $Q_6$
complex of Section~\ref{sec:individual-cycle-tests}, and
$[\Psi_+]-[\Psi_-]=\sqrt2[M_0]\ne0$ in that cohomology.  They nevertheless
give identical blocks for every diagonal current polarization and,
because they share the right Ramond ground, identical scalar correlators
of the form used above.  Collecting all these polarization blocks does
not recover their relative sign.  This is a degeneracy of this probe
family, not a statement about all observables at fixed $N$.  Nor is it a
pair with opposite continuation labels, since both combinations contain a
nonzero $F_6$ component.

}

\section{Discussion}
\label{sec:discussion-and-open-problems}

{The state-level and ensemble-level results answer different
questions, but they are linked by the same cycle observable: the first rules
out a deterministic relation between cycle geometry and fortuity, while the
second asks whether a typical cycle geometry nevertheless emerges after
counting all states at fixed charge.} The
matched states $F_6$ and $M_0$ have identical charges and cycle partition but
opposite continuation properties, while the fortuitous family
$F_{2k}^{\rm heavy}$ is built entirely from length-two cycles. These examples
exclude a state-by-state geometric criterion for fortuity. At fixed charge,
however, the $w^{-3}$ single-cycle factor places the free count in the
convergent weighted-partition regime
\cite{StuflerConvergent,StuflerUnlabelled}: in the open uncondensed region,
one cycle carries $N-O_{\Pr}(1)$ winding. This is stronger than the long-cycle
behavior associated with the $1/w$ weight of a uniformly random permutation
\cite{FordCycleType}.

The organization changes at short-string condensation. The aligned
winding-one occupation is $O(\sqrt N)$ on the critical line and extensive in
the condensed interior, where the excited long string remains on the
critical line and the condensate carries the excess winding and spin. This
reproduces the free-orbifold enigmatic phase of Ref.~\cite{BenaMoulting}.
The condensate fraction vanishes continuously at the transition; the entropy
and its first derivatives match there, while the second derivative jumps.
These occupation laws refer to the unsigned free count.
The modified-index seed instead contains the Jacobi factor that cancels the
double Bose pole, while the divisor formula retains the BMPV asymptotic
\cite{MaldacenaMooreStrominger,ChangLinZhang}. In the open uncondensed
region, the protected lower bound and unsigned upper bound have matching
exponential and polynomial scales. This transfers the long-string
concentration bound to a regularly transported exact BPS subspace and to an
exact non-graviton subspace with
$\log\dim(S^0\cap\mathcal G_N^\perp)=S_{\rm BH}-4\log S_{\rm BH}+O(1)$
\cite{HughesShigemori,ToweringGravitons}. At a fixed nonzero rotation ratio,
the logarithmic coefficient agrees with the BMPV one-loop result
\cite{SenLogBMPV}. At criticality and in the condensed interior, the index
still gives a black-hole-scale non-graviton population but does not determine
its full dimension or cycle statistics.

Identifying the entropy-carrying fortuitous sector requires additional
continuation input \cite{ChangLinFortuity,ChangLinZhang,ChangZhang}. The
explicit protected tower avoids every fixed-depth region without an
assumption on the full monotone space. A subexponential monotone dimension
makes the harmonic fortuitous complement carry the black-hole entropy and
long-string concentration, whereas separation of the full monotone and
fortuitous cycle distributions also requires the trace bound on monotone
representatives. Non-graviton and fortuitous are not the same class
\cite{GiustoInglisRusso}, and no protected analogue of the free Rayleigh or
Gaussian occupation law has been established in the condensed regimes.

The interaction and probe calculations resolve complementary data. In the
asymmetric limit, the exact twist kernels retain the finite winding of the
short cycle \cite{CarsonHamptonMathurTurtonTwist,CarsonHamptonMathurTurton},
although these kernels are not the normalized lifting operator. The neutral
scalar propagates through non-BPS intermediate states whose right-moving gap
is $m/n$, producing a dense near-BPS tower on a long cycle. This is parallel
to the soft modes of near-BPS AdS$_3$ backgrounds
\cite{BacCastroJainNearBPS}. The chiral finite-circle correction is
proportional to $K/N$, so the free-ensemble cycle laws give BTZ-like windows
of order $\log N$, $\tfrac12\log N$, and $O(1)$ in the uncondensed, critical,
and condensed regions. Exact-BPS long-string concentration is used for this
probe only in the open uncondensed region.

For the physical scalar in that region, finite winding becomes visible at
$t=O(\sqrt N)$ and the circle recurs at $t=O(N)$, sharpening the comparison
with symmetric-orbifold universality \cite{BelinBintanjaCastroKnop}. Both
times are well below the many-body Heisenberg scale
\cite{BarbonRabinovici}. Equal
charges, cycle geometry, mode lattice, and recurrence period do not fix the
response: $F_6$ and $M_0$ have different oscillator spectral weights, while
the equal-response pair $\Psi_\pm$ shows that diagonal current blocks need
not determine a cohomology class. A BTZ-like leading correlator therefore
does not fix the finite circle, the oscillator data, or the continuation
class. These distinctions motivate the open problems below.

\paragraph{Microscopic organization and interactions.}
It remains to determine which cycle organization survives
exact BPS protection beyond the open uncondensed region. Refined protected counts
\cite{Hughes:2026qqn} and the finite-coupling chiral algebra
\cite{BelinBPSChiral} provide complementary information, but the target is
the protected distribution across the critical and condensed regimes and a
derivation of the full monotone bounds.  A second, equally microscopic
question concerns interactions between fortuitous and monotone states.
The composite constructions of Ref.~\cite{ChangLinZhang} motivate computing
mixed operator products, normalized three- and four-point functions, and
joining or splitting amplitudes with both types of external state.  These
data would test which monotone dressings preserve a fortuitous class and
whether continuation imposes selection rules on the couplings.  The
oscillator coefficients derived here supply part of that calculation;
physical deformation amplitudes also require the supercurrent dressing,
permutation factors and BPS projections of the lifting formalism
\cite{GavaNarain,ChangLinZhang}.  Comparing these amplitudes with the
continuation maps would connect the state classification to its dynamics,
without assuming that the classification itself defines noninteracting
sectors.

\paragraph{Mesoscopic collective theory.}
A mesoscopic theory of cycle and spectral organization
should be derived from the microscopic CFT. The cycle occupations $n_w$ obey
$\sum_w w n_w=N$, while the matched-state comparison also requires
oscillator occupation data. Equal-response superpositions further indicate
that closure of the dynamics may require coherences not measured by these
blocks. A collective
description should derive the inner product, the finite-$N$ relations and
the action of the joining and splitting operators, rather than postulate
an independent model for $d$ or $K$.  Finite-$N$ collective constructions
for $S_N$-invariant models \cite{CollectiveFiniteNReduction} and regulated
bilocal singlet theories \cite{BilocalFiniteNHilbert,BilocalFiniteNAlgebra}
provide useful examples of implementing null relations and encoding the
physical Hilbert space in representations of an invariant operator
algebra.  The corresponding D1--D5 construction must additionally account
for twisted sectors and the supercharge cohomology.  Establishing whether
the cycle and occupation operators close under the deformation, or require
additional collective observables, would turn the statistical organization
found here into a derived mesoscopic theory.

\paragraph{Macroscopic geometry and horizons.}
Connecting these results to macroscopic geometry requires
transporting states and probes together along the exactly marginal
deformation toward a regime with controlled semiclassical gravity.
Conformal perturbation theory and normalized lifting data
\cite{GavaNarain,GaberdielGopakumarNairz} provide a starting point for
tracking changes in spectral weights and in the near-BPS excitation gap.
A quantitative comparison with near-BPS AdS$_3$ fluctuations must match
charges and boundary conditions \cite{BacCastroJainNearBPS}, rather than
identify softness alone with a bulk mode.  The present $\sqrt N$ and $N$
scales then become quantities to follow with coupling, not universal
geometric time scales.  The central question is whether, and under what
limits, the finite-circle response develops the causal and dissipative
structure of a semiclassical horizon.  Answering it requires more than a
BTZ-like leading two-point function \cite{BelinBintanjaCastroKnop}, including
phase-sensitive correlators and comparison with operator-algebraic
criteria for emergent time and causal structure
\cite{LeutheusserLiuEmergent,LeutheusserLiuCausal,GesteauLiuStringy}.
The aim is a geometric description derived from microscopic dynamics,
with its regime of validity fixed by the observables.
\section*{Acknowledgment}
The author used ChatGPT, including the GPT-5.6 Sol and GPT-6 Astro models, for parts of the calculations, literature exploration, and manuscript preparation. The research direction, selection of key literature and ideas, and refinement of the broader conceptual picture remained with the author, while the models were used to explore a large space of connections and intermediate possibilities across different results and subfields. This work also served as a personal experiment in exploring what an appropriate future role for AI in theoretical-physics research might be. By making such broad exploration more efficient, AI can help develop several related lines of inquiry together and organize them into a single large-scale scientific narrative---a form the author refers to as a \textit{cinematic} paper. In this mode, human researchers provide the conceptual direction, scientific judgement, and mathematical validation, while AI expands the range of connections that can be explored. All claims and results were examined and verified by the author, who takes full responsibility for the work.

\appendix

\section{Finite-rank calculations}
\label{app:finite-rank-calculations}

{This appendix supplies the finite-dimensional calculations
used in the main text.  Section~\ref{app:F3-source-census} enumerates the
complete rank-four source fibre of $F_3$ and reduces its closure equations to
a nonsingular $2\times2$ system. Section~\ref{app:F6-protected-projectors}
computes the nonzero $F_6\to\Xi_7$ residue and proves the descendant bounds,
while Section~\ref{app:F2-heavy-family} derives the local-to-global identity
for the length-two family.}

On a unit ground state annihilated by the conjugate modes, the algebra
in Eq.~\eqref{eq:review-oscillator-normalization} gives
\begin{align}
 \langle0|\alpha_r^n(\alpha_r^\dagger)^n|0\rangle
 &=nr\,\langle0|\alpha_r^{n-1}(\alpha_r^\dagger)^{n-1}|0\rangle
 =n!r^n, \nonumber \\
 \left\|\prod_{A,r}(\alpha_r^{A\dagger})^{n_{A,r}}
       \prod_{\alpha,s}(\psi_s^{\alpha\dagger})^{\epsilon_{\alpha,s}}
       |0\rangle\right\|^2
 &=\prod_{A,r}n_{A,r}!\,r^{n_{A,r}},\qquad \epsilon_{\alpha,s}\in\{0,1\}. \label{eq:appendix-fock-word-norm}
\end{align}
Zero modes act in the Ramond Clifford module. For orthogonal images
of a unit decorated word with stabilizer $H\subset S_N$,
\begin{equation}
\begin{aligned}
 \left\|\sum_{\sigma\in S_N/H}\sigma f\right\|^2
 &=\frac{N!}{|H|}, \\
 \left\|\sum_{\sigma\in S_N}\sigma f\right\|^2&=N!\,|H|.
\end{aligned}
 \label{eq:appendix-orbit-sum-norms}
\end{equation}
The signed stabilizer multiplicity gives the normalized state in
Eq.~\eqref{eq:setup-orbit-unit-state}.

The normalized states $F_2$, $F_3$, and $F_6$ are defined in
Section~\ref{sec:microscopic-fortuity}; the target $\Xi_7$ and its
normalization are given in Appendix~\ref{app:F6-protected-projectors}.

\subsection{Three-cycle continuation}
\label{app:F3-source-census}
\label{app:F3-dual-witness}
\label{app:F3-all-rank}

{The obstruction has three inputs: the charge census fixes
all possible rank-four precursors, the covering-space residues give their
six target overlaps, and a row combination annihilates every correction
column while remaining nonzero on the prescribed source.}

The seed normalization and NS charges are
\begin{align}
 \|\psi^-_{-1/2}\psi^-_{-1/6}
 (\alpha^1_{-1/3})^7|3_{--,--}\rangle\|^2
 &=7!\,3^{-7}=\frac{560}{243}, \nonumber \\
 \Delta h&=\frac12+\frac16+\frac73=3, \nonumber \\
 (h,j;\bar h,\bar j)&=(4,0;1,1).
\end{align}

Here $|\sigma_3\rangle_L$ denotes the bare length-three left NS twist
vacuum and $c_{\rm seed}=6$ is the central charge of one $T^4$ seed copy.
The bare-vacuum coordinates include the filled-sea hole,

\begin{align}
 |3_{--}\rangle_L
   &=\bar\psi^+_{-1/6}\psi^+_{-1/6}|\sigma_3\rangle_L, \nonumber \\
 h(\sigma_3)&=\frac{c_{\rm seed}}{24} \left.\left(w-\frac{1}{w}\right)\right\vert_{w=3}=\frac23, \nonumber \\
 \psi^-_{+1/6}|3_{--}\rangle_L
   &\propto\psi^+_{-1/6}|\sigma_3\rangle_L\ne0,
\end{align}
where $h(\sigma_3)=2/3$ and the two occupied fermions raise the left
chiral-ground weight to one. Use Cartan weights $R=2j$, $A=2J_a^3$, $B=2J_b^3$. In the character below,
$q$ counts left conformal weight while $x,y,z$ are fugacities for $R,A,B$,
respectively. The filled-sea hole is included by the following unprojected
character:
\begin{align}
 Z_w(q,x,y,z)&=q^{h_w}g_w(x,y)
 \prod_{n\ge1}\prod_{a,b=\pm1}(1-q^{n/w}y^az^b)^{-1}
 \prod_{r\in\mathcal R_w^+}\prod_{\epsilon,a=\pm1}
 (1+q^rx^\epsilon y^a), \nonumber \\
 (h_w,g_w,\mathcal R_w^+)
 &=\begin{cases}
 ((w^2-1)/(4w),1,(\mathbb Z_{\ge0}+1/2)/w),&w\text{ odd},\\
 (w/4,x^{-1}(1+xy)(1+xy^{-1}),\mathbb Z_{>0}/w),&w\text{ even}.
 \end{cases}
\end{align}

Let $n_{\rm even}(p)$ denote the number of even-length cycles in the
partition $p$. At $(h,R,A,B)=(4,0,9,7)$, parity and minimum energies give

\begin{equation}
 R+A+B\equiv n_{\rm even}(p)\pmod2,
\end{equation}

\begin{align}
 p=(4),(2,1,1)&:\quad0\not\equiv1\pmod2, \nonumber \\
 p=(2,2)&:\quad h\ge1+\frac72>4, \nonumber \\
 p=(1^4)&:\quad h\ge7>4. \label{eq:appendix-F3-partition-exclusion}
\end{align}
For the rank-three harmonicity test, one twist-two action
on a single three-cycle can only split it into the cycle shape $(2,1)$.
For the chosen supercharge component, the $\mathbf{10}$ factor is unchanged,
while the relevant internal $SU(2)$ doublet gives
$\mathbf 8\otimes\mathbf 2=\mathbf 9\oplus\mathbf 7$.  Hence the fixed
internal-charge block contains precisely the $({\bf10},{\bf9})$ and
$({\bf10},{\bf7})$ adjacent targets.  The following estimate tests whether
either allowed target fits inside the available conformal-energy budget.

{Here $\mathbf d$ denotes the $d$-dimensional irreducible
$SU(2)$ representation. Thus Cartan highest weights $A=9$ and $B=7$
correspond to the factors $\mathbf{10}$ and $\mathbf 8$, while the
supercharge doublet changes only the latter factor to $\mathbf9$ or
$\mathbf7$.}

The $(2,1)$ chiral ground contributes $1/2$, leaving
\(h_{\rm exc}^{\rm target}=4-\frac12=\frac72\).
The $({\bf10},{\bf9})$ target requires eight aligned bosons on the
two-cycle. The $({\bf10},{\bf7})$ target requires six bosons and three
aligned fermionic doublets; Pauli exclusion allows one zero mode,
with the other two fermions costing at least $1/2$ each.
Explicitly,
\begin{align}
 E_{({\bf10},{\bf9})}&\ge8\left(\frac12\right)=4, \nonumber \\
 E_{({\bf10},{\bf7})}&\ge6\left(\frac12\right)+0+\frac12+\frac12
 =4>\frac72.
\end{align}

{For the fixed $(h,R)$ and right-moving block under consideration,
$d_p(A,B)$ denotes the dimension of the Cartan-weight space in partition
$p$, after cyclic projection and graded symmetrization. We use $m_p(A,B)$
for the multiplicity of the internal $SU(2)_a\times SU(2)_b$ representation
with highest weights $(A,B)$. Each such representation contributes once
to each allowed weight, so highest-weight multiplicities are obtained by
taking one finite difference of step two in each Cartan variable. The
energy bounds above give}

\begin{align}
 d_{(2,1)}(9,6)&=d_{(2,1)}(9,8)=0, \nonumber \\
 (\operatorname{im}Q_3)_{h=4,R=0,A=9,B=7}&=0.
\end{align}
{For $B=6$, let $A_F$ denote the part of the Cartan charge
$A=2J_a^3$ carried by fermionic excitations. The missing charge requires
$A_F\ge3$, with at most one aligned zero mode.} The protected-descendant bound derived in
Appendix~\ref{app:F6-protected-projectors} gives
$n_B\le2(h-h_0)\le6$ for a length-three chiral seed at $h=4$, since
$h_0\ge1$. The seven-boson state $F_3$ is therefore orthogonal to
$\mathcal T_3$, as recorded in Table~\ref{tab:finite-rank-summary}.

{For the $(3,1)$ census, $n_B$ and $n_F$ count the aligned
bosonic and fermionic excitations, while $E_B$ and $E_F$ are their excitation
energies in units of $1/6$. In the ledger below, the last three columns count
words with the length-one strand unexcited, words with it excited, and their
sum.}

In $(3,1)$, use energy units $1/6$:
\begin{align}
 E_B+E_F&=6(4-2/3)=20, \nonumber \\
 n_B&\in\{7,9\}, \nonumber \\
 n_F&\equiv0\pmod2,
\end{align}

\begin{align}
 E_{B,3}&\in2\mathbb Z_{>0}, \nonumber \\
 E_{F,3}&\in2\mathbb Z_{\ge0}+1, \nonumber \\
 E_{B,1}&\ge6, \nonumber \\
 E_{F,1}&\ge3.
\end{align}

The occupation ledger is

\begin{equation}
 \begin{array}{ccc|rrr}
 n_B&n_F&E_F&\text{vacuum length-one strand}&\text{excited length-one strand}&\text{total}\\\hline
 9&0&0&2&0&2\\
 9&2&2&4&0&4\\
 7&2&2&2&1&3\\
 7&2&4&2&2&4\\
 7&2&6&3&3&6\\
 7&4&6&2&2&4\\\hline
 &&&15&8&23
 \end{array}.
\end{equation}

Its bosonic partitions and fermion placements are

\begin{align}
 (9,0,0)&:\quad 2\ \text{choices for the raised boson's $B$ sign}, \nonumber \\
 (9,2,2)&:\quad2\ \text{zero-$A$ pairs}+2\ \text{aligned-pair placements}, \nonumber \\
 (7,2,2)&:\quad(6),\ (4,4)\ \text{on the long cycle, or one bosonic excitation on the length-one strand}, \nonumber \\
 (7,2,4)&:\quad2\ \text{fermion signs}\times2\ \text{cycles}, \nonumber \\
 (7,2,6)&:\quad(1,5),(3,3),(5,1)\ \text{on the long cycle};\ 3\ \text{placements on the length-one strand}, \nonumber \\
 (7,4,6)&:\quad2\ \text{repeated fermion signs}\times2\ \text{cycles}.
\end{align}

The length-one strands satisfy $h-j\in\mathbb Z$; total $h-j=4$ fixes
the three-cycle projection. The neighboring coefficients give

\begin{equation}
 \begin{array}{c|rrrr}
 p&d(9,7)&d(11,7)&d(9,9)&d(11,9)\\\hline
 (3)&15&1&4&1\\
 (3,1)&23&1&4&1
 \end{array},
\end{equation}

\begin{align}
 m_p(9,7)&=d_p(9,7)-d_p(11,7)-d_p(9,9)+d_p(11,9), \nonumber \\
 m_{(3)}(9,7)&=11, \nonumber \\
 m_{(3,1)}(9,7)&=19.
\end{align}
\begin{table}[!htbp]
\centering\small
\setlength{\tabcolsep}{5pt}
\begin{tabular}{c|ccccc|cc}
$p$ & $(4)$ & $(3,1)$ & $(2,2)$ & $(2,1,1)$ & $(1^4)$
& $\rank\pi_{3,4}$ & $\dim\ker\pi_{3,4}$ \\ \hline
$\dim V_p$ & $0$ & $23$ & $0$ & $0$ & $0$ & $15$ & $8$
\end{tabular}
\caption{Complete rank-four source census at the Cartan weight of $F_3$.
The adjacent lower cochains are empty, so the affine fibre is one prescribed
projection plus the eight-dimensional exclusion kernel. The explicit words and neighboring-weight subtraction are given below.}
\label{tab:F3-source-census}
\end{table}
Vacuum removal has the source and kernel dimensions shown in the table;
Eq.~\eqref{eq:F3-main-explicit-fibre}
fixes the required projection. The dimensions $15$ and $23$ count Cartan
weight spaces, while $11$ and $19$ are highest-weight multiplicities.
The full census includes filled-sea holes and half-integer fermion modes on the length-one strand satisfying the cyclic condition.

The source words $x_0,\ldots,x_8$ defined in
Section~\ref{sec:F3-all-rank-fortuity} exhaust the affine fibre by the census
above; the residue calculation below verifies the matrix elements used in
Eq.~\eqref{eq:F3-residue-matrix}.

For $\widehat x_0=f_L^\dagger Z$, $\widehat x_4=f_S^\dagger Z$,
with $f_L^\dagger=\psi^{-,(3)}_{-1/2}$, $f_S^\dagger=\psi^{-,(1)}_{-1/2}$
and $f_LZ=f_SZ=0$, define
\begin{equation}
 \begin{aligned}
 D_{\rm in}^\dagger&=\frac{\sqrt3f_L^\dagger+f_S^\dagger}{2}, \\
 R^\dagger&=\frac{f_L^\dagger-\sqrt3f_S^\dagger}{2}, \\
 \{D_{\rm in},D_{\rm in}^\dagger\}&=1, \\
 \{D_{\rm in},R^\dagger\}&=0.
\end{aligned}
\end{equation}

For each outgoing partition $p=(w_1,\ldots,w_c)$, the projected Ward
identity of Eq.~\eqref{eq:left-affine-Q-exact-commutator} reads

\begin{equation}
 \begin{aligned}
 \mathcal Q_p&=P_pQ_4P_{(3,1)}, \\
 D^{(p)\dagger}&=\sum_{r=1}^c\sqrt{w_r/4}\,f_r^\dagger, \\
 D^{(p)}\mathcal Q_p&=-\mathcal Q_pD_{\rm in}, \\
 D^{(p)\dagger}\mathcal Q_p&=-\mathcal Q_pD_{\rm in}^\dagger.
\end{aligned}
\end{equation}
Put $u_1=\mathcal Q_pD_{\rm in}^\dagger Z$,
$u_0=\mathcal Q_pR^\dagger Z$ and
$\widehat n_D^{(p)}=D^{(p)\dagger}D^{(p)}$. Then
\begin{align}
 \widehat n_D^{(p)}u_1&=u_1, \nonumber \\
 \widehat n_D^{(p)}u_0&=0,
\end{align}

\begin{align}
 \mathcal Q_p\widehat x_0&=\frac{\sqrt3u_1+u_0}{2}, \nonumber \\
 \mathcal Q_p\widehat x_4&=\frac{u_1-\sqrt3u_0}{2}, \nonumber \\
 \mathcal Q_p\widehat x_4
 &=\frac{4\widehat n_D^{(p)}-3}{\sqrt3}\mathcal Q_p\widehat x_0.
\end{align}
The same bounds, with bare energy $1/2$, give
\begin{equation}
 d_{(2,1,1)}(9,6)=d_{(2,1,1)}(9,8)=0.
\end{equation}

For the $3+1\to4$ join, take

\begin{align}
 \Gamma(t)&=t^3(4-3t), \nonumber \\
 (t_3,t_1,t_{\rm def},t_{\rm out})&=(0,4/3,1,\infty),
\end{align}
and bosonize
\begin{align}
 \psi^+&=e^{i\phi_1}, \nonumber \\
 \bar\psi^-&=-e^{-i\phi_1}, \nonumber \\
 \bar\psi^+&=e^{i\phi_2}, \nonumber \\
 \psi^-&=e^{-i\phi_2}.
\end{align}

For the $2J_b^3=-1$ supercharge component \cite{ChangZhang},

\begin{align}
 (v_1,v_2)&=(\tfrac12,-\tfrac12), \nonumber \\
 q_{\rm out}&=(-\tfrac12,-\tfrac12), \nonumber \\
 q_\infty&=-q_{\rm out},\qquad
 \partial X_{\rm def}\leftrightarrow\alpha^1.
\end{align}
The other descendant term has zero bosonic contraction. Strip the
common nonzero right correlator and permutation factors.

For charge $\epsilon=\pm1$ and signed mode $r$, define
\begin{align}
 g^{(a)}_{r,\epsilon}(t)
 &=\sqrt{\Gamma'(t)}\,\Gamma(t)^{r-1/2}
                  (t-1)^{\epsilon v_a}, \nonumber \\
 K_{ij}^{(a)}
 &=\oint_{\mathcal C_i}\frac{dt}{2\pi i}
   \oint_{\mathcal C_j}\frac{du}{2\pi i}
   \frac{g^{(a)}_{r_i,-}(t)g^{(a)}_{r_j,+}(u)}{t-u}, \nonumber \\
 u_i^{(1)}&=\oint_{\mathcal C_i}\frac{dt}{2\pi i}
                           g^{(1)}_{r_i,-}(t). \label{eq:fermion-residues}
\end{align}

Here $r<0$ on incoming legs and $r\ge0$ on bra legs. Order
negative insertions as rows, positive insertions as columns, with
incoming modes first. The $\phi_1$ imbalance gives one unpaired
column $u^{(1)}$; the $\phi_2$ numbers are equal.
For incoming energy $e_j$ and outgoing energy $e_i'$,

\begin{align}
 H_{ij}&=\oint_\infty\frac{dt}{2\pi i}
            \oint_{\mathcal C_j}\frac{du}{2\pi i}
       \frac{\Gamma(t)^{e_i'}\Gamma(u)^{-e_j}}{(t-u)^2}, \nonumber \\
 h_j&=\oint_{\mathcal C_j}\frac{du}{2\pi i}
                 \frac{\Gamma(u)^{-e_j}}{(u-1)^2}, \nonumber \\
 M_{ij}^{\rm cover}
 &\propto\operatorname{per}\begin{pmatrix}h\\H\end{pmatrix}
       \det(K^{(1)}\,|\,u^{(1)})\det K^{(2)}.
\end{align}
The seven-column permanent pairs one incoming boson with the
descendant and six with the target. Retain incoming--incoming
fermion contractions. After stripping individual nonzero mode constants, set
\begin{align}
 \mathcal B_n(\alpha,\beta;A,B)
 &=\sum_{j=0}^{n}\binom\alpha j\binom\beta{n-j}A^jB^{n-j},
 \quad n\ge0; \nonumber \\
 \mathcal B_n&=0,\qquad (n<0).
\end{align}

For $g=z^L(1+Az)^\alpha(1+Bz)^\beta$ at finite $c$, with $z=t-c$,
and $g=t^L(1+A/t)^\alpha(1+B/t)^\beta$ at infinity,

\begin{equation}
 [g]_p=
 \begin{cases}
 \mathcal B_{p-L}(\alpha,\beta;A,B),&c\text{ finite},\\
 \mathcal B_{L-p}(\alpha,\beta;A,B),&c=\infty.
 \end{cases}
\end{equation}

For $\gamma=4/3$ and $\nu=\epsilon v_a$, the fermion parameters are

\begin{equation}
 \begin{array}{c|rrrrr}
 c&L&\alpha&\beta&A&B\\\hline
 0&-3e-\tfrac12&-e-\tfrac12&\tfrac12+\nu&-\tfrac34&-1\\
 \gamma&-e-\tfrac12&-3e-\tfrac12&\tfrac12+\nu&\tfrac34&3\\
 \infty&4e-\tfrac12+\nu&e-\tfrac12&\tfrac12+\nu&-\tfrac43&-1
 \end{array}.
\end{equation}

Here $e>0$ is the creation energy ($e\ge0$ at infinity), and
$\epsilon$ is the inserted charge, reversed on bra legs. Fractional
moding makes $L\in\mathbb Z$. For example,

\begin{equation}
 g(t)\propto t^{-3e-1/2}
 (1-3t/4)^{-e-1/2}(1-t)^{1/2+\epsilon v_a},
\end{equation}

by substitution in Eq.~\eqref{eq:fermion-residues}. The bosonic
factors $\Gamma^{-e}$ and $\Gamma^e$ have parameters

\begin{equation}
 \begin{array}{c|rrrrr}
 c&L&\alpha&\beta&A&B\\\hline
 0&-3e&-e&0&-\tfrac34&0\\
 \gamma&-e&-3e&0&\tfrac34&0\\
 \infty&4e&e&0&-\tfrac43&0
 \end{array}.
\end{equation}

For a local Laurent series $g$, we write $[g]_m$ for the coefficient of
power $m$ in the relevant local coordinate; $L_g$ and $L_h$ below denote the
limiting Laurent powers that make the sums finite. Expanding the Cauchy
kernel at the outer contour gives, for $c=0,\gamma$,
\begin{equation}
 \mathcal P(g_\infty,h_c)
 =\sum_{k\ge0}\sum_{j=0}^{k}
   \binom kj c^{k-j}[g_\infty]_k[h_c]_{-j-1}.
\end{equation}

For distinct incoming punctures,

\begin{equation}
 \mathcal P(g_0,h_\gamma)
 =-\sum_{i,j\ge0}(-1)^j
   \binom{i+j}{i}\gamma^{-i-j-1}[g_0]_{-i-1}[h_\gamma]_{-j-1}.
\end{equation}

At one puncture, with the first contour outermost,

\begin{equation}
 \mathcal P(g_c,h_c)=\sum_{k\ge0}[g_c]_k[h_c]_{-k-1}.
\end{equation}

Reversing insertion order changes the sign. The upper limits are

\begin{equation}
 \begin{array}{c|c}
 \text{contraction}&\text{nonzero summation range}\\\hline
 \mathcal P(g_\infty,h_c)&k\le L_g\\
 \mathcal P(g_0,h_\gamma)&i\le-L_g-1,\quad j\le-L_h-1\\
 \mathcal P(g_c,h_c),\ c\text{ finite}&k\le-L_h-1\\
 \mathcal P(g_\infty,h_\infty)&k\le L_g
 \end{array};\qquad \text{empty sums}=0.
\end{equation}

The double pole and the derivative at $t=1$ give
\begin{equation}
 \begin{aligned}
 \mathcal H(g_\infty,h_c)
 &=\sum_{k\ge0}\sum_{j=0}^{k}(k+1)\binom kj
    c^{k-j}[g_\infty]_{k+1}[h_c]_{-j-1}, \\
 \mathcal D(h_c)
 &=\sum_{j\ge0}(-1)^j(j+1)(c-1)^{-j-2}[h_c]_{-j-1}.
\end{aligned}
\end{equation}

The bounds are $k\le L_g-1$ and $j\le-L_h-1$. In the permanent below,
$b$ denotes the local factor of the descendant-contracted incoming boson,
$o_i$ ($i=1,\ldots,6$) the six target-contracted bosonic factors, and $b_*$
the modified local factor when the distinguished incoming boson is used.
For seven identical incoming bosons with factor $b$,

\begin{equation}
 \operatorname{per}(h;H)=7!\,\mathcal D(b)
                    \prod_{i=1}^{6}\mathcal H(o_i,b).
\end{equation}

For six copies of $b$ and one $b_*$, set $v_0=\mathcal D(b)$,
$v_i=\mathcal H(o_i,b)$ and $v_i^*=v_i|_{b\mapsto b_*}$:

\begin{equation}
 \operatorname{per}(h;H)
 =6!\sum_{i=0}^{6}v_i^*\prod_{\substack{j=0\\j\ne i}}^{6}v_j.
\end{equation}
The $i=0$ term contracts $b_*$ with the descendant.

The six targets $T_1,\ldots,T_6$ are defined in
Section~\ref{sec:F3-all-rank-fortuity}. Their ground is the bare four-cycle NS vacuum
$|\sigma_4;--\rangle_L$ with $(h,j)=(1,-1/2)$ and covering Ramond moding.

We now implement the finite-rank obstruction
\eqref{eq:finite-rank-target-matrix}. Order the rows as
$T_1,\ldots,T_6$ and the columns as $x_0,\ldots,x_8$, following the
definitions in Section~\ref{sec:F3-all-rank-fortuity}. Column zero
is the prescribed source and columns one through eight are the complete
projection kernel. The physical overlap matrix $M$ is related to the working coordinates by
invertible diagonal row and column rescalings:
\begin{align}
 P&=D_{\rm row}M D_{\rm col}, \nonumber \\
 \ker P&=D_{\rm col}^{-1}\ker M, \nonumber \\
 y&=D_{\rm col}c,\qquad y_0\ne0\Longleftrightarrow c_0\ne0.
\end{align}

They preserve solvability and the condition $c_0\ne0$; the entries are
overlap data, not lifting eigenvalues. For the stripped residue matrix
$B$, the explicit row factors are

\begin{equation}
 P=\operatorname{diag}\!\left(
 \frac3{140},-\frac9{20},\frac9{80},-\frac9{40},\frac3{80},-\frac9{20}
 \right)B.
\end{equation}

The finite sums reproduce the matrix printed in
Eq.~\eqref{eq:F3-residue-matrix}.  The first column is the image of $x_0$
and the last eight columns contain all allowed cancellation images.
The following entries illustrate the residue evaluation.

For $(T_1,x_0)$, put $b=t^{-1}(1-3t/4)^{-1/3}$ and
$o_n=t^n(1-4/(3t))^{n/4}$. Then
\begin{align}
 \mathcal D(b)&=\mathcal H(o_1,b)=1, \nonumber \\
 \mathcal H(o_3,b)&=-\frac16, \nonumber \\
 \operatorname{per}(h;H)&=7!\left(-\frac16\right)=-840.
\end{align}

The two fermionic matrices in the prescribed ordering are

\begin{align}
 &(K^{(1)}\,|\,u^{(1)})=\begin{pmatrix}0&1\\1&1/3\end{pmatrix}, \nonumber \\
 K^{(2)}&=\begin{pmatrix}-1/2&-1&0\\-3/16&1/4&-1\\1&1/3&0\end{pmatrix}.
\end{align}

\begin{equation}
 \begin{aligned}
 \det(K^{(1)}\,|\,u^{(1)})&=-1, \\
 \det K^{(2)}&=\frac56, \\
 B_{1,0}&=700, \\
 P_{1,0}&=\frac3{140}(700)=15,
\end{aligned}
\end{equation}

\begin{equation}
 \begin{aligned}
 \bigl[\det(K^{(1)}\,|\,u^{(1)})\det K^{(2)}\bigr]_{1,2}
 &=-\frac49, \\
 P_{1,2}&=8,
\end{aligned}
\end{equation}

\begin{equation}
 \begin{aligned}
 \mathcal D(b_*)&=\frac52, \\
 \mathcal H(o_1,b_*)&=\frac12, \\
 \mathcal H(o_3,b_*)&=-\frac{25}{12}, \\
 \operatorname{per}(h;H)_{1,8}&=-2100, \\
 P_{1,8}&=20.
\end{aligned}
\end{equation}

The row reduction can now be checked directly. Numbering rows from one
to six gives
\begin{align}
 \frac{(Pc)_2+5(Pc)_3-5(Pc)_5+(Pc)_6}{420}
   &=c_0-4c_4-\frac47c_8, \nonumber \\
 \frac{14(Pc)_1+2(Pc)_4}{420}
   &=3c_0+4c_4+\frac47c_8.
\end{align}

All correction coefficients except $c_4,c_8$ cancel. With
$\rho=c_4+c_8/7$, these conditions become

\begin{equation}
 \boxed{\begin{pmatrix}1&-4\\3&4\end{pmatrix}
     \binom{c_0}{\rho}=0,\qquad \det\begin{pmatrix}1&-4\\3&4\end{pmatrix}
     =16\ne0,\qquad c_0\ne0.} \label{eq:main-two-sector-obstruction}
\end{equation}
The determinant forces $c_0=\rho=0$.
{Equivalently, the covector
$\ell=(14,1,5,2,-5,1)$ is the corresponding unscaled combination of the
six target rows. Direct multiplication gives
$\ell P=(1680,0,0,0,0,0,0,0,0)$: all eight correction columns are
annihilated, whereas the prescribed-source column is not.} The full matrix
has rank six and its correction submatrix has rank five. \begin{figure}[!htbp]
\centering
\begin{tikzpicture}[
 >=Latex,
 node distance=7mm and 13mm,
 every node/.style={font=\small,align=center,inner sep=1.2pt},
 lab/.style={font=\scriptsize,fill=white,inner sep=1pt}
]
  \node (src) {$V_{(3,1)}$\\[-1mm]$\dim=23$};
  \node[right=20mm of src] (aff)
    {$x_0+\operatorname{span}\{x_1,\ldots,x_8\}$\\[-1mm]
     $\rank\pi_{3,4}=15,\quad \dim\ker\pi_{3,4}=8$};

  \node[below=10mm of aff] (q)
    {$Q_4\!\left(x_0+\operatorname{span}\{x_1,\ldots,x_8\}\right)$};
  \node[below left=10mm and 23mm of q] (split)
    {$p=(2,1,1)$\\[-1mm]$0$};
  \node[below right=10mm and 23mm of q] (single)
    {$p=(4)$};

  \node[below left=8mm and 10mm of single] (n0)
    {$\widehat n_D=0$};
  \node[below right=8mm and 10mm of single] (n1)
    {$\widehat n_D=1$};

  \node[below=11mm of single] (targets)
    {$T_1,\ldots,T_6$};
  \node[below=8mm of targets] (reduce)
    {$\ell_0,\ell_1,\qquad \rho=c_4+\dfrac{c_8}{7}$};
  \node[below=8mm of reduce] (cert)
    {$c_0-4\rho=0,\qquad 3c_0+4\rho=0,
      \qquad \det W=16$};
  \node[below=8mm of cert] (nolift)
    {$c_0=0\quad\text{but}\quad c_0\neq0\ \text{from projection}$\\[-1mm]
     $\Longrightarrow\quad \text{no rank-four lift}$};

  \draw[->] (src) -- node[lab,above] {$\pi_{3,4}$} (aff);
  \draw[->] (aff) -- node[lab,right] {$Q_4$} (q);
  \draw[->,densely dashed] (q) -- (split);
  \draw[->] (q) -- (single);
  \draw[->] (single) -- (n0);
  \draw[->] (single) -- (n1);
  \draw[->] (n0) |- (targets);
  \draw[->] (n1) |- (targets);
  \draw[->] (targets) -- node[lab,right] {row reduction} (reduce);
  \draw[->] (reduce) -- (cert);
  \draw[->] (cert) -- (nolift);
\end{tikzpicture}
\caption{Complete rank-four lift test for $F_3$. Projection leaves eight
correction directions. The $(2,1,1)$ branch vanishes, while the $(4)$ branch
splits into $\widehat n_D=0,1$. Six target covectors reduce the affine problem
to Eq.~\eqref{eq:main-two-sector-obstruction}.}
\label{fig:F3-lift-obstruction}
\end{figure}
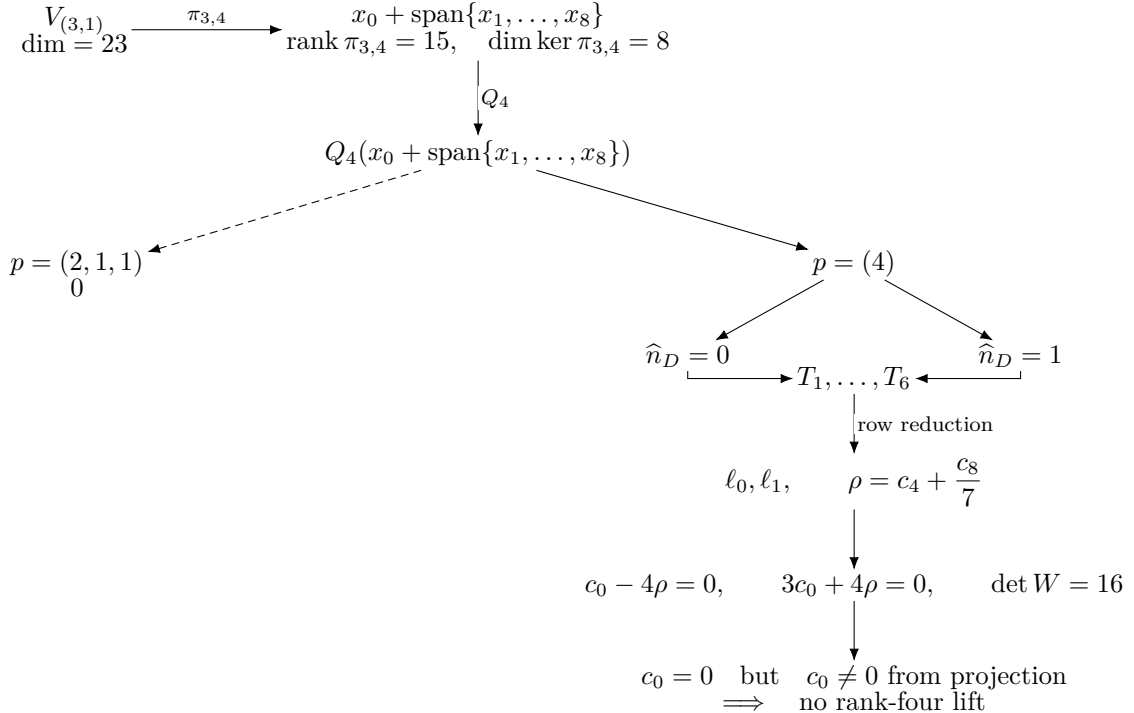

The ranks and row combinations follow directly from the printed matrix.

The all-rank extension is given in
Section~\ref{sec:F3-all-rank-fortuity}; the only additional ingredients are
vacuum-spectator orthogonality and the standard $\sqrt{M-3}$ occupation
factor.

\subsection{Six-cycle residues and descendants}
\label{app:F6-protected-projectors}

{The harmonicity and source-uniqueness arguments for $F_6$,
together with the rank-seven source and target conventions, are given in
Section~\ref{sec:individual-cycle-tests}. Here we record the normalization
and charge checks, evaluate the nonzero $F_6\to\Xi_7$ matrix element term by
term, and then use superconformal Ward identities to construct the monotone
descendant tower and prove that $F_6$ is orthogonal to it.}

\begin{align}
 \|F_6^{\rm raw}\|^2&=(8!)^2\,6^{-16}, \nonumber \\
 \|\Xi_7^{\rm raw}\|^2&=7!8!\,7^{-15},
\end{align}

\begin{align}
 N_p(F_6)&=\frac{16+1+1}{6}=3, \nonumber \\
 N_p(\Xi_7)&=\frac{15+4+1+1}{7}=3.
\end{align}
For the $F_6$ Ramond join the covering map is
\begin{equation}
 \Gamma(t)=t^6(7-6t),\qquad a=\frac76,
 \label{eq:F6-app-cover}
\end{equation}
with the two incoming Ramond fields at $t=0,a$, the supercharge
insertion at $t=1$, and the conjugate outgoing Ramond field at infinity.
The corresponding bosonized charges are
\begin{equation}
 \begin{array}{c|cccc}
 t&0&a&1&\infty\\\hline
 (q_1,q_2)&(-\frac12,-\frac12)&(-\frac12,-\frac12)
 &(\frac12,-\frac12)&(\frac12,\frac12).
 \end{array}
 \label{eq:F6-app-spin-charges}
\end{equation}

We use the covering-space prescription directly.  A fractional
weight-$1/2$ fermion mode lifts according to
\begin{equation}
 \oint_{z_0}\frac{dz}{2\pi i}\,
 \psi(z)(z-z_0)^{r-1/2}
 \longrightarrow
 \oint_{t_0}\frac{dt}{2\pi i}\,
 \psi(t)\sqrt{\Gamma'(t)}
 \bigl(\Gamma(t)-z_0\bigr)^{r-1/2},
 \label{eq:F6-app-standard-lift}
\end{equation}
up to the nonzero normalization of the twisted state
\cite{LuninMathurOrbifold,LuninMathurThreePoint,GaberdielGopakumarNairz}.
The fermion is then contracted in the background of the four spin
fields in Eq.~\eqref{eq:F6-app-spin-charges}.  We use
\begin{equation}
 \psi^+=e^{i\phi_1},\qquad
 \bar\psi^-=-e^{-i\phi_1},\qquad
 \bar\psi^+=e^{i\phi_2},\qquad
 \psi^-=e^{-i\phi_2}.
 \label{eq:F6-app-bosonization}
\end{equation}

The incoming state has two strands but only the length-six strand carries
fermion oscillators.  The added length-one strand is the $R_{--}$ Ramond
vacuum at $t=7/6$ and therefore contributes a spin field but no
fermion-mode contour.  Moreover, the fermionic correlator factorizes into
its $\phi_1$ and $\phi_2$ species.  The $\phi_1$ sector has only the
incoming $\bar\psi^-_{-1/6}$ mode and its single conjugate outgoing
partner, so there is only one Wick contraction.  The nontrivial sector is
$\phi_2$: it contains the incoming six-cycle mode $\psi^-_{-1/6}$, the
local $\psi^-$ supplied by $Q_7$ at $t=1$, and the two outgoing conjugate
$\bar\psi^+$ modes.

The incoming six-cycle mode becomes
\begin{equation}
 \oint_0\frac{du}{2\pi i}\,
 \psi^-(u)\,
 u^{-1}\left(1-\frac{6u}{7}\right)^{-1/6},
 \label{eq:F6-app-incoming-mode}
\end{equation}
and the two outgoing ket modes $\psi^-_{-4/7}$ and $\psi^-_{-1/7}$ appear
in the bra as
\begin{align}
 &\oint_\infty\frac{dt_1}{2\pi i}\,
 \bar\psi^+(t_1)(t_1-1)t_1^2
 \left(1-\frac{7}{6t_1}\right)^{-3/7},
 \label{eq:F6-app-outgoing-47}\\
 &\oint_\infty\frac{dt_2}{2\pi i}\,
 \bar\psi^+(t_2)(t_2-1)t_2^{-1}
 \left(1-\frac{7}{6t_2}\right)^{-6/7}.
 \label{eq:F6-app-outgoing-17}
\end{align}
These powers follow by inserting Eq.~\eqref{eq:F6-app-cover} into
Eq.~\eqref{eq:F6-app-standard-lift} and multiplying by the powers produced
by the bosonized spin-field correlator.  Common nonzero constants have
been stripped, but all coordinate dependence relevant to the residues is
retained.

{The $\phi_2$ contour integral is the one displayed in
Eq.~\eqref{eq:F6-main-integrated-phi2-correlator}.} Using
\begin{equation}
 \bar\psi^+(t)\psi^-(u)\sim\frac{1}{t-u},
 \label{eq:F6-app-fermion-ope}
\end{equation}
Wick's theorem gives
\begin{equation}
 \langle
 \bar\psi^+(t_1)\bar\psi^+(t_2)\psi^-(1)\psi^-(u)
 \rangle
 =\underbrace{\frac{1}{t_1-1}\frac{1}{t_2-u}}
 _{(4/7\ \text{with }Q_7)(1/7\ \text{with incoming }1/6)}
 -\underbrace{\frac{1}{t_1-u}\frac{1}{t_2-1}}
 _{(4/7\ \text{with incoming }1/6)(1/7\ \text{with }Q_7)}.
 \label{eq:F6-app-four-fermion-wick}
\end{equation}
{The two Wick terms reduce the calculation to the four
pairwise residues evaluated below.}

For the contractions with the fermion at $t=1$, the pole
$1/(t-1)$ in Eq.~\eqref{eq:F6-app-fermion-ope} cancels the explicit
factor $(t-1)$ in Eqs.~\eqref{eq:F6-app-outgoing-47}--\eqref{eq:F6-app-outgoing-17}.
Using
\begin{align}
 t^2\left(1-\frac{7}{6t}\right)^{-3/7}
 &=t^2+\cdots+\frac5{12}+\frac{85}{216}t^{-1}+O(t^{-2}),
 \label{eq:F6-app-series-47}\\
 t^{-1}\left(1-\frac{7}{6t}\right)^{-6/7}
 &=t^{-1}+O(t^{-2}),
 \label{eq:F6-app-series-17}
\end{align}
{In the bra-contour convention used throughout this
appendix, $\oint_\infty dt/(2\pi i)$ denotes extraction of the $t^{-1}$
coefficient of the transformed bra-mode integrand. With this convention,
the two residues are}
\begin{align}
 \oint_\infty\frac{dt}{2\pi i}\,
 t^2\left(1-\frac{7}{6t}\right)^{-3/7}
 &=\frac{85}{216},\nonumber\\
 \oint_\infty\frac{dt}{2\pi i}\,
 t^{-1}\left(1-\frac{7}{6t}\right)^{-6/7}
 &=1.
 \label{eq:F6-app-Q-column}
\end{align}

For the contractions with the incoming $1/6$ mode, expand
$1/(t-u)=t^{-1}\sum_{k\ge0}(u/t)^k$ for the nested contours
$|u|<|t|$.  Since
\begin{equation}
 u^{-1}\left(1-\frac{6u}{7}\right)^{-1/6}
 =u^{-1}+O(u^0),
\end{equation}
only the $k=0$ term contributes to the $u$ residue.  The two double
contours therefore reduce to
\begin{align}
 &\oint_\infty\frac{dt}{2\pi i}\,
 \frac{t-1}{t}\,
 t^2\left(1-\frac{7}{6t}\right)^{-3/7}
 =\frac{85}{216}-\frac5{12}=-\frac5{216},
 \nonumber\\
 &\oint_\infty\frac{dt}{2\pi i}\,
 \frac{t-1}{t}\,
 t^{-1}\left(1-\frac{7}{6t}\right)^{-6/7}
 =1-0=1.
 \label{eq:F6-app-incoming-column}
\end{align}
Equivalently, the first number is the difference between the
$t^{-1}$ and $t^0$ coefficients in Eq.~\eqref{eq:F6-app-series-47}; the
second uses the corresponding coefficients $1$ and $0$ in
Eq.~\eqref{eq:F6-app-series-17}.

Ordering the rows by outgoing energies $4/7,1/7$ and the columns by the
$Q_7$ fermion and the incoming $1/6$ mode, the actual Wick contractions
are therefore
\begin{equation}
 \begin{array}{c|cc}
 &Q_7\text{ fermion at }t=1&\psi^-_{-1/6}\text{ at }t=0\\\hline
 4/7&85/216&-5/216\\
 1/7&1&1
 \end{array},
 \qquad
 \det=\frac{85}{216}+\frac5{216}=\frac5{12}\ne0.
 \label{eq:F6-app-explicit-wick-determinant}
\end{equation}
{This reproduces
Eq.~\eqref{eq:F6-main-fermion-determinant}.}

The other fermion sector contains only one incoming--outgoing pair.
Its corresponding double contour is
\begin{equation}
 \oint_\infty\frac{dt}{2\pi i}\oint_0\frac{du}{2\pi i}\,
 \frac{(1-1/t)(1-7/(6t))^{-6/7}
       u^{-1}(1-6u/7)^{-1/6}}{t-u}=1,
\end{equation}
so this sector is also nonzero.

For each propagated boson, use

\begin{align}
 b(t)&=t^{-1}(1-6t/7)^{-1/6}, \nonumber \\
 o_B(t)&=t(1-7/(6t))^{1/7}, \nonumber \\
 \mathcal D(b)&=\mathcal H(o_B,b)=1.
\end{align}
{For the $\alpha^2$ oscillators, there are eight choices for
the incoming boson contracted with the descendant and $7!$ bijections from
the remaining incoming bosons to the seven outgoing slots, giving
$8\cdot7!$.  All eight $\bar\alpha^2$ bosons propagate to outgoing slots,
giving $8!$ bijections.} Combining these assignments with the
fermion determinant gives the stripped left-moving residue
\begin{equation}
 \pm(8!)^2\frac5{12}\ne0.
\end{equation}

The right-moving charges are those displayed in the cover table above. With
$(t_0,t_a,t_1,t_\infty)=(0,a,1,T)$ and $a=7/6$, their spin-field
correlator has magnitude
\begin{equation}
 \lim_{T\to\infty}T^{1/2}
 a^{1/2}T^{-1/2}(a-1)^{-1/2}(T-a)^{-1/2}(T-1)^{1/2}
 =\sqrt{\frac a{a-1}}=\sqrt7.
\end{equation}
Thus the product of the stripped left- and right-moving factors has
magnitude $(8!)^2 5\sqrt7/12>0$. Restoring the nonzero cover, norm and
orbit factors gives Eq.~\eqref{eq:F6-finite-w-amplitude}.

Let $\Pi_{0,M}$ be the projector onto the free right-BPS space at rank $M$.
Expand the exact right supercharge and a total left generator as
\begin{align}
 \mathcal Q_M(g)&=\mathcal Q_{0,M}+g\mathcal Q_{1,M}+O(g^2), \nonumber \\
 \mathcal O_M(g)&=\mathcal O_{0,M}+g\mathcal O_{1,M}+O(g^2).
\end{align}

Left and right superconformal generators graded-commute in the deformed CFT.
Taking the term linear in $g$ gives

\begin{equation}
 [\mathcal Q_{1,M},\mathcal O_{0,M}\}
 =-[\mathcal Q_{0,M},\mathcal O_{1,M}\}.
\end{equation}

Now project both sides to the free right-BPS space.  Since
$\mathcal Q_{0,M}$ annihilates this space from either side,

\begin{align}
 \mathcal Q_{0,M}\Pi_{0,M}&=\Pi_{0,M}\mathcal Q_{0,M}=0, \nonumber \\
 \Pi_{0,M}[\mathcal Q_{0,M},\mathcal O_{1,M}\}\Pi_{0,M}&=0.
\end{align}

The zeroth-order total left generators preserve the free BPS space and its
orthogonal complement \cite{ChangLinZhang}, so they commute with
$\Pi_{0,M}$.  With
$Q_M=\Pi_{0,M}\mathcal Q_{1,M}\Pi_{0,M}$, the projected Ward identity
therefore becomes

\begin{equation}
 [Q_M,\Pi_{0,M}\mathcal O_{0,M}\Pi_{0,M}\}
 =\Pi_{0,M}[\mathcal Q_{1,M},\mathcal O_{0,M}\}\Pi_{0,M}=0.
\end{equation}

Applying the same reasoning to the adjoint identity gives

\begin{align}
 [Q_M,\mathcal O_r^{({\rm T},M)}\}&=0, \nonumber \\
 [Q_M^{\dagger},\mathcal O_r^{({\rm T},M)}\}&=0
 \quad\text{on }V_M. \label{eq:left-affine-Q-exact-commutator}
\end{align}
Thus total left creation operators map harmonic BPS representatives to
harmonic BPS representatives.

The chiral--chiral scalar seeds themselves lack the spin-shifted
recombination partner in the $T^4$ sector, and the same symmetry protects
their Clifford descendants \cite{ChangLinZhang}.  After adding unexcited length-one NS vacuum strands and symmetrizing, choose representatives $\Omega_M$ such that
\begin{align}
 Q_M\Omega_M&=Q_M^{\dagger}\Omega_M=0, \nonumber \\
 \pi_{N,M}\Omega_M&=\Omega_N.
\end{align}

For a fixed word $A$, let $A_M$ denote the same ordered word in total
rank-$M$ generators.  Equation~\eqref{eq:left-affine-Q-exact-commutator}
then gives Eq.~\eqref{eq:tower-representatives-harmonic}.
It remains to check that removing the added vacuum strands recovers the
desired lower-rank word.  Split every total generator into a piece acting on the
original $N$ strands and a piece acting on the $M-N$ added vacuum strands,
$\mathcal O^{({\rm T},M)}=\mathcal O^{({\rm T},N)}+\mathcal O^{({\rm sp})}$.
Any negative mode acting on an added vacuum strand creates
positive NS excitation energy, so its overlap with the vacuum used in the
projection is zero.  The allowed zero-mode lowerings also annihilate the NS
vacuum.  Hence every term in which the total generator acts nontrivially on
an added strand disappears when that strand is removed:

\begin{equation}
 \langle0|^{\otimes(M-N)} A_{\rm sp}|0\rangle^{\otimes(M-N)}=0
 \quad\text{for every nonempty added-strand creation word}.
\end{equation}
Only the identity action on the added strands survives, giving
Eq.~\eqref{eq:intrinsic-tower-explicit-lift}.
Equations~\eqref{eq:tower-representatives-harmonic} and
\eqref{eq:intrinsic-tower-explicit-lift} provide an explicit closed lift at
every $M>N$, proving Eq.~\eqref{eq:intrinsic-tower-definition}.

With $|R\rangle=|6;R_{--}\rangle$, define
\begin{align}
 f_r^\dagger&=\psi^-_{-r/6}, \nonumber \\
 \bar f_r^\dagger&=\bar\psi^-_{-r/6},\qquad r>0.
\end{align}

The current convolution is

\begin{align}
 J^-_{-3}|R\rangle
 &=\sum_{r\in\mathbb Z}
       :\!\psi^-_{-r/6}\bar\psi^-_{-(18-r)/6}\!:\,|R\rangle
 =\sum_{r=1}^{17}f_r^\dagger\bar f_{18-r}^\dagger|R\rangle, \nonumber \\
 \psi^-_0|R\rangle&=\bar\psi^-_0|R\rangle=0, \nonumber \\
 \{f_r,f_s^\dagger\}&=\{\bar f_r,\bar f_s^\dagger\}=\delta_{rs}.
\end{align}

The zero modes kill the endpoints; normal ordering kills $r<0$ and $r>18$. Hence

\begin{equation}
 \begin{aligned}
 \langle R|\bar f_{18-r}f_r
       f_s^\dagger\bar f_{18-s}^\dagger|R\rangle
 &=\delta_{rs}, \\
 \|J^-_{-3}|R\rangle\|^2
 &=\sum_{r,s=1}^{17}\delta_{rs}=17, \\
 \frac r6+\frac{18-r}{6}&=3.
\end{aligned}
\end{equation}

{Here $k=6$ is the affine current-algebra level carried by
the length-six strand.}
\begin{equation}
 \|J^-_{-3}R^{--}_6\|^2
 =\langle R^{--}_6|[J^+_3,J^-_{-3}]|R^{--}_6\rangle
 =3k+J_L(R^{--}_6)=18-1=17.
\end{equation}
The current adds $(N_p,J_L)=(3,-2)$ to the same ground as $F_6$.
Spectral flow takes the descendant to
$17^{-1/2}J^-_{-2}\Omega_6^{\rm NS}$, establishing its membership in
$\mathcal T_6$. The continuation of total generators above then gives
Eq.~\eqref{eq:M0-explicit-every-rank-lift}.

For the independent projection of $F_6$, total generators preserve cycle
shape. Only length-six seeds contribute, with
\begin{equation}
 h_0=j_0\in\left\{\frac52,3,\frac72\right\}.
\end{equation}

At $h=9/2$ their descendant level is $\ell=9/2-h_0\le2$.
The nonzero creation modes and their maximal bosonic degrees obey

{In the following bounds, $n_B(\mathcal O_{-r})$ denotes
the largest number of elementary bosonic oscillators appearing when the
generator mode $\mathcal O_{-r}$ is expanded in free fields.}
\begin{align}
 r&\in\mathbb Z_{>0}\quad\text{for }\alpha_{-r},J_{-r},L_{-r}, \nonumber \\
 r&\in\mathbb Z_{\ge0}+\tfrac12\quad\text{for }\psi_{-r},G_{-r}.
\end{align}

\begin{align}
 n_B(\alpha_{-r})&\le1, \nonumber \\
 n_B(G_{-r})&\le1, \nonumber \\
 n_B(L_{-r})&\le2, \nonumber \\
 n_B(J_{-r})&=n_B(\psi_{-r})=0{.}
\end{align}

Thus $n_B(\mathcal O_{-r})\le2r$; contractions can only reduce this
number, so every such descendant satisfies

\begin{equation}
 n_B\le2\ell\le4.
\end{equation}
The sixteen-boson state $F_6$ is orthogonal to this entire space.

\subsection{Length-two composites}
\label{app:F2-heavy-family}

{This subsection reduces the all-$k$ obstruction to the local
length-two calculation. It first fixes the diagonal and relative Clifford
modes, then lists
all source channels that can reach the chosen target row, and finally proves
that the sum of the local $N=3$ projection overlaps equals the global
$2k\to2k+1$ projection overlap.}

The complete local $2\to3$ obstruction, including the unique source and
nonzero covering-space matrix element, is given in
Eqs.~\eqref{eq:F2-local-source-ledger}--\eqref{eq:F2-local-nonzero-amplitude}.

For the $k$-copy family, let $\eta_r=\widetilde\psi_r^+$ and
$\bar\eta_r=\widetilde{\bar\psi}{}^+_r$ be normalized right Clifford
creators on the $r$th two-cycle. One orthonormal choice is
\begin{align}
 \eta_D&=\frac1{\sqrt{k}}\sum_{r=1}^k\eta_r, \nonumber \\
 \eta_{\perp,a}&=\frac{\sum_{r=1}^a\eta_r-a\eta_{a+1}}
                        {\sqrt{a(a+1)}},\qquad 1\le a<k,
\end{align}

with $\bar\eta_D=k^{-1/2}\sum_r\bar\eta_r$.
Up to a common phase,
$\eta_D\prod_{a=1}^{k-1}\eta_{\perp,a}=\eta_1\cdots\eta_k$.
Including $\bar\eta_D$ gives the state in
{Eq.~\eqref{eq:heavy-allF-state}. For $k=2$ and $k=3$, the representatives are}

\begin{align}
 F_4^{\rm heavy}
 &\propto\sum_{\sigma\in S_4}\sigma\!\left[
   \frac{\bar\eta_1+\bar\eta_2}{\sqrt2}\eta_1\eta_2
   (F_2)_1(F_2)_2\right], \nonumber \\
 F_6^{\rm heavy}
 &\propto\sum_{\sigma\in S_6}\sigma\!\left[
   \frac{\bar\eta_1+\bar\eta_2+\bar\eta_3}{\sqrt3}
   \eta_1\eta_2\eta_3(F_2)_1(F_2)_2(F_2)_3\right].
\end{align}

The composite construction of Ref.~\cite{ChangLinZhang} fills the
relative modes at maximal right outer charge. Joining targets are then
excluded by charge conservation, and splitting or its adjoint acts on
one harmonic factor at a time. This gives the harmonicity in
Eq.~\eqref{eq:heavy-main-family-ledger}.
The stabilizer and orbit norm are

\begin{align}
 |H|&=2^kk!, \nonumber \\
 \left\|\sum_{\sigma\in S_{2k}}\sigma f\right\|^2
 &=(2k)!\,2^kk!,
\end{align}
fixing the prefactor in Eq.~\eqref{eq:heavy-allF-state}.

Let $\eta_r,\bar\eta_r$ denote right Clifford creators with outer charges
$+1/2,-1/2$, and let $\widetilde K$ denote the corresponding right outer
$SU(2)$ Cartan grading used in the composite construction. For $c$ cycles of total winding $n$ and $r$ creators,
\begin{align}
 \bar j_0&=\frac{n-c}{2}, \nonumber \\
 \bar j&=\frac{n-c+r}{2}, \nonumber \\
 \bar j=\frac n2&\Longrightarrow r=c, \nonumber \\
 \widetilde K(\eta_D\bar\eta_D)&=0, \nonumber \\
 |\widetilde K|&\le\frac{c-2}{2}.
\end{align}

{The last inequality follows because the neutral diagonal
pair $\eta_D\bar\eta_D$ uses two of the $c$ Clifford creators, while each of
the remaining $c-2$ creators contributes at most $1/2$ in magnitude to
$\widetilde K$.}

At $n=2k+1$, $\widetilde K=(k-1)/2$ implies $c\ge k+1$; this is the
charge bound used in the precursor reduction displayed in the main text.

Define the local harmonic states $\chi=\eta F_2$ and
$\varphi=\bar\eta\eta F_2$, and choose the ordered-cycle phase so that
\begin{align}
 R_k&=\bar\eta_{d_0}\Omega_k, \nonumber \\
 \Omega_k&=\eta_1\wedge\cdots\wedge\eta_k, \nonumber \\
 d_0&=\frac{(1,\ldots,1)}{\sqrt{k}}.
\end{align}

{Let $C_{k,i}$ contract the $k-1$ spectator two-cycles with
$\bra\chi$ and project the retained $(i,s)$ subsystem to its diagonal
top. Work first with ordered cycles and test $Q_{2k+1}X$ against a
length-three target tensored with $\chi^{\otimes(k-1)}$; afterwards
sum over the orbifold orbit. A transposition that joins the active
length-two cycle and the singleton gives the local row
$\langle T^{(2)}_3|Q_3C_{k,i}$. A transposition joining two singletons
instead creates one of the spectator cycles and is killed by
$\bra\chi Q_2=0$. The two split precursors have $k-1$ source cycles
and are absent by the charge bound above. These exhaust the ways to
reach $(3,2^{k-1})$; all other source partitions pair to zero.}

{
Both the diagonal Clifford creation and annihilation operators
supercommute with the local deformation supercharge
\cite{ChangLinZhang}. Their occupation projectors therefore commute
with it, so selecting the matching top component in the active pair
commutes with this local row calculation.
The spectator $\chi$ is even: moving the odd supercharge past any
spectator produces no Koszul sign. Permuting equal spectator cycles
therefore gives the same local row, multiplied only by positive orbit
and transposition counts. After absorbing that common nonzero factor
into the target normalization, the required identity is
\begin{equation}
 \langle\{T^{(2)}_3,\chi^{k-1}\}|Q_{2k+1}|X\rangle
 =\sum_{i=1}^k\langle T^{(2)}_3|Q_3C_{k,i}|X\rangle.
\end{equation}
Here $T^{(2)}_3$ denotes the matching local Clifford component, normalized
so that its amplitude on the unique local source is
$24C_{2\to3}$ as in Eq.~\eqref{eq:F2-local-nonzero-amplitude}.
The uniqueness of that source converts its coefficient to
$\langle\varphi|\pi_{2,3}C_{k,i}|X\rangle$. This proves precisely
the row reduction used in Eq.~\eqref{eq:heavy-target-row-witness}; it
makes no assertion that the combined contraction
$C_k:=\sum_{i=1}^kC_{k,i}$ satisfies
$C_kQ_{2k+1}=Q_3C_k$ on other channels.
}

{For the surviving $(2^k,1)$ source, first fix the
left-moving weight sector selected by the projection onto
$F_{2k}^{\rm heavy}$ and by the spectator contractions with $\bra\chi$.
The local weight space is $\mathbb C F_2$, as in
Eq.~\eqref{eq:F2-local-source-ledger}; the other local and spectator
components give zero in both overlaps below. It remains to resolve the
right-moving Clifford top in this selected sector. Let $s$ denote the
added length-one strand. With the fixed left-moving factors suppressed,
a general vector in this top space is}
{
\begin{equation}
 X(v)=\bar\eta_{\boldsymbol d}\,\iota_v
 (\eta_1\wedge\cdots\wedge\eta_k\wedge\eta_s)|0\rangle,
 \qquad
 v^{(a)}=e_a-\sqrt2e_s,\quad a=1,\ldots,k.
 \label{eq:heavy-top-source-basis}
\end{equation}
}
{Here $e_1,\ldots,e_k,e_s$ is the orthonormal basis that
labels the $k$ two-cycles and the singleton, $\iota_v$ is interior
contraction by $v$, and
\begin{equation*}
 \Omega_{k+1}:=\eta_1\wedge\cdots\wedge\eta_k\wedge\eta_s,
 \qquad
 \boldsymbol d:=\frac{(\sqrt2,\ldots,\sqrt2,1)}{\sqrt{2k+1}}.
\end{equation*}
The coefficients in $\boldsymbol d$ are the square roots of the component
windings, so $\bar\eta_{\boldsymbol d}$ is the normalized diagonal barred
creator at total winding $2k+1$.}
{The vectors $v^{(a)}$ form a complete basis of the
selected right-moving top space, so it suffices to check the overlaps on
this basis.} Contract the $k-1$ spectator two-cycles and let $C_{k,i}$
retain the $i$th two-cycle together with the length-one strand.  With
$\varphi=\bar\eta\eta F_2$, the following exterior-algebra calculation
relates the global projection to the local $N=3$ projection.

With $P_s$ the Clifford-vacuum projection on the length-one strand, removal and
recreation of the diagonal pair gives
\begin{equation}
 \pi_{\rm top}X(v)
 =\bar\eta_{d_0}\eta_{d_0}P_s\iota_{\boldsymbol d}\iota_v\Omega_{k+1}.
\end{equation}

{A general vector in the span of the $v^{(a)}$ has
$v=\sum_{i=1}^k v_i e_i+v_se_s$. Setting $S=\sum_i v_i$, the basis
definition gives $v_s=-\sqrt2S$.} The surviving $(i,s)$ minors give

\begin{align}
 &\begin{aligned}
 m_i&=(\boldsymbol d)_sv_i-(\boldsymbol d)_iv_s=\frac{v_i+2S}{\sqrt{2k+1}}, \\
 \sum_i m_i&=\sqrt{2k+1}\,S, \\
 P_s\iota_{\boldsymbol d}\iota_v\Omega_{k+1}
 &=(-1)^{k-1}\iota_m\Omega_k, \\
 \eta_{d_0}\wedge\iota_m\Omega_k
 &=\frac{\sum_i m_i}{\sqrt{k}}\Omega_k.
\end{aligned} \nonumber \\
 \shortintertext{Since $\|R_k\|=1$,}
 \langle R_k|\pi_{\rm top}X(v)\rangle
 &=\frac{(-1)^{k-1}}{\sqrt{k}}\sum_i m_i
 =(-1)^{k-1}\sqrt{\frac{2k+1}{k}}S. \label{eq:appendix-global-exterior-overlap}
\end{align}

For the retained local pair, the normalized diagonal vector is
$d_{\rm loc}=(\sqrt2,1)/\sqrt3$. The restricted barred diagonal
and unbarred form give

{
\begin{equation}
 \begin{aligned}
 \bar\eta_{\boldsymbol d}\big|_{(i,s)}
 &=\sqrt{\frac3{2k+1}}\bar\eta_{d_{\rm loc}}, \\
 \iota_{\rm spectators}\iota_v\Omega_{k+1}
 &\propto-v_s\eta_i+v_i\eta_s, \\
 \sqrt{\frac3{2k+1}}\,
 d_{\rm loc}\cdot(-v_s,v_i)
 &=\frac{v_i-\sqrt2v_s}{\sqrt{2k+1}}=m_i.
 \end{aligned}
\end{equation}
}
The common spectator-ordering sign gives the local overlap
\begin{equation}
 \langle\varphi|\pi_{2,3}C_{k,i}|X(v)\rangle=(-1)^{k-1}m_i.
\end{equation}

Evaluating the two overlaps on the basis $v^{(a)}$ gives
\begin{align}
 \langle F_{2k}^{\rm heavy}|\pi_{2k,2k+1}|X(v^{(a)})\rangle
 &=(-1)^{k-1}\sqrt{\frac{2k+1}{k}}, \nonumber\\
 \langle\varphi|\pi_{2,3}C_{k,i}|X(v^{(a)})\rangle
 &=\frac{(-1)^{k-1}}{\sqrt{2k+1}}(2+\delta_{ia}).
 \label{eq:heavy-global-local-overlaps}
\end{align}
{For each basis vector $v^{(a)}$,
$\sum_i(2+\delta_{ia})=2k+1$. Hence the sum of the local overlaps in
Eq.~\eqref{eq:heavy-global-local-overlaps} is
$(-1)^{k-1}\sqrt{2k+1}$, which equals $\sqrt{k}$ times the global overlap.
Linearity, together with the vanishing of the other source components
established above, therefore gives for every source $X$ in the prescribed
charge block}
\begin{equation}
 \sum_{i=1}^k
 \langle\varphi|\pi_{2,3}C_{k,i}|X\rangle
 =\sqrt{k}\,
 \langle F_{2k}^{\rm heavy}|\pi_{2k,2k+1}|X\rangle.
 \label{eq:main-all-k-identity}
\end{equation}
\section{Uniform bounds for the fixed-charge asymptotics}
\label{app:fixed-charge-technical-bounds}

{This appendix supplies the uniform estimates that permit
the coefficient limits in Section~\ref{sec:long-strings-lifting} to be
summed over the finite-winding remainder. We first control repeated
occupations and transfer the $r^{-3}$ single-cycle tail through the Fock
exponential. We then establish the two-dimensional local limit in energy
and charge and use it to bound the large-remainder tail uniformly in $N$.}

\subsection{Finite-rank census and index conventions}
\label{app:finite-N-partition-census}

The finite-$N$ example quoted in Section~\ref{sec:sector-ensemble-construction}
is recorded here as a check on the partition-resolved positive counting. At
$N=6$ and $(N_p,J_L)=(3,-3)$, the full eleven-partition census is listed in
Table~\ref{tab:N6-partition-census}.
\begin{table}[!htbp]
\centering\small
\setlength{\tabcolsep}{4pt}
\begin{tabular}{@{}crr|rr@{}}
$p$ & $w_{\max}$ & $d$ & $\dim V_{p,\Gamma}$ & fraction of $V_{6,\Gamma}$ \\ \hline
$(6)$ & 6 & 0 & $53\,245\,800$ & 5.131856\% \\
$(5,1)$ & 5 & 1 & $145\,811\,168$ & 14.053351\% \\
$(4,2)$ & 4 & 2 & $36\,531\,872$ & 3.520959\% \\
$(4,1^2)$ & 4 & 2 & $192\,381\,344$ & 18.541807\% \\
$(3^2)$ & 3 & 3 & $12\,410\,880$ & 1.196167\% \\
$(3,2,1)$ & 3 & 3 & $115\,627\,136$ & 11.144199\% \\
$(3,1^3)$ & 3 & 3 & $176\,187\,552$ & 16.981041\% \\
$(2^3)$ & 2 & 4 & $8\,386\,196$ & 0.808266\% \\
$(2^2,1^2)$ & 2 & 4 & $113\,200\,640$ & 10.910332\% \\
$(2,1^4)$ & 2 & 4 & $149\,238\,912$ & 14.383718\% \\
$(1^6)$ & 1 & 5 & $34\,532\,960$ & 3.328303\% \\ \hline
Total & & & $1\,037\,554\,460$ & $100\%$
\end{tabular}
\caption{Exact free right-BPS counts at $N=6$ and $(N_p,J_L)=(3,-3)$,
summed over the right Ramond ground multiplet.  The graded symmetric-product
Fock space already implements permutation invariance; no additional orbit
multiplicity is inserted.}
\label{tab:N6-partition-census}
\end{table}
As an independent check, the right-charge helicity trace gives
\begin{align}
 \frac12\sum_t(-1)^{-3+t}t(t-1)\dim V_{6,(3,-3);J_R=t}
 &=13\,311\,567, \nonumber\\
 &=\widehat c(18,-3)+3\widehat c(2,-1),\qquad
 \widehat c(2,-1)=39.
 \label{eq:N6-index-check-main}
\end{align}
The $(2^3)$ entry is obtained from the graded symmetric cube of the
length-two single-cycle spectrum, providing an independent check that the
Fock-space product is counting indistinguishable component strings rather
than labeled copies.

\subsection{Uniform asymptotic estimates}

Only the uniform estimates used to justify the coefficient limits of
Section~\ref{sec:long-strings-lifting} are collected in this subsection.
The counting identities and the use of these estimates are given in the
main text.

{
At the uncondensed saddle, Eq.~\eqref{eq:positive-product-from-character}
separates the first occupation from the higher odd repetitions:
\begin{align}
 &\log Z(ze^{-\beta_*\nu},e^{-\beta_*},e^{\mu_*})
       -\sum_{w\ge1}a_wz^w\nonumber\\
 &\qquad=\sum_{w\ge1}\sum_{\substack{k\ge3\\k\ {\rm odd}}}
       \frac{z^{kw}}{k}e^{-k\beta_*\nu w}
       Z_w^{\rm phys}(k\beta_*,k\mu_*).
\end{align}
All coefficients are nonnegative. Applying
Eq.~\eqref{eq:no-pole-chamber} to each species gives
\begin{equation}
 e^{-k\beta_*\nu w}Z_w^{\rm phys}(k\beta_*,k\mu_*)
 \le e^{-(k-1)\beta_*(\nu-|\jmath|/2)w}a_w.
\end{equation}
For total degree $r=kw$ with $k\ge3$, the extra exponent is at least
$2\beta_*(\nu-|\jmath|/2)r/3$. Summing over divisors is harmless since
$\sum_w wa_w<\infty$. The logarithmic remainder above, and its
exponential, are therefore analytic in a disk of radius greater than one.
Their coefficients are consequently bounded by $CR^{-r}$ for some $R>1$,
so they cannot modify the algebraic $r^{-3}$ tail of the $k=1$ term.
}

{
The sequence $a_w$ is positive for every integer $w\ge1$ and satisfies
$a_w\sim(\delta w^3)^{-1}$. Thus its support is not confined to multiples
of an integer larger than one, and all coefficient limits below are taken
along the full integer sequence.
Splitting a two-fold convolution into its two ends and its middle gives
\begin{align}
 \sum_{j=L}^{r-L}a_ja_{r-j}
 &\le\frac{Ca_r}{(1+L)^2},\qquad 1\le L<r/2,\nonumber\\
 \sum_{j=1}^{r-1}a_ja_{r-j}
 &\sim2a_r\sum_{j\ge1}a_j.
\end{align}
The second line follows by taking $r\to\infty$ at fixed $L$, using
$a_{r-j}/a_r\to1$ at each end, and then taking $L\to\infty$.
Induction gives the analogous $k$-fold coefficient
$k(\sum_j a_j)^{k-1}a_r$. For a uniform bound, one part in every
composition of $r$ is at least $r/k$. There are at most $k$ choices for
that part, and $a_m\le Ck^3a_r$ for $m\ge r/k$ follows from the uniform
$m^{-3}$ bound. Summing each of the other $k-1$ parts over
$\sum_w a_w$ yields
\begin{equation}
 [z^r]\left(\sum_{w\ge1}a_wz^w\right)^k
 \le Ck^4\left(\sum_{w\ge1}a_w\right)^{k-1}a_r.
\end{equation}
Division by $k!$ makes this bound summable. Dominated convergence in
the exponential, followed by convolution with the analytic repeated-
occupation factor, proves Eq.~\eqref{eq:remainder-subexponential} and
the uniform coefficient bound $C(1+r)^{-3}$. In other words, the
coefficient of the full Fock exponential has the same $r^{-3}$ tail as
the one-cycle coefficient, multiplied by the finite normalization from
the remaining cycles. This is the convergent weighted-partition estimate
of Ref.~\cite{StuflerConvergent}, verified here directly for the paired
graded Fock product.
}

{
At the critical saddle, remove the two unit-activity bosons and factor
out their fermionic partners as $(1+z)^2$. The remaining product is over
Bose--Fermi pairs with strictly positive cost per unit winding:
$C(w,E,J)\ge\pi^2w/\beta-\beta/(4w)$ controls large $w$, while
$C(w,E,J)\ge(\beta-2|\mu|)E$ and the nonzero remaining ground-state
costs control the finitely many small windings. This positive gap makes
higher repetitions exponentially small in their total degree.

The first-occupation coefficients of this remaining paired product
differ from $e^{-|\mu|w}Z_w^{\rm phys}(\beta,\mu)$ only at $w=1$.
By Eq.~\eqref{eq:projected-canonical-asymptotic} and
$(\pi^2+\mu^2)/\beta=|\mu|$, their large-$w$ tail is therefore
$\beta^2/(\pi^2w^3)$. Regular Ramond ground states with $E=J=0$
remain at every winding, so these coefficients are positive at every
positive integer winding. The logarithmic coefficients retain the same tail because
the higher repetitions are exponentially small. {The preceding
convolution proof therefore applies to the remaining paired product.} Restoring
$(1+z)^2$ multiplies its finite normalization by $(1+1)^2=4$ and gives
\begin{equation*}
 \begin{aligned}
 &[z^r](1-z)^2Z(ze^{-|\mu|},e^{-\beta},e^\mu)\\
 &\qquad\sim\frac{\beta^2}{\pi^2r^3}
 \lim_{z\to1^-}(1-z)^2Z(ze^{-|\mu|},e^{-\beta},e^\mu).
 \end{aligned}
\end{equation*}
The same proof gives the uniform bound $C(1+r)^{-3}$ and a finite first
winding moment. This supplies the residual coefficient transfer used
in Section~\ref{sec:phase-and-moments}.
}

{Applied to the unsigned oscillator character at \(q=e^{-\beta/w}\),}
Euler--Maclaurin on the fermionic sums and the Dedekind product on the
bosonic sums give
\begin{align}
 2\sum_{m\ge1}\!\left[
 \log(1+e^{\mu-\beta m/w})
 +\log(1+e^{-\mu-\beta m/w})\right]
 &=
 \frac{w}{\beta}\left(\frac{\pi^2}{3}+\mu^2\right)
 -\log\!\left[4\cosh^2\!\frac{\mu}{2}\right]
 +O(w^{-1}), \nonumber\\
 -4\sum_{m\ge1}\log(1-e^{-\beta m/w})
 &=
 \frac{2\pi^2w}{3\beta}
 +2\log\frac{\beta}{2\pi w}
 +O(w^{-1}),
 \label{eq:positive-character-EM-main}
\end{align}
which combine to the prefactor \((\beta/2\pi w)^2\) used in
Section~\ref{sec:sector-ensemble-construction}.  The second derivatives of
the same exponent give the charge covariance per unit winding,
\begin{equation}
 \begin{pmatrix}
 2(\pi^2+\mu_*^2)/\beta_*^3&2\mu_*/\beta_*^2\\
 2\mu_*/\beta_*^2&2/\beta_*
 \end{pmatrix},
 \qquad
 \det=\frac{4\pi^2}{\beta_*^4}>0,
\end{equation}
so both charge directions have Gaussian width \(O(\sqrt w)\).

In the fixed-charge divisor formula, expanding the double pole as
\((1-x)^{-2}x=\sum_{k\ge1}kx^k\) and extracting the coefficient of
\(p^Nq^{N_p}y^{J_L}\) gives
\begin{equation}
 I_N(N_p,J_L)
 =\sum_{k\mid\gcd(N,N_p,|J_L|)}
 k\,\widehat c\!\left(\frac{NN_p}{k^2},\frac{J_L}{k}\right).
 \label{eq:divisor-extraction-main}
\end{equation}
Each divisor obeys \(k\le N\), so the prefactor of the \(k\)-th term is
polynomial in \(k\) and \(N\) and contributes \(O(\log N)\) to its
logarithm, as does the number of divisors.  The exponent of the \(k\)-th
term is \(S_{\rm BH}/k\le S_{\rm BH}/2\) for \(k>1\).
{The sum of all terms with $k>1$ is therefore bounded in
absolute value by $e^{S_{\rm BH}/2+O(\log N)}$}, and
\begin{equation}
 I_N(N_p,J_L)
 =\widehat c(NN_p,J_L)
 \left[1+O\!\left(e^{-S_{\rm BH}/2+O(\log N)}\right)\right].
\end{equation}

The center-of-mass factor quoted in Section~\ref{sec:core-counting} is
\begin{align}
 (e^{\pi iv_*}-e^{-\pi iv_*})^4
 &=16\cosh^4\!\left(\frac{\pi J}{4\sqrt{Q_1Q_5n-J^2/4}}\right),
 \nonumber\\
 v_*&=\frac12-\frac{iJ}{4\sqrt{Q_1Q_5n-J^2/4}},
 \label{eq:BMPV-cm-factor-main}
\end{align}
which stays finite at the rotating saddle and therefore affects only the
\(O(1)\) term.

The supergravity envelope of
Section~\ref{sec:core-counting} follows by Legendre transform: minimizing
\(ht+8\pi^4/(3t^3)\) over \(t>0\) at \(t_*=(8\pi^4/h)^{1/4}\) gives
\(\tfrac43\pi\,2^{3/4}h^{3/4}\).

For a nonidentity cyclic insertion, let
\(\ell\in\{1,\ldots,w-1\}\), and set
\(r_m=e^{-\beta m/w}\) and \(\theta_m=2\pi\ell m/w\). A single bosonic
factor obeys
\begin{equation}
 \left|\frac{1-r_m}{1-r_me^{i\theta_m}}\right|
 =\left[1+\frac{2r_m}{(1-r_m)^2}(1-\cos\theta_m)\right]^{-1/2}.
\end{equation}
For $w\le m<2w$, $r_m$ stays in a compact subinterval of $(0,1)$, so
there exists \(c_\beta>0\), depending only on \(\beta\), such that
\begin{align}
 \log\left|\frac{1-r_m}{1-r_me^{i\theta_m}}\right|
 &\le-c_\beta(1-\cos\theta_m),\nonumber\\
 \sum_{m=w}^{2w-1}(1-\cos(2\pi\ell m/w))&=w,
 \qquad 0<\ell<w.
\end{align}
All remaining normalized oscillator factors have modulus at most one, so
\begin{equation}
 \left|\frac{P_+(e^{(-\beta+2\pi i\ell)/w},e^\mu)}
 {P_+(e^{-\beta/w},e^\mu)}\right|\le e^{-c_\beta w}.
 \label{eq:cyclic-nonidentity-uniform-bound}
\end{equation}
This proves the uniform suppression used in
Eq.~\eqref{eq:projected-canonical-asymptotic}.

{
For the joint charge extraction, the exact cyclic projection gives
\begin{equation}
 \Psi_w(t,\phi)=
 \frac{g_R(e^{\mu_*+i\phi})}{g_R(e^{\mu_*})}
 \frac{\displaystyle\sum_{\ell=0}^{w-1}
 P_+(e^{(-\beta_*+i(t+2\pi\ell))/w},e^{\mu_*+i\phi})}
 {\displaystyle\sum_{\ell=0}^{w-1}
 P_+(e^{(-\beta_*+2\pi i\ell)/w},e^{\mu_*})},
 \qquad (t,\phi)\in[-\pi,\pi]^2.
\end{equation}
To control this entire torus, take oscillator levels $w\le m<2w$.
Their positive occupation weights stay in compact intervals away from
zero and, for bosons, from one. Besides the bosonic estimate above, a
fermionic factor with positive weight $a$ obeys
\begin{equation}
 \left|\frac{1+ae^{i\theta}}{1+a}\right|^2
 =1-\frac{2a}{(1+a)^2}(1-\cos\theta).
\end{equation}
{Its logarithm is bounded above by $-c(1-\cos\theta)$},
uniformly for the two charged species in this oscillator window.
For the identity cyclic term the phases are $tm/w$ for neutral bosons
and $tm/w\pm\phi$ for the charged fermions. The elementary inequality
\begin{equation}
 (1-\cos\theta)+(1-\cos(\theta+\phi))+(1-\cos(\theta-\phi))
 \ge\tfrac12\bigl[(1-\cos\theta)+(1-\cos\phi)\bigr]
\end{equation}
and $\sum_{m=w}^{2w-1}(1-\cos(tm/w))\ge cw\min(t^2,1)$
give a bound $\exp[-cw\min(t^2+\phi^2,1)]$ for that normalized term.
The latter sum follows from $1-\cos x\asymp x^2$ near zero and the
strictly positive Riemann-sum limit away from zero on $[-\pi,\pi]$.
}

{
For $\ell\ne0$, reduce $s=(t+2\pi\ell)/w$ modulo $2\pi$ to
$[-\pi,\pi]$. Then $|s|\ge\pi/w$, and for $w\ge2$ the geometric sum gives
\begin{equation}
 \left|\sum_{m=w}^{2w-1}e^{ism}\right|
 \le\frac{1}{|\sin(s/2)|}\le\frac{w}{\sqrt2}.
\end{equation}
It follows explicitly that
\begin{equation}
 \sum_{m=w}^{2w-1}(1-\cos sm)
 =w-\operatorname{Re}\sum_{m=w}^{2w-1}e^{ism}
 \ge\left(1-\frac1{\sqrt2}\right)w.
\end{equation}
The neutral bosons alone therefore suppress every nonidentity term
by $e^{-cw}$, uniformly in $t$ and $\phi$. The ground-state factor and
all unused normalized oscillator factors have modulus at most one.
The denominator is its identity term times $1+O(we^{-cw})$.
Combining the terms, and enlarging constants for finitely many small $w$,
proves
\begin{equation}
 |\Psi_w(t,\phi)|\le C\exp[-cw\min(t^2+\phi^2,1)].
 \label{eq:app-uniform-characteristic-bound}
\end{equation}
{The origin is therefore the only unsuppressed Fourier point
in this torus.} The explicit charge differences $(1,0)$ and $(0,1)$ recorded
in the main text generate the full lattice $\mathbb Z^2$. Hence Fourier
inversion does not split into separate even--odd charge sublattices.
}

{
Near the origin, the same oscillator expansion is analytic and gives
\begin{align}
 \log\Psi_w(t,\phi)
 &=iw(\nu t+\jmath\phi)
 -w\left[
 \frac{\pi^2+\mu_*^2}{\beta_*^3}t^2
 +\frac{2\mu_*}{\beta_*^2}t\phi
 +\frac{1}{\beta_*}\phi^2\right]\nonumber\\
 &\quad+O(|t|+|\phi|)+O(w|(t,\phi)|^3),\nonumber\\
 \det\Sigma_w/w^2&=\frac{4\pi^2}{\beta_*^4}+O(w^{-1}).
\end{align}
Fourier inversion on the integer charge lattice, with
Eq.~\eqref{eq:app-uniform-characteristic-bound} as the dominating bound,
now proves Eq.~\eqref{eq:main-one-string-local-limit}, uniformly for
bounded offsets $(u,v)$. Integrating the same absolute bound gives
$C/w$ for every charge coefficient, including those away from the saddle.
}

{
For $r<N/2$, a cycle of length $N-r$ is unique. Its tilted
single-cycle weight is $O((N-r)^{-3})$, and its conditional probability
at any specified $(E,J)$ is $O((N-r)^{-1})$. {This gives:}
\begin{equation}
 e^{-S_{\rm BH}}\Tr_VP_r
 \le CN^{-4}[z^r]Z(ze^{-\beta_*\nu},e^{-\beta_*},e^{\mu_*}).
\end{equation}
The $C/w$ coefficient bound permits dominated convergence over all
remainder charges at fixed winding $r$, not merely those with bounded
energy. Summing the $r^{-3}$ coefficient bound for $r>R$ gives
$CN^{-4}(1+R)^{-2}$. The no-giant estimate
\eqref{eq:no-giant-fixed-charge-bound} supplies the remaining $O(N^{-6})$
term. Finally the fixed-$r$ limits and this uniform tail give
Eq.~\eqref{eq:main-full-fixed-charge-asymptotic}, without an interchange
of limits unsupported by a tail estimate.
}

\subsection{Projector constructions}
\label{app:projector-constructions}

For a finite-dimensional subspace \(S\), diagonalize
\(P_SP_{\le R}P_Su_i=\lambda_iu_i\).  Since \(0\le\lambda_i\le1\),
\begin{align}
 \sum_i(1-\lambda_i)
 &=\Tr[P_S(1-P_{\le R})P_S],
 \nonumber\\
 \epsilon\,\#\{i:\lambda_i<1-\epsilon\}
 &\le\Tr[P_S(1-P_{\le R})P_S],
 \nonumber\\
 \left\|(1-P_{\le R})\sum_{\lambda_i\ge1-\epsilon}c_iu_i\right\|^2
 &\le\epsilon\sum_i|c_i|^2 .
 \label{eq:general-spectral-bound-main}
\end{align}
{For
$v=\sum_{\lambda_i\ge1-\epsilon}c_iu_i$, diagonalization gives
$\|(1-P_{\le R})v\|^2=\sum_i(1-\lambda_i)|c_i|^2$; each retained
eigenvalue obeys $1-\lambda_i\le\epsilon$, which is the third line. The
second line is the same estimate applied to the number of eigenvalues below
$1-\epsilon$.}
The span of the eigenvectors with \(\lambda_i\ge1-\epsilon\) therefore has
codimension at most \(\epsilon^{-1}\Tr[P_S(1-P_{\le R})P_S]\) in \(S\),
and every unit vector in it satisfies
\(\langle\psi|P_{\le R}|\psi\rangle\ge1-\epsilon\).

For adjacent cochain spaces
$V_-\xrightarrow{Q_-}V_0\xrightarrow{Q_+}V_+$ with Gram matrices $G_-$,
$G_0$, and $G_+$, pass to orthonormal coordinates by
\begin{align}
 q_-&=G_0^{1/2}Q_-G_-^{-1/2},&
 q_+&=G_+^{1/2}Q_+G_0^{-1/2}. \nonumber
\end{align}
{The product $q_-q_-^+$ is the orthogonal projector onto
$\operatorname{im}q_-$, while $q_+^+q_+$ projects onto
$\operatorname{im}q_+^\dagger$. Nilpotence gives $q_+q_-=0$, so these two
image spaces are orthogonal. The complement is the harmonic subspace, and
its projector is}
\begin{equation}
 P_H=1-q_-q_-^+-q_+^+q_+,
 \label{eq:Hodge-projector-main}
\end{equation}
where $+$ denotes the Moore--Penrose inverse.  If the columns of $B$ are the
protected-seed descendants before harmonic projection, their orthogonal
projector inside cohomology is
\begin{equation}
 P_{\mathcal T}=P_HG_0^{1/2}B
 \left[(P_HG_0^{1/2}B)^\dagger(P_HG_0^{1/2}B)\right]^+
 (P_HG_0^{1/2}B)^\dagger.
 \label{eq:protected-tower-projector-main}
\end{equation}
{Indeed, if $X=P_HG_0^{1/2}B$, then
$X(X^\dagger X)^+X^\dagger$ is the orthogonal projector onto
$\operatorname{im}X$. The Moore--Penrose inverse discards zero singular
values, so the formula remains valid when the seed columns are linearly
dependent.}

\end{document}